\documentclass{aa}
\usepackage{graphicx}
\usepackage{txfonts}
\usepackage{natbib}
\usepackage[colorlinks=true, linkcolor=blue, urlcolor=blue, citecolor=blue]{hyperref}
\def\cm#1{\ifmmode {\,{\rm cm^{-#1}}}                  
        \else \hbox{$\,${\rm cm$^{\rm -#1}$}}\fi}
\def\raw {\ifmmode\rightarrow\else$\rightarrow$\fi}
\def\ex#1{\ifmmode {\times 10^{#1}}         
        \else \hbox{{$\times 10^{\rm #1}$}}\fi}

\newcommand{\peryr}{\mbox{yr$^{-1}$}}
\newcommand{\iram}{IRAM-30\,m}
\newcommand{\hso}{{\it Herschel}}
\newcommand{\lsim}{\raisebox{-.4ex}{$\stackrel{<}{\scriptstyle \sim}$}}

\newcommand{\farc}{\mbox{$.\!\!^{\prime\prime}$}}
\newcommand{\s}{\mbox{$''$}}
\newcommand{\mloss}{\mbox{$\dot{M}$}}
\newcommand{\my}{\mbox{$M_{\odot}$~yr$^{-1}$}}
\newcommand{\ls}{\mbox{$L_{\odot}$}}
\newcommand{\msun}{\mbox{$M_{\odot}$}}
\newcommand{\md}{\mbox{$M_{\rm d}$}}

\newcommand{\rs}{\mbox{$R_{\star}$}}
\newcommand{\rd}{\mbox{$r_{\rm d}$}}

\newcommand{\rin}{\mbox{$R_{\rm in}$}}
\newcommand{\rout}{\mbox{$R_{\rm out}$}}
\newcommand{\kms}{\mbox{km\,s$^{-1}$}}
\newcommand{\los}{\mbox{\sl los}}

\newcommand{\vt}{\mbox{$\upsilonup_{\rm t}$}}
\newcommand{\vr}{\mbox{$\upsilonup_{\rm r}$}}
\newcommand{\vorb}{\mbox{$\upsilonup_{\rm orb}$}}
\newcommand{\iorb}{\mbox{$i_{\rm orb}$}}
\newcommand{\ic}{\mbox{$i_{\rm c}$}}
\newcommand{\vgrad}{\mbox{$\nabla \upsilonup$}}
\newcommand{\vexp}{\mbox{$V_{\rm exp}$}}
\newcommand{\vs}{\mbox{$V_{\rm s}$}}
\newcommand{\vrot}{\mbox{$V_{\rm rot}$}}
 
\newcommand{\vsys}{\mbox{$V_{\rm sys}$}} 
\newcommand{\vlsr}{\mbox{$V_{\rm LSR}$}} 
\newcommand{\porb}{\mbox{$P_{\rm orb}$}}

\newcommand{\h}{$^{\rm h}$}
\newcommand{\m}{$^{\rm m}$}

\newcommand{\td}{\mbox{$T_{\rm d}$}}
\newcommand{\teff}{\mbox{$T_{\rm eff}$}}
\newcommand{\trot}{\mbox{$T_{\rm rot}$}}
\newcommand{\tkin}{\mbox{$T_{\rm kin}$}}      
\newcommand{\tdyn}{\mbox{$t_{\rm kin}$}}      
\newcommand{\tgro}{\mbox{$t_{\rm growth}$}}      
\newcommand{\ttra}{\mbox{$t_{\rm travel}$}}      
\newcommand{\dens}{\mbox{$n_{\rm H_2}$}}
\newcommand{\nh}{\mbox{$N_{\rm H_2}$}}

\newcommand{\ntotcs}{\mbox{$N^{\rm tot}_{\rm CS}$}}
\newcommand{\ntotch}{\mbox{$N^{\rm tot}_{\rm CH_3OH}$}}

\newcommand{\nc}{\mbox{$n_{\rm crit}$}}
\newcommand{\hal}{\mbox{H$\alpha$}}

\newcommand{\dv}{\mbox{$\Delta \upsilonup$}}
\newcommand{\eu}{\mbox{E$_{\rm u}$}}

\newcommand{\ta}{\mbox{$T^*_{\rm A}$}}
\newcommand{\tmb}{\mbox{$T_{\rm MB}$}}
\newcommand{\kb}{\mbox{$k_{\rm B}$}}

\newcommand{\qx}{\mbox{QX\,Pup}}
\newcommand{\cs}{\mbox{clump {\sl S}}}
\newcommand{\ohs}{\mbox{OH\,231.8}}
\newcommand{\oh}{\mbox{OH\,231.8$+$4.2}}
\newcommand{\lbol}{\mbox{$L_{\rm bol}$}}
\newcommand{\nbw}{\mbox{$FB_{\rm N}$}}
\newcommand{\sbw}{\mbox{$FB_{\rm S}$}}
\newcommand{\lhg}{\mbox{large hg}}

\newcommand{\water}{\mbox{H$_2$O}}
\newcommand{\docem}{$^{12}$CO}
\newcommand{\trecem}{$^{13}$CO}
\newcommand{\docet}{$^{12}$CO\,($J$=3-2)}
\newcommand{\trecet}{$^{13}$CO\,($J$=3-2)}

\newcommand{\somas}{SO$^+$}

\newcommand{\htresomas}{H$_3$O$^+$}
\newcommand{\metanol}{CH$_3$OH}

\newcommand{\sodos}{SO$_2$}
\newcommand{\tsodos}{$^{34}$SO$_2$}
\newcommand{\tso}{$^{33}$SO}
\newcommand{\sisv}{SiS\,$v$=1}
\newcommand{\tsiov}{$^{30}$SiO\,$v$=1}
\newcommand{\tsio}{$^{30}$SiO}
\newcommand{\vsio}{$^{29}$SiO}
\newcommand{\vsiot}{$^{29}$SiO\,($J$=8-7)}
\newcommand{\siot}{SiO\,($J$=7-6)}
\newcommand{\nacl}{Na$^{37}$Cl}
\newcommand{\ocst}{OCS\,($J$=25-24)}
\newcommand{\sodost}{SO$_2$\,(11$_{6,6}$-12$_{5,7}$)}
\newcommand{\htresomast}{H$_3$O$^+$\,($J^p_K$=1$^-_1$ - 2$^+_{1}$)}
\newcommand{\tsodost}{$^{34}$SO$_2$\,(3$_{3,1}$-2$_{2,0}$)} 
\def\snu#1{\ifmmode {S_\nu\,\propto\,\nu^{#1}}
          \else \hbox{$S_\nu$\,$\propto$\,$\nu^{#1}$}\fi}

\hypersetup{draft}  
\begin{document}



\title{Through the magnifying glass: ALMA acute viewing of the intricate nebular architecture of OH231.8+4.2}


   \author{ C.~S\'anchez Contreras\inst{1},
	  J.~Alcolea\inst{2} 
          \and 
          V.~Bujarrabal\inst{3}
          \and
          A.~Castro-Carrizo\inst{4}
          \and
          L.~Velilla Prieto\inst{5}
          \and
          M.~Santander-Garc\'ia\inst{2}
          \and
          G.~Quintana-Lacaci\inst{5}
          \and
         J.~Cernicharo\inst{5}
   }

       \institute{Centro de Astrobiolog{\'i}a (CSIC-INTA), Postal address:
  ESAC, Camino Bajo del Castillo s/n, Urb. Villafranca del Castillo,
  E-28691 Villanueva de la Ca\~nada, Madrid, Spain\\ \email{csanchez@cab.inta-csic.es}
  \and Observatorio Astron\'omico Nacional (IGN), Alfonso XII
  No 3, 28014 Madrid, Spain
  \and Observatorio Astron\'omico Nacional
  (IGN), Ap 112, 28803 Alcal\'a de Henares, Madrid, Spain
  \and Institut de Radioastronomie Millimetrique, 300 rue de la Piscine, 38406 Saint Martin d’Heres, France
    \and Instituto de Fisica Fundamental (CSIC), C/ Serrano, 123, E-28006, Madrid, Spain  
       }


\abstract{We present continuum and molecular line emission ALMA
  observations of \oh, a well studied bipolar nebula around an
  asymptotic giant branch (AGB) star. The high angular resolution
  ($\sim$0\farc2-0\farc3) and sensitivity of our ALMA maps provide the
  most detailed and accurate description of the overall nebular
  structure and kinematics of this object to date. We have identified
  a number of outflow components previously unknown. Species studied
  in this work include \docem, \trecem, CS, SO, SO$_2$, OCS, SiO, SiS,
  \htresomas, \nacl, and \metanol. The molecules \nacl\ and
  \metanol\ are first detections in \oh, with \metanol\ being also a
  first detection in an AGB star.  Our ALMA maps bring to light the
  totally unexpected position of the mass-losing AGB star (\qx)
  relative to the large-scale outflow. \qx\ is enshrouded within a
  compact ($\lesssim$60\,AU) parcel of dust and gas (clump S) in expansion
  (\vexp$\sim$5-7\,\kms) that is displaced by $\sim$0\farc6 to the
  south of the dense equatorial region (or waist) where the bipolar
  lobes join. Our SiO maps disclose a compact bipolar outflow
  that emerges from \qx's vicinity. This outflow is oriented
  similarly to the large-scale nebula but the expansion velocities are
  about ten times lower (\vexp$\lesssim$35\,\kms). We deduce short
  kinematical ages for the SiO outflow, ranging from $\sim$50-80\,yr,
  in regions within $\sim$150\,AU, to $\sim$400-500\,yr at the lobe
  tips ($\sim$3500\,AU). Adjacent to the SiO outflow, we identify a
  small-scale hourglass-shaped structure (mini-hourglass) that is
  probably made of compressed ambient material formed as the SiO
  outflow penetrates the dense, central regions of the nebula. The
  lobes and the equatorial waist of the mini-hourglass are both
  radially expanding with a constant velocity gradient
  (\vexp$\propto$\,$r$). The mini-waist is characterized by extremely
  low velocities, down to $\sim$1\,\kms\ at $\sim$150\,AU, which
  tentatively suggest the presence of a stable structure. The
  spatio-kinematics of the large-scale, high-velocity lobes (HV
    lobes) and the dense equatorial waist (large waist) known from
  previous works are now precisely determined, indicating that both
  were shaped nearly simultaneously about $\sim$800-900\,yr ago. We
  report the discovery of two large ($\sim$8\arcsec$\times$6\arcsec),
  faint bubble-like structures (fish bowls) surrounding the central
  parts of the nebula . These are relatively old structures although
  probably slightly ($\sim$100-200\,yr) younger than the large waist
  and the HV lobes. We discuss the series of events that may have
  resulted in the complex array of nebular components found in \oh\ as
  well as the properties and locus of the central binary system. The
  presence of $\lesssim$80\,yr bipolar ejections indicate that the
  collimated fast wind engine is still active at the core of this
  outstanding object.}



\titlerunning{ALMA observations of OH\,231.8+4.2}
\authorrunning{S\'anchez Contreras et al.}

   \maketitle
%

\section{Introduction}
\label{intro}

The very short ($\approx$10$^3$\,yr) evolutionary transition from the
asymptotic giant branch (AGB) to the planetary nebula (PN) phase is
accompanied by significant morphological and dynamical changes that
are not fully understood: the roughly round circumstellar envelopes
(CSEs) around AGB stars that result from their slow
($\sim$\,5-15\,\kms), isotropic stellar winds evolve into post-AGB
nebulae (or pre-PNe=pPNe) with high-speed ($\approx$100\,\kms)
outflows and a dazzling variety of shapes and intriguing symmetries.
The primary agent for breaking the spherical symmetry is unknown but
most theories require a central binary system to assist
in the production of collimated fast winds (CFW) or jets \citep[see,
  e.g., the review paper on PN-shaping by][]{bal02}. Together with
pPNe, observationally recognized by prominent aspherical nebulosities
surrounding a central post-AGB star, there are a few objects that show
pPN-like morphologies and fast outflows but have AGB central stars,
indicating that the onset of asymmetry can begin while the central
star is still on the\,AGB.

The best example of this class of objects is \oh\ (hereafter, \ohs): a
fast, $\approx$0.04$\times$0.4\,pc-sized bipolar outflow around a
Mira-type pulsating M\,8-10\,III star (QX\,Pup). (V Hya and
$\pi^1$\,Gruis are other AGB stars with fast bipolar outflows, see
\citealt*{hir04,doa17}.) \ohs\ has been extensively studied at many
wavelengths by several authors, including our team, and is believed to
be a PN precursor
\citep[e.g.,][]{rei87,mor87,kas92,san97,kas98,alc01,buj02,mea03,san04}.
Before ALMA came into play (this work), two major large-scale
components had been identified in its massive ($\sim$\,1\,\msun) and
predominantly cold ($\sim$10-40\,K) molecular envelope: ($i$) a dense,
equatorial structure with an angular diameter of $\lsim$6\s\ expanding
at low velocity, $<$30\,\kms, and ($ii$) a highly collimated
$\lsim$6\s$\times$57\s\ bipolar outflow with deprojected expansion
velocities that increase linearly with the distance from the center up
to $\sim$180 and 400\,\kms\ at the tips of the north and south lobes,
respectively.\,It has been proposed that most of this nebular material
was ejected by the central AGB star during a period of very high
mass-loss rate, \mloss$\sim$1.6\ex{-4}\,\my, that started
$\la$4000\,yr ago \citep{alc01}.
The bipolar flow, with a total linear momentum of $\sim$15\,\msun\kms,
has been interpreted in the general framework of pPN and PN shaping, that is,   
as the result of a sudden interaction between CFWs on the
pre-existing, slowly expanding AGB envelope. The linear
distance-velocity relation observed in the CO outflow
suggests that such an interaction took place $\sim$\,800\,yr ago 
  and lasted less than $\sim$150\,yr.

Ground-based images of \ohs\ in the near-IR revealed a faint,
quasi-spherical halo surrounding the central parts of the nebula
\citep{alc01}.  This halo
is detected up to radial distances of $\sim$20\arcsec\ and is thought
to be a relic of an ancient wind
that was ejected at a constant rate of \mloss$\la$2\ex{-6}\,\my. The
extent of the halo indicates that this mass-loss process started at
least $\sim$11000\,yr ago and ended about 6500\,yr later. This was
followed by the period of highest mass-loss rate mentioned above
during which most of the mass of the molecular outflow of \ohs\ was
ejected.

The detection of a binary companion (A0\,V) to the central AGB star
QX\,Pup\footnote{The light from the central sources inside \ohs's core
  is highly obscured by dust along the line-of-sight and is only seen
  indirectly scattered by the dust in the lobe walls.} is a key to
understand the formation of this nebula and, in particular, the
mechanism for producing the underlying jets; the latter are proposed
to be launched by the compact companion, powered by mass accretion
from the mass-losing AGB star through an FU Orionis-like outburst
\cite[see][]{san04}.  These outbursts are short periods of increased
disk-to-companion accretion rates, reaching values up to
10$^{-3}$\,\msun\peryr. During the quiescent phases when the outburst
activity is switched off,
matter from the AGB wind piles up in the disk steadily at a relatively
low rate until a rapid (lasting $\lsim$100\,yr) and massive accretion
event occurs, resulting in the kind of energetic fast wind seen in
\ohs. At present, \ohs\ is most likely in a period of steady accretion
at a low rate, given the lack of any strong signature of current
accretion (e.g., \hal\ emission from the core, but see also \S\,\ref{res-sio}).

 Several SiO\,$v$=1 and $v$=2 maser transitions have been observed in
 \ohs\ \citep{mor87,san97,san00b,san02,des07,dod18}. These masers are known to
 arise from the pulsating layers of \qx\ ($\sim$2.4\rs) interior to the
 dust condensation and wind-acceleration region. 
The presence of SiO masers in the stellar vicinity indicates that
\qx\ is currently undergoing substantial mass-loss.
The mass-loss rate of the present-day wind of \qx\ appears to
 be significantly smaller than it was when the bulk of the nebula was
 ejected. 
 An upper limit to the mass-loss rate of the present-day wind of
 \qx\ of \mloss$<$2\ex{-5}\,\my\ was estimated by \cite{san02}. 
 This limit is consistent with the mass-loss rate of the warm,
 inner-envelope layers ($\sim$few\ex{15}\,cm) recently derived from a
 molecular line study of mid- to high-$J$ CO transitions (with
 upper-level energies \eu$\sim$500-1000\,K) observed with \hso/PACS
 toward this object, \mloss$\approx$10$^{-5}$\,\my\ (Ramos-Medina et
 al., private communication). Water masers are also present at the base of
 the bipolar flow of \ohs\ (within 100\,AU from the central star)
 and expanding along the symmetry axis at moderate velocities of
 $\sim$19\,\kms\ \citep{des07,choi12,leal12,dod18}.
  
\ohs\ not only exhibits an outstanding nebular structure and
kinematics, but it is also the chemically richest envelope amongst
O-rich AGB stars and post-AGB objects
\citep[e.g.,][]{mor87,san97,san00b}. Recent studies based on spectral
surveys at millimeter and far-IR wavelengths have led to the discovery
of $>$30 new species (including \somas, \htresomas, HNCO, HNCS, NO,
etc) and to an unprecedented detailed description of the global
physico-chemical structure of this object \citep[][Velilla Prieto et
  al., in preparation]{vel15,san15}.

Unlike for most pPNe, the distance to \ohs\ and the inclination of the
polar axis of the large-scale nebula are well known from several
independent works. The distance to \ohs\ has been most accurately
determined from a trigonometric parallax measurement of the
\water\ masers, leading to $d$=1.54$^{+0.02}$$^{(-0.01)}$\,kpc \citep{choi12}.
The inclination of the bipolar lobes with respect to the plane of the sky is
$i$$\sim$35\degr-37\degr, with the south lobe being further away
\citep{kas92,bow84,shu95}. Here we have adopted $d$=1500\,pc and
$i$=35\degr, which are rounded values commonly used in the literature.

The stellar radius of \qx\ is \rs$\sim$2.1\,AU, considering its total
luminosity and effective stellar temperature,
\lbol$\sim$7000$\times$(1.5/$d$[kpc])$^2$\,\ls\ and
\teff$\sim$2500\,K. The luminosity is derived by us by integrating the
spectral energy distribution (SED) of \ohs\ from the optical to the
mm-wavelength domain (see Fig.\,\ref{f-cont}) and after correcting by
interstellar extinction $A_V$=0.4665. This value of $A_V$ corresponds
to a color excess of E(B-V)$\sim$0.15 measured toward the
$\sim$0.25\,Gyr old cluster to which \ohs\ belongs
\citep[M46=NGC\,2437; e.g.,][]{jur85,dav13} and adopting a standard value
of $R_v$=3.1.

Here we present a continuum and molecular line study
based on $\sim$0\farcs2-0\farcs3-angular resolution ALMA observations
in the $\sim$\,294-345\,GHz frequency range.  These data have revealed
a collection of substructures previously unknown, which are the focus
of this publication.
In Sect.\,\ref{obs} the observations and data reduction steps are
described. The observational results derived from the continuum and
molecular line maps are presented in \S\,\ref{res-cont} and
\S\,\ref{res-molecules}, respectively.
In \S\,\ref{s-anal}, we present the data
analysis, namely, density and temperature maps of the central regions
of the molecular outflow (\S\,\ref{s-rd}) and estimates of the
abundances of some molecules (\S\,\ref{s-abun}). A final discussion
and a summary of our main conclusions are offered in \S\,\ref{dis} and
\S\,\ref{summ}, respectively.

\section{Observations} 
\label{obs}

%
\begin{table}
  \caption{Observing log. Spectral windows in project 2015.1.00256.S}
\label{t1}      
\centering          
\begin{tabular}{c c c}     
\hline\hline       
 Center     &     Bandwidth    &    Velocity width  \\
 (GHz)   &        (MHz)     &      (m/s) \\   
\hline                    
\multicolumn{3}{c}{\sl Science goal SG1: 5-point mosaic (2016-07)}\\ 
344.246   &	468.750   &	212.614    \\ 
345.732   &	468.750   &	211.700    \\ 
342.819   &	468.750   &	426.998    \\ 
330.526   &	468.750   &	442.879    \\ 
331.438   &	937.500   &	883.320    \\ 
\multicolumn{3}{c}{\sl Science goal SG2: 1-pointing at center (2016-07)}\\ 
293.859  &	468.750  &	498.140   \\ 
294.447  &	234.375  &	248.573   \\ 
304.070  &	937.500  &	481.412   \\ 
307.045  &	468.750  &	953.496   \\ 
305.945  &	468.750  &	956.924   \\ 
\hline                  
\end{tabular}
\end{table}
%

\ohs\ was observed with the ALMA\,12-m array as part of project
2015.1.00256.S on July 2016. Five different spectral windows (SPWs)
within band 7 ($\sim$294-345\,GHz) were observed in each of our two
science goals, SG1 and SG2, to map the emission from different
molecular transitions as well as the continuum (Table\,\ref{t1}).
SG1-observations were performed as a five-pointing mosaic covering a
$\sim$19\arcsec$\times$54\arcsec\ area along the main symmetry axis of
\oh's molecular outflow, at position angle PA=21\degr;
SG2-observations were done as a single-pointing toward the nebula
center, using a nominal position of R.A.(ICRS)=07\h:42\m:16\fm894
Dec.(ICRS)=$-$14\degr42\arcmin49\farc836.  The field of view (FoV) of
the SG2-observations has the same size as the ALMA\,12-m antennas'
primary beam (i.e., $\sim$21\arcsec\ at 300\,GHz at half-intensity).
SG1 and SG2 observations were split in three different sessions (or
blocks) executed within a window of a few days. The total number of
antennas ranged between 34 and 39, depending on the science goal and
the session. The minimum and maximum baseline length in our SG1 and
SG2 observations were 15.1 and 867.2\,m and 16.7 and 1100\,m,
respectively, leading to a spatial resolution of
$\sim$0\farcs2-0\farcs3.
The total time spent on the science target, \oh, was 35$\times$3 and
44$\times$3 minutes for SG1 and SG2, respectively (a total of about
4\,hours). Following the standard calibration procedure, a
number of sources (J0522-3627, J0730-1141, J0740-1351 and
J0750+1231) were also observed as passband, complex gain, and flux
calibrators.


The visibility data for each session (execution block) were calibrated
by the automated pipeline of the Common Astronomy Software
Applications (CASA) version 4.7.0.
This includes the standard corrections from the initial MeasurementSet
(based on system temperature and water vapor radiometer measurements,
antenna positions, derived calibrations for the bandpass, phase, and
amplitude, flagging tables, etc) to the fully calibrated data, which
are ready for imaging.
For each of our science goals, two continuum images were made using
the line-free channels from the upper and lower sideband (USB and LSB)
spectral windows, respectively. This has lead to a total of four
different continuum images centered at $\sim$344\,GHz (SG1-USB),
$\sim$330\,GHz (SG1-LSB), $\sim$304\,GHz (SG2-USB), and
$\sim$294\,GHz(SG2-LSB). The continuum images at these four
representative frequencies have been restored using a common clean
beam, with half-power beam width HPBW=0\farc31$\times$0\farc25 and
oriented along PA=$-$84.5\degr\ (Fig.\,\ref{f-cont}), which is the
beam with lowest resolution of the four sideband combinations
  (i.e., SG1-USB).

Line emission cubes were created after subtracting the continuum from
the visibility data in the spectral baseband containing the line (USB
or LSB).  After this, we created spectral cubes for the different
transitions using the CASA task {\tt tclean} using a spectral
resolution of \dv=1-3\,\kms, depending on the signal-to-noise ratio
(S/N) of the maps. By default, we used Briggs weighting with a
robust parameter of 0.5, which results in angular resolution of
$\sim$0\farc31$\times$0\farcs25 and $\sim$0\farc28$\times$0\farc22 for
SG1 and SG2, respectively.  For certain transitions, we have created
additional velocity-channel maps with uniform weighing (robust=$-$2)
for an improved angular resolution (e.g., down to
0\farc22$\times$0\farc18 in the USB of SG2).
The typical rms noise level in the line-free channels of our spectral
cubes is 1.6-6\,mJy/beam. The pixels have a size of 0\farc04 in
all our maps. All images are corrected for the primary beam,
that is, the dependence of the instruments sensitivity on direction
within the field of view, using the CASA task {\tt impbcor}.

\subsection{Flux losses}
\label{floss}

For the ALMA configurations used in this project, the angular size of
the largest smooth structure to which our observations are sensitive
(or maximum recoverable scale) is $\sim$3\arcsec. The largest angular
scales in most regions of the highly-structured CO outflow of
\ohs\ are known to be $\la$3\arcsec\ \citep{alc01} and, thus, 
significant interferometric flux losses in our ALMA
maps are not expected.

We confirm negligible interferometric flux losses in the continuum
emission maps by comparison with single-dish measurements (see
details in \S\,\ref{res-cont}). In Appendix \ref{res-30m}, we confront
line integrated profiles from our ALMA maps with single-dish spectra
from our mm-wavelength survey of \ohs\ with the IRAM\,30m telescope
(Velilla Prieto et al., in preparation) -- see Fig.\,\ref{f-30m}. For
the majority of the transitions, the ALMA line fluxes are larger than
the single-dish values. This is expected for a number of reasons. In
the case of molecules with an emission distribution comparable to, or
more extended than, that of the continuum (e.g.\,CS) this is partially
due to the relatively small beam  of our
single-pointed \iram\ spectra ($\sim$7\arcsec\ at 340\,GHz) compared with the full extent of the
emitting region. This together with the typical \iram\ pointing
errors, which can easily be of up to 2\arcsec-3\arcsec\ (i.e., about
1/3 of the beam) is probably the main reason for the
inferior single-dish line fluxes. Absolute calibration uncertainties
($\sim$30\%\ in the \iram\ spectra at 0.9\,mm) also partially
contribute to the differences observed.  Finally, in the case of high
excitation transitions with compact emission, like $v$=1 lines but
also $v$=0 \siot\ and \vsiot\, intrinsic line variability, which may
result from variable excitation conditions modulated by the stellar
near-to-mid IR variability, is also possible as observed, e.g., in
IRC+10216 and IRC+10420 \citep{cer14,quin16}.

For the reasons explained above, significant interferometric flux
losses in our maps are not expected, except maybe for some of the
large-scale \docem/\trecem\, emission of the fossil AGB envelope that
surrounds the low-velocity core. These will be addressed in detail in
a future publication as part of a comprehensive study of the
ALMA maps of these transitions.


\section{Continuum emission by dust}
\label{res-cont}

\begin{figure*}[htbp]
   \centering 
   \includegraphics*[bb=40 0 375 600,clip, width=0.256\hsize]{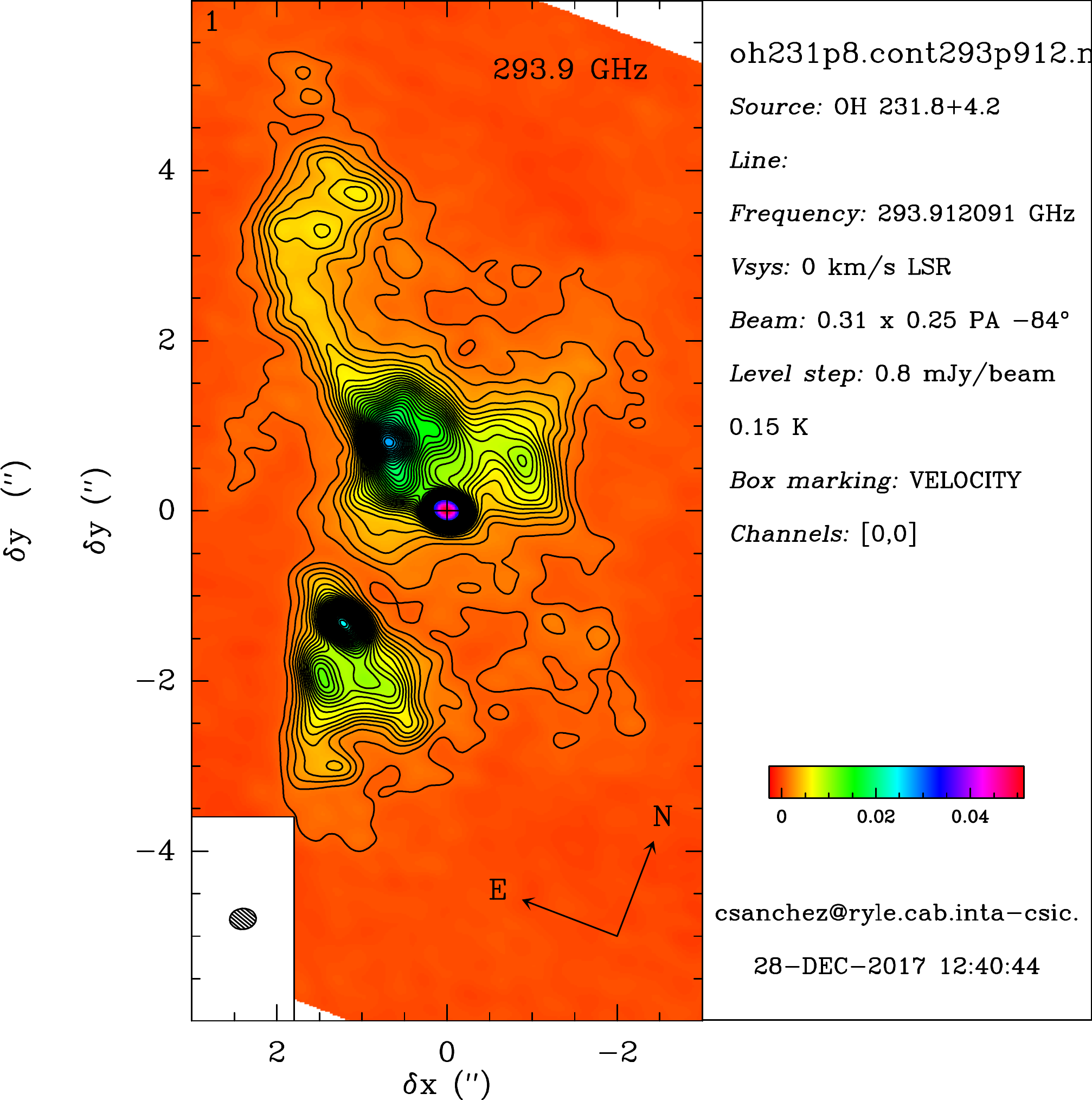}
   \includegraphics*[bb=100 0 375 600,clip,width=0.21\hsize]{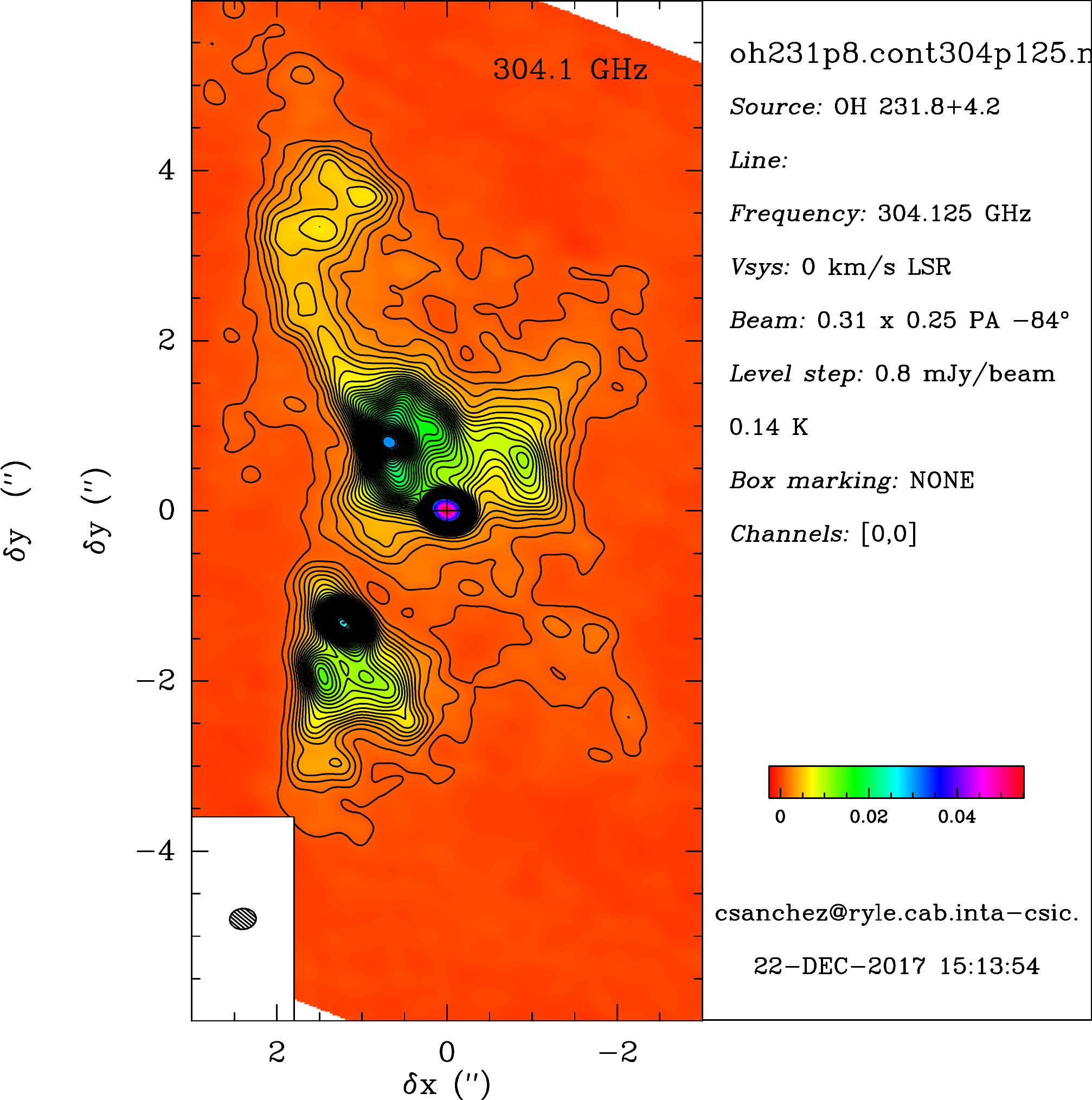}
   \includegraphics*[bb=100 0 375 600,clip,width=0.21\hsize]{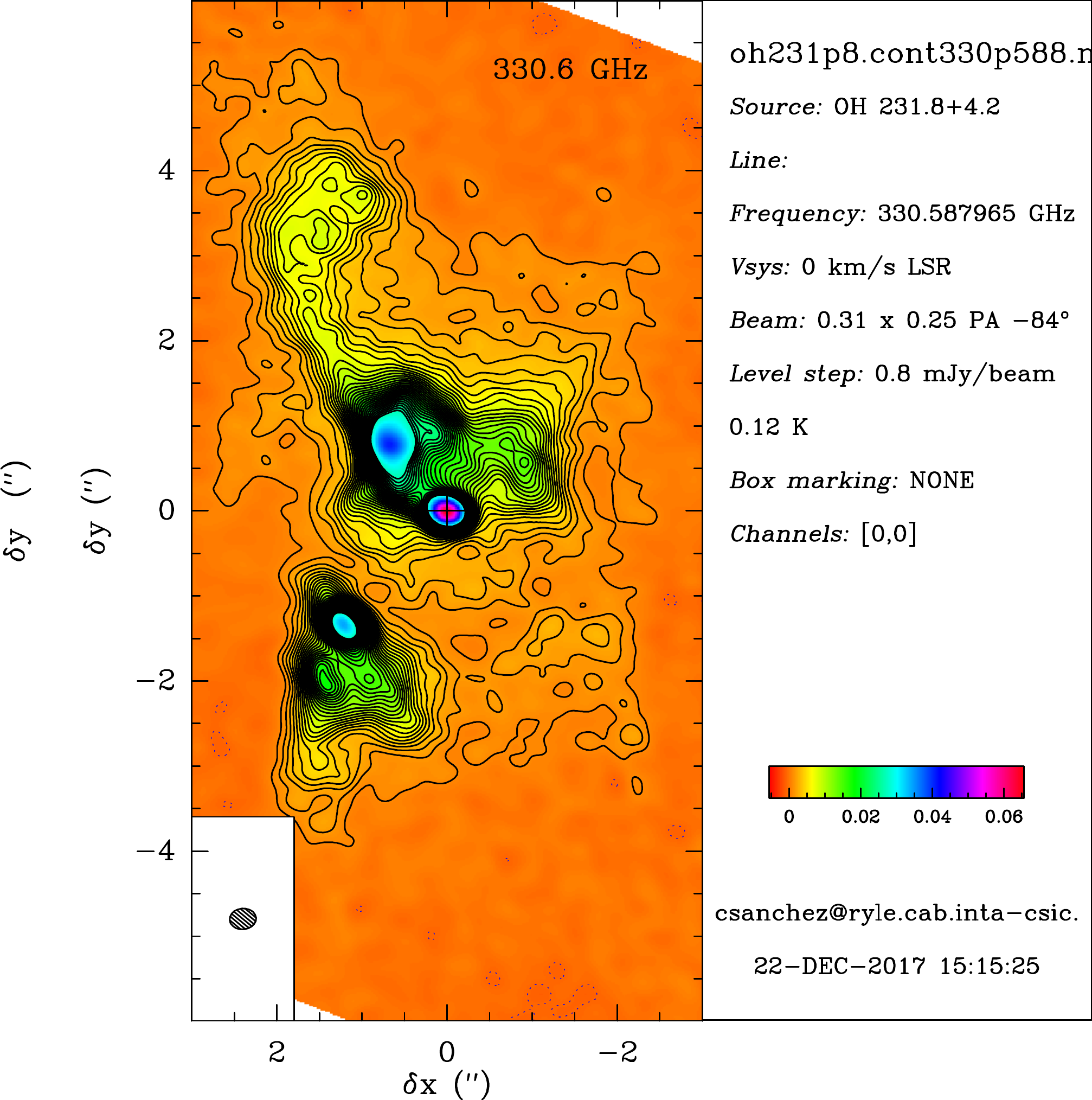}
   \includegraphics*[bb=100 0 375 600,clip,width=0.21\hsize]{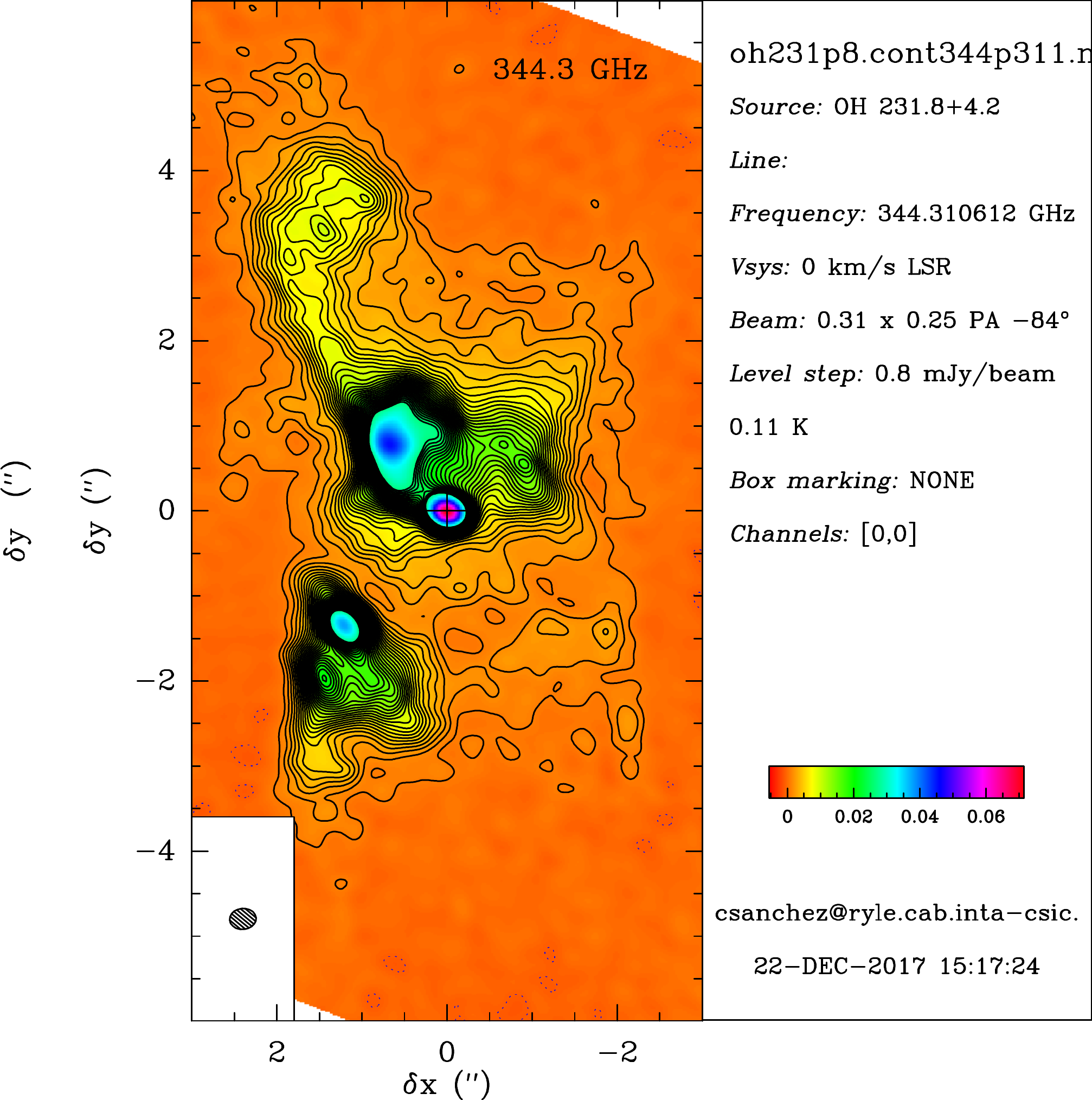} \\
   \vspace{0.25cm}
   \includegraphics[width=0.6\hsize]{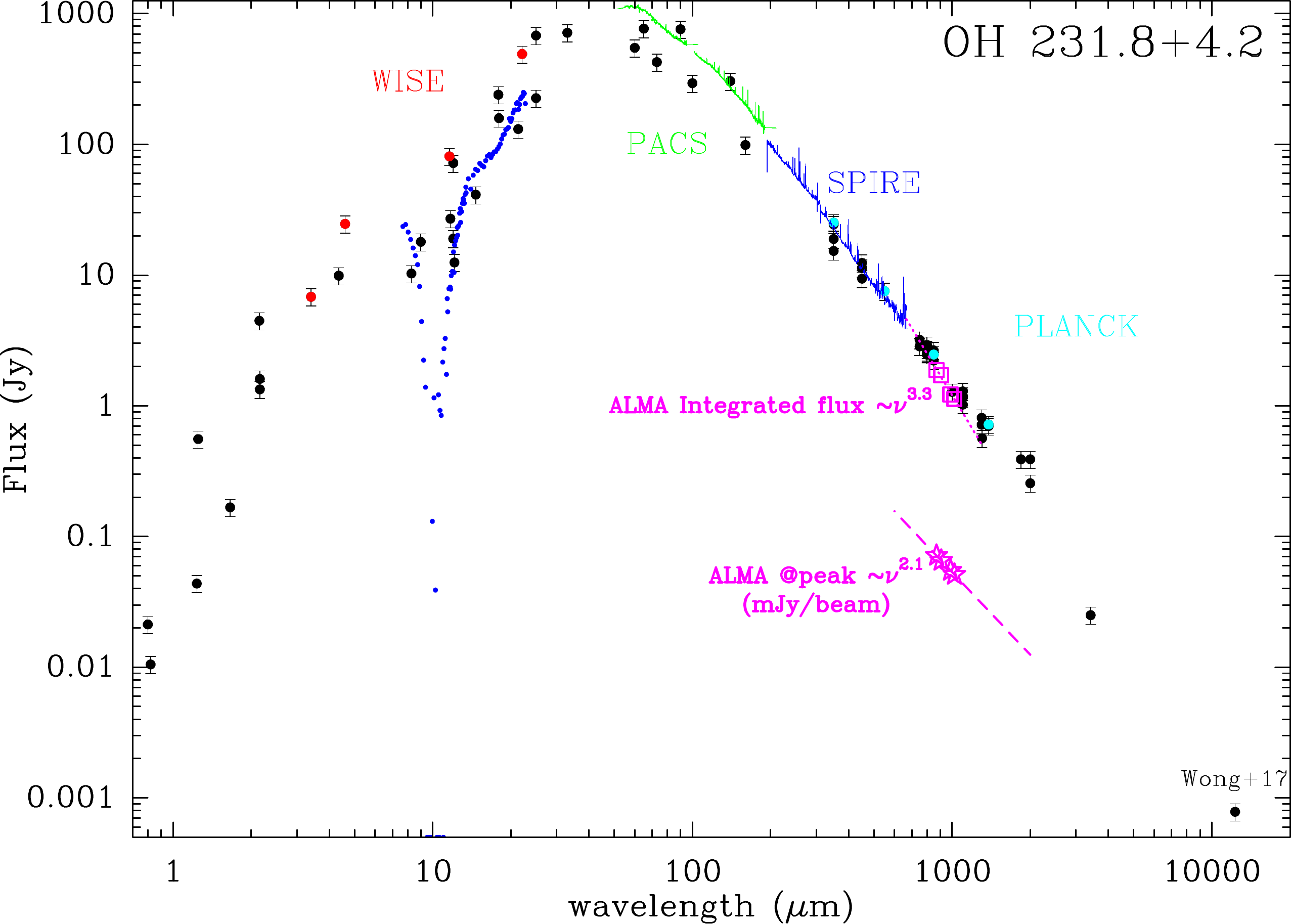}
   \caption{{\bf Top:} ALMA continuum emission maps of \oh\ at 293,
     304, 330, and 344\,GHz rotated clockwise
     by 21\degr\ so the symmetry axis is vertical. Countour
     level spacing is 0.8\,mJy/beam. The clean beam
     (HPBW=0\farc31$\times$0\farcs25, PA=$-$84.5\degr) is plotted at
     the bottom-left corner of each panel.  The compact region at the
     center (offset 0\arcsec,0\arcsec) where the continuum emission
     peaks is referred to as \cs\ and indicates the location of the
     AGB star \qx\ (\S\,\ref{res-cont}). The center of the
     maps has equatorial coordinates
     $\alpha$(J2000)=07\h\,42\s\,16\fs915,
     $\delta$(J2000)=$-$14\degr\,42\arcmin\,50\farc06.
     {\bf Bottom:} Spectral energy distribution (SED) of \ohs. Pink
     symbols are ALMA continuum measurements from this work, as given
     in Table\,\ref{t-cont} (squares= integrated flux; stars=
     peak-flux surface density at \cs) and the dotted and dashed lines
     are $\propto \nu^{3.3}$ and $\propto \nu^{2.1}$ fits to these
     ALMA data points, respectively. The rest of the continuum fluxes
     are from the literature, mainly from the compilation by
     \cite{san98}, but also including PACS and SPIRE spectroscopy
     (from the THROES catalogue,
     \citealt{ram18}), 2MASS,
     MSX, WISE, and PLANK photometry (from their respective mission
     archives), one radio-continuum datapoint from \cite{wong17}, and
     mid-infrared photometry from \cite{jur02}. Relative errors of
     15\%\ (plotted as errorbars) are adopted for all non-ALMA
     data. Errorbars of the ALMA data ($<$5-10\%) are smaller than the
     symbols. }
         \label{f-cont}
   \end{figure*}

The ALMA continuum emission maps of \ohs\ at 294, 304, 330 and 344\,GHz
are shown in Fig.\ref{f-cont}.
The surface brightness distribution of
the continuum emission, which is very similar at these four
frequencies, appears as an extended, incomplete
hourglass-like structure of dimensions
$\sim$8\arcsec$\times$4\arcsec\ (at the rms level) formed by two
opposing, dense-walled
lobes roughly oriented along PA$\sim$21\degr\ (i.e., the symmetry axis
of the large-scale nebula). The lobes join at the nebula center in a
pinched ($\sim$2\farc5-wide) equatorial waist.
The hourglass is notably asymmetric with respect to the
PA$\sim$21\degr\ axis, with the east side of the lobes being
much brighter than the west. The nebular asymmetry with respect to the equatorial plane is also appreciable.

%
\begin{table}
\caption{Continuum fluxes.}      
\label{t-cont}      
\centering          
\begin{tabular}{c c c c }     
\hline\hline       

Frequency    &   Total Flux      &   Peak-Flux    &    rms  \\ 
(GHz)        &      (Jy)         &   (mJy/beam)   &    (mJy/beam) \\ 
\hline                     
293.912	 &	1.13    &      51.1     &    0.30 \\ 
304.125	 &	1.22    &      54.9     &    0.25 \\ 
330.588	 &	1.71    &      65.5     &    0.50 \\ 
344.310	 &	1.88    &      71.4     &    0.50  \\ 
\hline                  
\end{tabular}
\tablefoot{Absolute flux calibration uncertainties are $\sim$5\%\ and
  10\%\ at [294-306] and [330-345]\,GHz, respectively. The beam of the
  continuum maps is always HPBW=0\farc31$\times$0\farc25, PA=$-$84.5\degr.}
\end{table}
%


The continuum emission distribution is not uniform
but it shows a number of bright compact condensations or clumps. The
peak of the continuum maps is attained at one of these clumps referred
to as \cs, which has coordinates R.A.=07\h42\m16\fs915
and Dec.=$-$14\degr42\arcmin50\farc06 (J2000). As we will show below,
\cs\ marks the position of the central star \qx\ (for epoch
2016.6). These coordinates coincide to within 0\farc01 with the
absolute position of the SiO\,$v$=1,2 masers at 43\,GHz recently
measured by \cite{dod18}. Throughout this paper, the position of
  \cs\ has been adopted in defining the (0\arcsec, 0\arcsec)
  positional offsets in all figures illustrating continuum (and line)
  image data. We note that \cs\ does not lie on the equatorial plane of
the central waist, but it is clearly displaced along the axis toward
the south by $\sim$0\farc6.

Using high-angular resolution $K$- and $L'$-band images of \oh,
  \cite{kas98} report the detection of a point-like source at the
  midpoint of the nebula waist with the reddest $K-L'$ colors, which
  is tentatively attributed by the authors to the central illuminating
  star, QX Pup. The latter is presumably deeply buried in a
  high-density, very optically thick ($A_V$$\sim$100\,mag) dusty
  region. Unfortunately, the absolute coordinates of these NIR images
  are not provided by \cite{kas98} and, therefore, we cannot assess
  whether or not the reddest $K-L'$ region coincide with \cs.

The total continuum flux, obtained by integrating the surface
brightness over the emitting region, as well as the continuum peak
surface brightness at \cs\ at the four frequencies observed with ALMA are
given in Table\,\ref{t-cont}. The good agreement of the ALMA total
fluxes with measurements from single-dish observations in the
submm-to-mm wavelength range (Fig.\,\ref{f-cont}, bottom panel)
indicate negligible flux losses of the extended emission
in our interferometric continuum maps.


Except for \cs, the spectral index of the continuum shows no
significant deviations from a \snu{3.3} power-law across the different
regions. This implies a dominant contribution by optically thin
$\sim$75\,K dust with an emissivity index of $\alpha\sim$1.3,
consistent with previous studies (\citealt{san98}; see SED in
Fig.\,\ref{f-cont}).
At \cs, which is the brightest condensation, the continuum follows a
less steep \snu{2.1} frequency dependence approximately consistent with
blackbody emission.

The compact emission \cs\ appears to be partially resolved in our
maps, although an accurate determination of its angular size is
difficult due to the presence (and poor subtraction) of the
non-uniform underlying emission. In order to mitigate this effect, we
have fitted a single component source model to the $uv$ continuum data
using only the longest baselines (length$>$250\,m). This enables us to
filter out partially the underlying extended emission from the dust
and to better isolate the emission from \cs. To do that, we used the
CASA task {\tt uvmodelfit}. The fit of the $\sim$306\,GHz-continuum
visibility data adopting an elliptical gaussian model for \cs\ yields
beam-deconvolved dimensions of $\sim$0\farc054$\times$0\farc043, with
the major axis oriented along PA=53\degr. By fitting a disk source
model, we find somewhat larger dimensions
$\sim$0\farc09$\times$0\farc07 and similar orientation.
Within uncertainties, both results suggest an elongated structure with
a typical radius of $\sim$40-70\,AU.
We stress that this value
is very uncertain and it could well represent an upper limit to the
real size given that \cs\ is marginally resolved in our maps.

%
\begin{table}[htbp]
\caption{Molecular transitions detected}        
\label{t-lines}      

\begin{tabular}{l c c r}     
\hline\hline       
Molecule & Transition & Rest Frequency & E$_{\rm u}$~~ \\ 
         & QNs$_{\rm (u \raw l)}$    &     (GHz)      &   (K) \\ 
\hline                    
CS                &     6 0 -- 5 0     &      293912.086 &     49.4 \\
\tsiov            &     7 1 -- 6 1     &      294539.577 &   1804.3 \\
SiO               &     7 0 -- 6 0     &      303926.812 &     58.3 \\
OCS               &      25 -- 24      &      303993.262 &    189.7 \\
SO                &     7 8 -- 6 7     &      304077.844 &     62.1 \\
\metanol          & 2  1  1 -0 -- 2  0  2 +0 &304208.350 &    21.6  \\
\tsodos           &   3 3 1 -- 2 2 0   &      304332.030 &     26.8 \\
\sisv             &    17 1 -- 16 1    &      307014.349 &   1203.8 \\
\htresomas        & 1$^-_1$ - 2$^+_{1}$ &      307192.410 &     79.5 \\
\metanol          & 4  1  3 -0 -- 4  0  4 +0 &307165.940 &     38.0 \\
\trecem           &       3 -- 2       &      330587.965 &     31.7 \\
\nacl             &    26 0 -- 25 0    &      330805.749 &    214.5 \\
\sodos            &  11 6 6 -- 12 5 7  &      331580.244 &    149.0 \\ 
CS                &     7 0 -- 6 0     &      342882.850 &     65.8 \\
$^{29}$SiO         &     8 0 -- 7 0     &      342980.847 &     74.1\\
\tso              &   8 9 8 -- 7 8 7   &      343086.102 &     78.0 \\  
\tsodos           &  10 4 6 -- 10 3 7  &      344245.346 &     88.4 \\  
SO                &     8 8 -- 7 7     &      344310.612 &     87.5 \\ 
\tsodos           &   7 4 4 -- 7 3 5   &      345519.656 &     63.6 \\  
\tsodos           &   5 4 2 -- 5 3 3   &      345651.293 &     51.7 \\   
\docem            &       3 -- 2       &      345795.990 &     33.2 \\ 
\hline                                                              
\end{tabular}
\tablefoot{Spectroscopic information and quantum number (QN) notation of the transitions from the {\sl Cologne Database for Molecular
    Spectroscopy} (CDMS, \citealt{mul05}).}
\end{table}
%

The continuum emission flux from \cs\ is much larger than that
expected from the photosphere of \qx. In fact, the effective
temperature and luminosity of \qx\ (\teff$\sim$2500\,K and
\lbol$\sim$7000\,\ls, \S\,\ref{intro}) imply a main-beam brightness
temperature of
\tmb=2500\,K$\times$($\frac{\rs=2.1\,AU}{(beam/2)=208\,AU}$)$^2$=0.25\,K=1.8\,mJy
at 344\,GHz (considering the ALMA beam=0\farc31$\times$0\farc25 and
mJy-to-K=7.3 conversion factor at this frequency) which is much
smaller than the observed flux (Table\,\ref{t-cont} and
Fig.\,\ref{f-cont}).  This implies that the continuum flux from
\cs\ is dominated by thermal emission from a source other than
  the star, most likely, by dust distributed in its vicinity (within
$\sim$100\,AU). The spectral index of the continuum indicates that the
dust emission is either {\it i)} optically thick or {\it ii)}
optically thin with emissivity $\alpha$$\sim$0 (i.e., produced by
large solid particles). We explore these two possibilities.

{\it i)} In the case of optically thick dust, the 344\,GHz-continuum
flux at \cs, which is equivalent to a main-beam brightness temperature
of \tmb$\sim$9.6\,K, would imply a dust temperature of $T_{\rm
  d}$$\sim$9.6$\times$($\frac{(beam/2)=208\,AU}{r_{\rm d}[AU]}$)$^2$\,K.
Considering the half-intensity radius of
\cs\ estimated above, $r_{\rm d}$$\la$40-70\,AU, we derive a
characteristic dust temperature of $T_{\rm
  d}$$\ga$260-85\,K.
Using this temperature and the observed flux from
\cs\ (Table\,\ref{t-cont}), we obtain a lower limit for the dust mass
using the equation:

\begin{equation}
  \label{md-thick}
S_{\nu}=\frac{2\kb\nu^2}{c^2} \td \md \kappa_{\nu} /d^2, 
\end{equation}

\noindent
where \kb\ and $c$ are the Boltzmann constant and the speed of light,
as usual, and adopting a dust absorption coefficient of
$\kappa_{\nu}$=0.33\,cm$^2$/g at 1\,mm \citep{li01} and the
  Rayleigh-Jeans approximation. We deduce a very high value for the
mass of \md$\ga$0.008-0.0025\,\msun\ (for \td=85 and 260\,K,
respectively), which, for a typical gas-to-dust mass ratio in O-rich
AGB stars of $\sim$100-200 \citep[e.g.,][]{kna85}, implies an
unrealistically high amount of material in \cs\ (even larger
than\footnote{For a dust optical depth of $\tau_{\rm
      1mm}$$\sim$3.5, as suggested by the \snu{2.1} law, the mass
  correction factor would be $\sim$3.6.}  all the mass contained in
the dense equatorial regions of the large-scale nebula,
$\sim$0.6\,\msun, \S\,\ref{intro}). This result rules out the
optically-thick dust scenario to explain \snu{2.1} in \cs.

{\it ii)} In the case of optically-thin thermal emission produced by large
($\ga$\,100\,$\mu$m-sized) grains, we can
obtain a rough estimate of the mean dust temperature and mass in
\cs\ as follows.  For big grains, a flat emissivity law ($\alpha$=0)
is expected\footnote{The grain emissivity law flattens, $\alpha$\raw0,
typically at wavelengths smaller than $\sim$4$\pi$$a$, with $a$ being
the grain radius.} and, in this case, assuming radiative equilibrium
(i.e., that the energy absorbed by the grains is totally reemitted at
longer wavelengths) and that the only dust heating mechanism is the
stellar radiation, the dust temperature can be expressed as:

\begin{equation}
\label{equ-td}
\td=(\frac{\lbol}{16\pi\sigma\,\rd^{2}})^{1/4}, 
\end{equation}

\noindent
where, \lbol\ is the stellar luminosity, \rd\ is the distance of the
dust to the central star, and $\sigma$ is the Stefan-Boltzmann
constant \citep[see e.g.,][]{her86}. Using this simplified formula, we
derive a dust temperature of $\sim$400-300\,K at
\rd$\sim$40-70\,AU. Adopting the average value \td$\sim$350\,K, and
using the formulation described, for example, in \cite{san98}, we find
that the dust mass in \cs\ is \md$\sim$2\ex{-5} (2\ex{-4})\,\msun\ for
a grain radius of $a$=100 (1000)\,$\mu$m.
As expected, this represents a very small fraction of the total dust
content of the nebula, $M^{\rm total}_{\rm d}$ $\approx$
5\ex{-3}-10$^{-2}$\,\msun, which resides predominantly in
the large-scale hourglass nebula \citep{san98,kas95}.

\section{Molecular line emission: nebular structure and kinematics}
\label{res-molecules}

The list of (identified) molecular transitions that have been detected
toward \ohs\ within the spectral windows observed with ALMA as part
of project 2015.1.00256.S is given in Table\,\ref{t-lines}.
The vibrationally excited ($v$=1) lines of SiS and \tsio\ as well as
the \nacl\ and \metanol\ molecules are first time detections in this
object.  The detection of \metanol\ is particularly important since
this is the
first AGB star 
in which this species has been
discovered. We confirm the presence of weak
\mbox{p-\htresomas\,($J^p_K$=1$^-_1$ - 2$^+_{1}$)} emission, which was
previously reported by \cite{san15}.



In addition to adding detailed information on the morphology and
dynamics of the two major components of the molecular envelope of
\ohs\ already known from previous works (namely, the fast bipolar
outflow and the slowly expanding equatorial waist of the large-scale
nebula -- \S\,\ref{intro}), our ALMA observations unveil a series of
new nebular components. In the following subsections,
we describe the different (small-to-large scale) nebular structures
traced by ALMA, starting from the most compact ones (the
inner-envelope layers within a few 10$^{15}$\,cm from the mass-losing
star) and ending with the most extended zones (reaching out to
$\sim$few\ex{17}\,cm). The different envelope components of \ohs\ are
selectively (and, in some cases, exclusively) traced by certain
molecules or transitions included in these observations, which
facilitates disentangling the intricate nebular architecture of \ohs.

%
\begin{table*}[htbp]
  \caption{Nebular components of \ohs\ schematically represented in
    Fig.\,\ref{f-sketch}.}
\label{t-sketch}      
\centering          
\small 
\renewcommand{\arraystretch}{1.1} 
\begin{tabular}{l | c c c c l }     
\hline\hline       

Name        &       molecular  &  projected      &  \vlsr\ range   &  Sections &  Figures  \\
            &        tracer\tablefootmark{1}    &  size (\arcsec)      &   (\kms)      &          &           \\
\hline
clump S            & \nacl, \sisv, \tsiov & $\sim$0\farc08$\times$0\farc08\tablefootmark{\dag} & [25:45] & \S\,\ref{res-clumps} & Fig.\,\ref{f-compact} \\ 
SiO outflow        & SiO, \vsio & $\sim$1\arcsec$\times$4\arcsec, PA$\sim$17\degr\tablefootmark{\ddag} & [18:53] & \S\,\ref{res-sio} & Figs.\,\ref{f-sio}, \ref{f-29sio}  \\ 
mini-hourglass     &
\begin{tabular}{c} \docem, \trecem, CS, SO, SO$_2$, OCS\tablefootmark{*} \\ \metanol, \htresomas\tablefootmark{*}   \end{tabular} & 
'' & 
[20\tablefootmark{\ddag}:50] & \S\,\ref{res-minihg} & Figs.\,\ref{f-pv-SiO-13co}, \ref{f-miniw}, \ref{f-ch3oh} \\
large hourglass\tablefootmark{2}  &
\begin{tabular}{c}as in mini-hourglass \\ (\htresomas\ and \metanol\ are marginal)\end{tabular} & 
$\sim$4\arcsec$\times$8\arcsec, PA$\sim$21\degr& [$-$13:80] & \S\,\ref{res-lsnebula}, \S\,\ref{res-largehg}  & Figs.\,\ref{f-cs76}, \ref{f-largew}, \ref{f-waist} \\
HV lobes & \docem,\trecem, CS\tablefootmark{\#}, SO\tablefootmark{\#}, SO$_2$\tablefootmark{\#}, SiO\tablefootmark{\#} &  
\begin{tabular}{c}
$\sim$8\arcsec$\times$17\arcsec, PA$\sim$21\degr\ (N)\\$\sim$8\arcsec$\times$30\arcsec,  PA$\sim$21\degr\ (S)
\end{tabular} & 
\begin{tabular}{c}
$[$$-$80:$-$13$]$ (N)\\ $[$$+$80:$+$225$]$ (S) 
\end{tabular} & 
\S\,\ref{res-lsnebula}, \S\,\ref{res-hv} & Figs.\ref{f-12co}, \ref{f-co-pvs} \\
  fish bowls         & \docem, \trecem\  &
  \begin{tabular}{c}
    $\sim$6\arcsec$\times$8\arcsec, PA$\sim$125\degr\ (N)\\$\sim$6\arcsec$\times$8\arcsec, PA$\sim$100\degr\ (S)
  \end{tabular} &
  \begin{tabular}{c}
    $[-$13:59$]$ (N)\\ $[$+17:+77$]$ (S)
  \end{tabular} & 
\S\,\ref{res-calamardos} & Fig.\ref{f-calamardos}  \\
\hline                  
\end{tabular}
\tablefoot{
  \tablefoottext{1}{Transitions are listed in Table\,\ref{t-lines}.} 
  \tablefoottext{2}{Includes the equatorial waist and base of the large-scale lobes.} 
  \tablefoottext{\dag}{Barely resolved.} 
  \tablefoottext{\ddag}{Uncertain.} 
  \tablefoottext{*}{Predominantly in the waist.}
  \tablefoottext{\#}{Detected in selected regions of the HV-lobes -- not discussed (shown) in this paper.}
}
\end{table*}
%
%

We have focussed on the new major features exposed by these ALMA observations.
A brief report on the ALMA view of the principal
envelope components of \ohs\ already known is also given but we defer to a future
publication a more detailed description and comprehensive study.

To facilitate visualizing the complex array of nebular structures in
\ohs\ we offer a schematic view in Fig.\,\ref{f-sketch}. We also
summarize the main identificative properties of the components in
Table\,\ref{t-sketch} together with references to sections and Figures
where the individual structures are described in great detail.


\subsection{The close surroundings of QX Pup: \cs}
\label{res-clumps}

\begin{figure*}[htpb]
   \centering 
   \includegraphics[width=0.325\hsize]{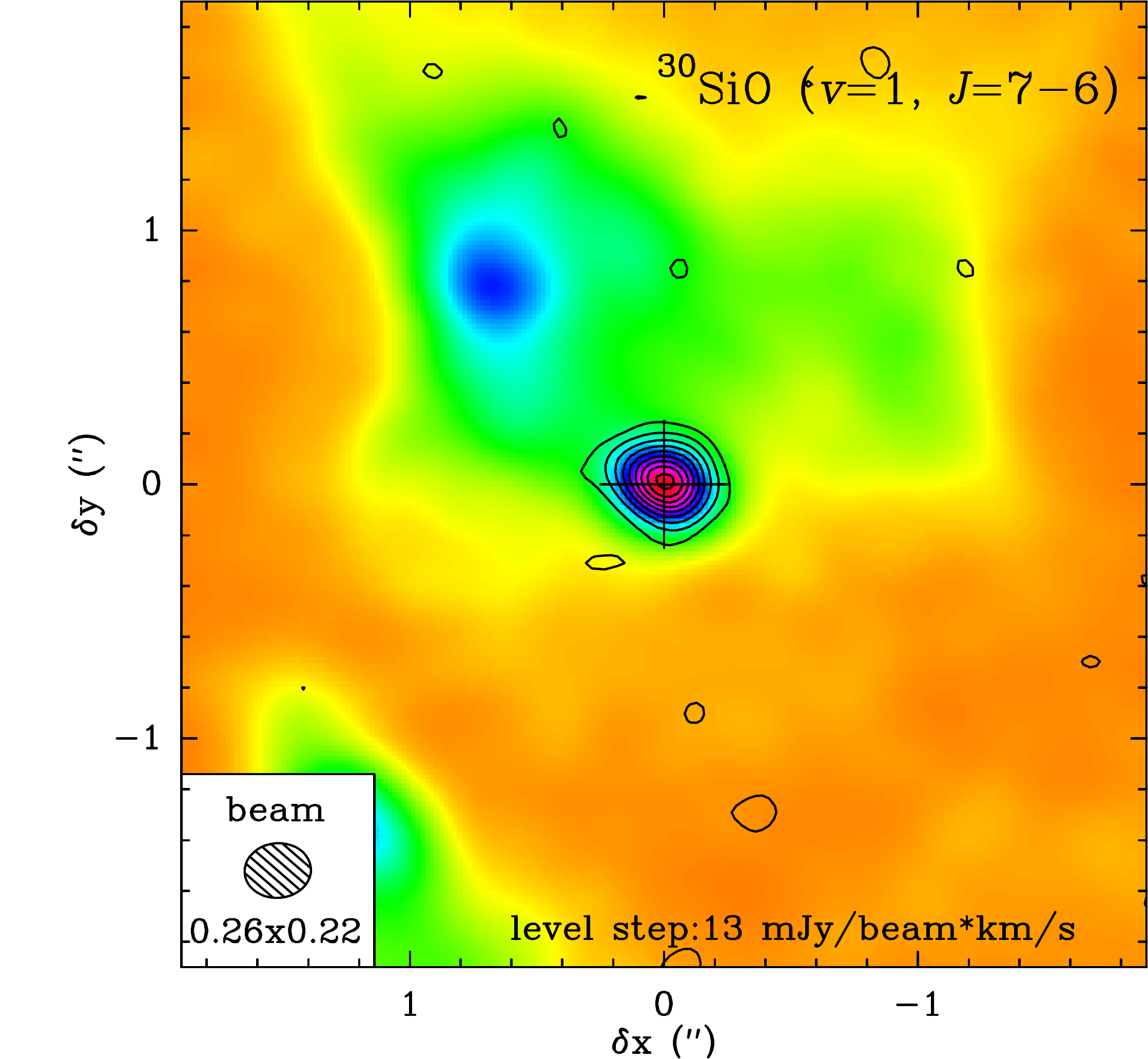} 
   \includegraphics[width=0.325\hsize]{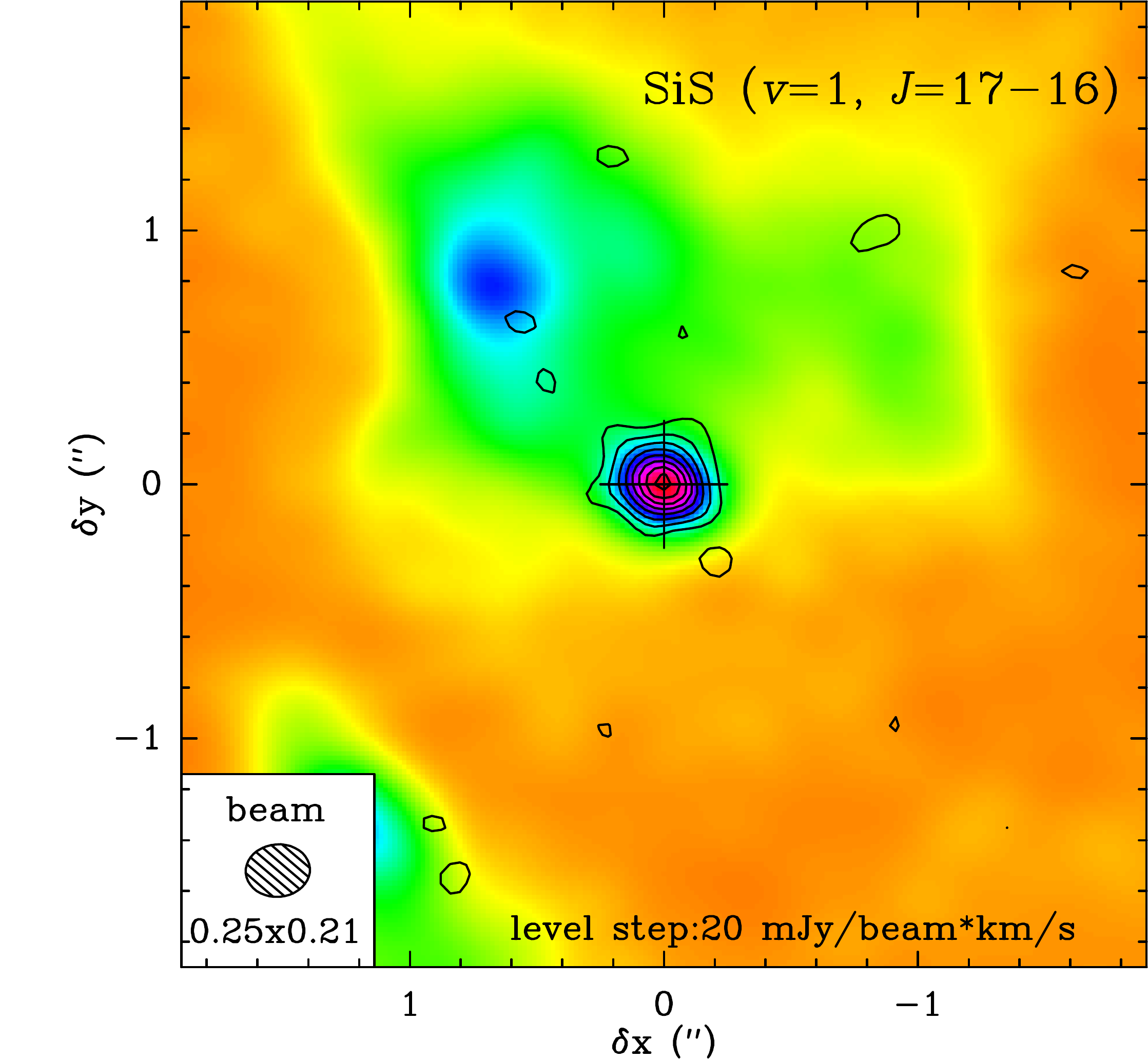}   
   \includegraphics[width=0.325\hsize]{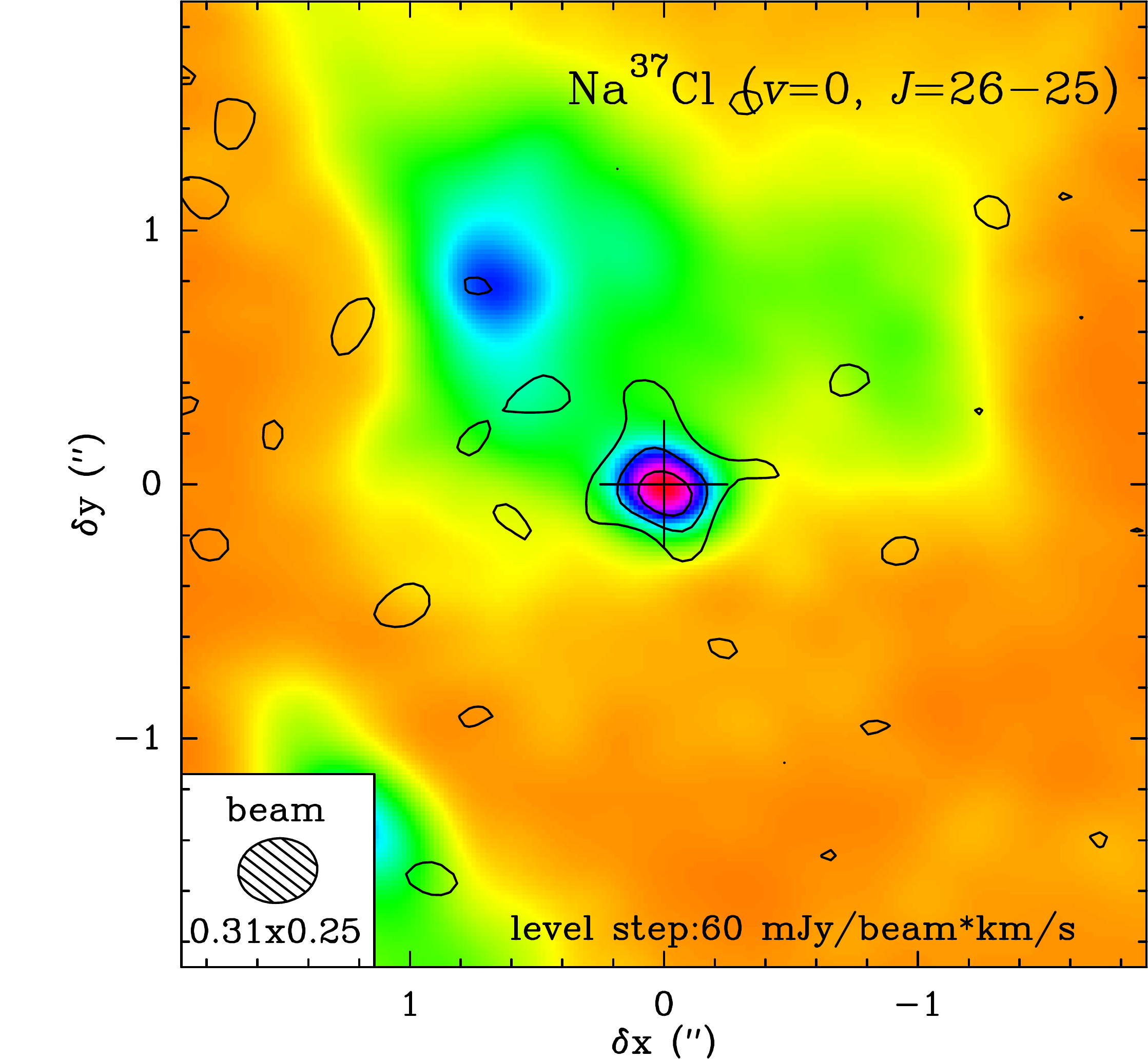} \\ 
   \vspace{0.35cm}
   \includegraphics[width=0.325\hsize]{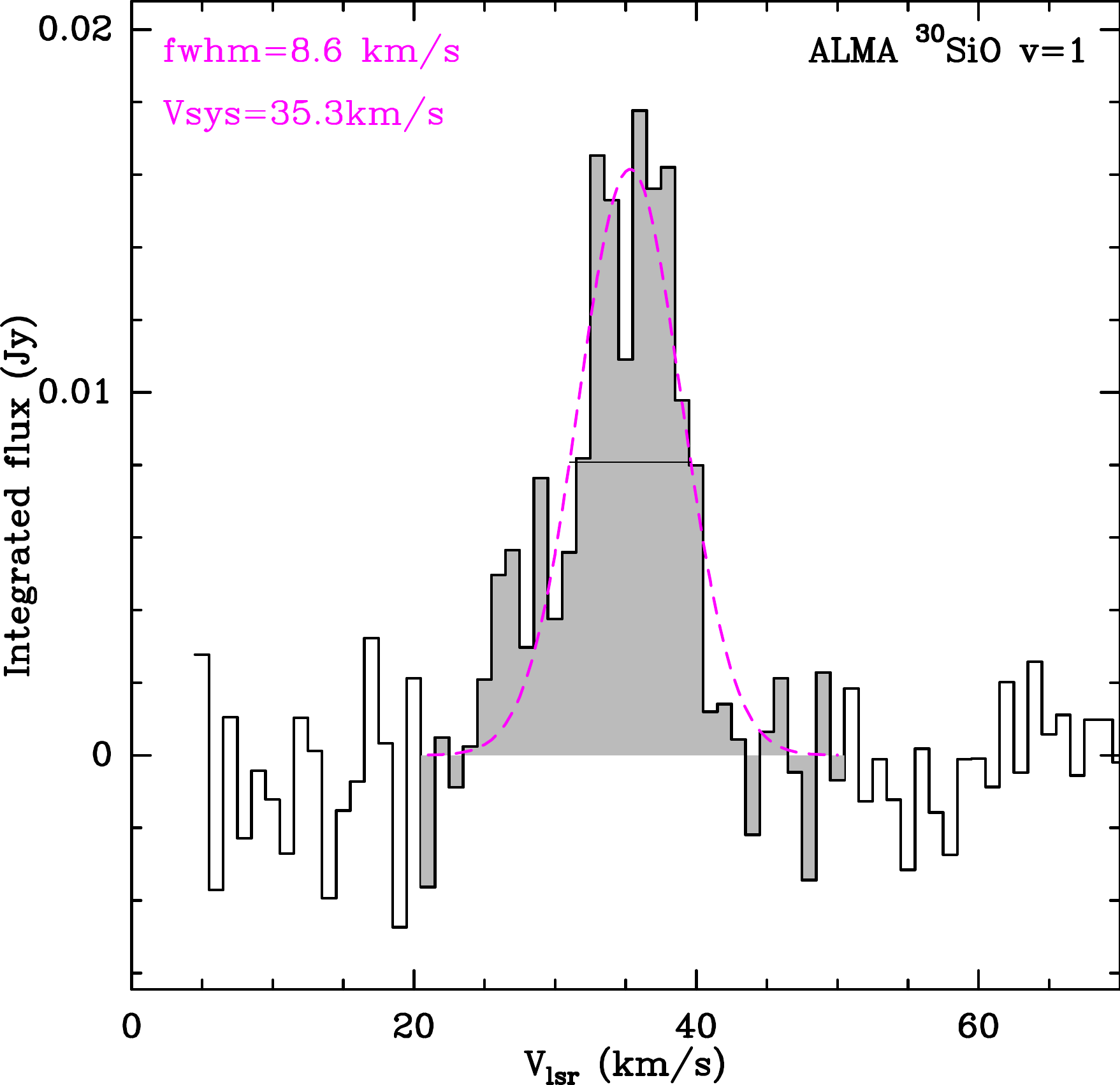}   
   \includegraphics[width=0.325\hsize]{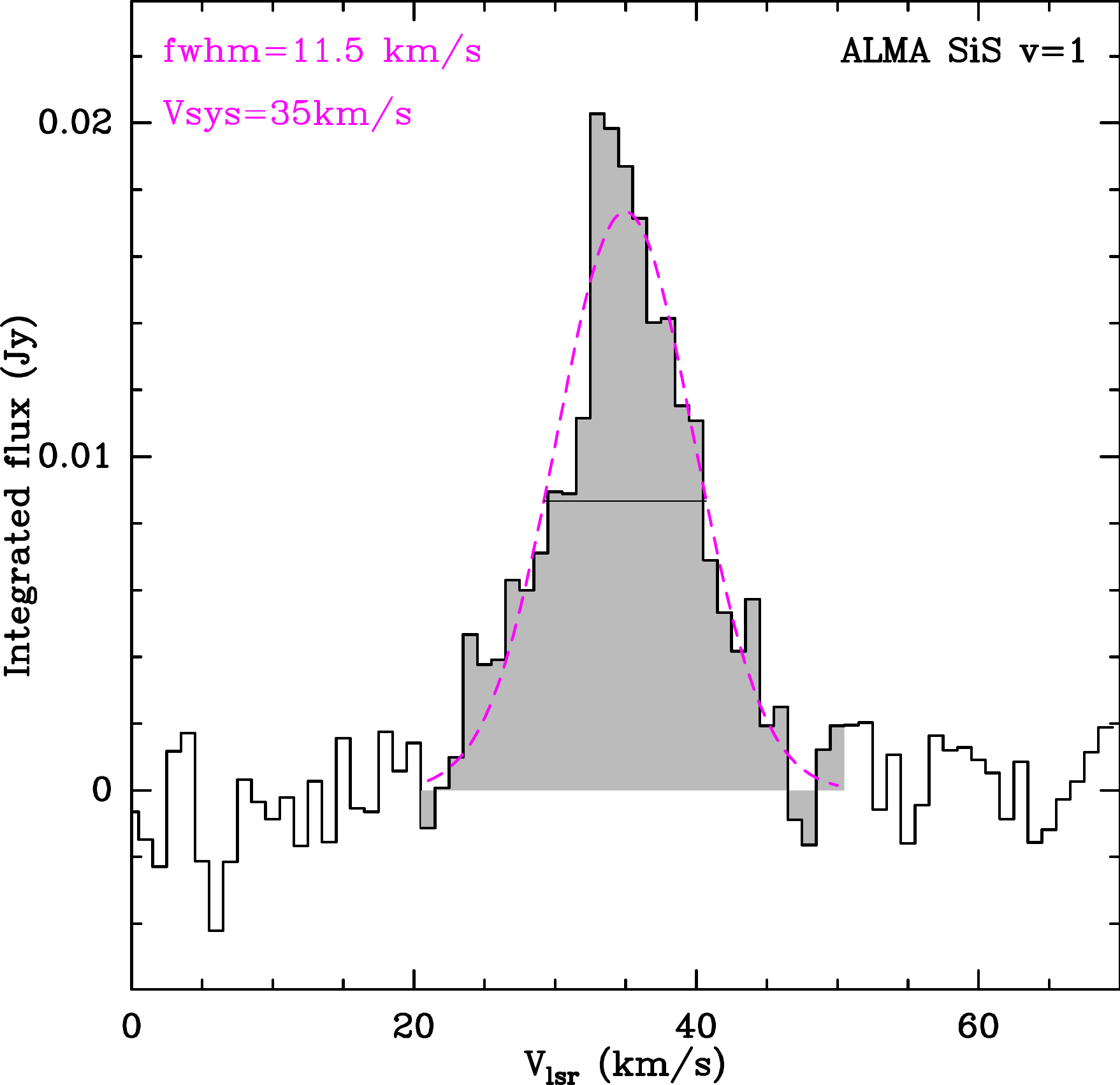}   
   \includegraphics[width=0.325\hsize]{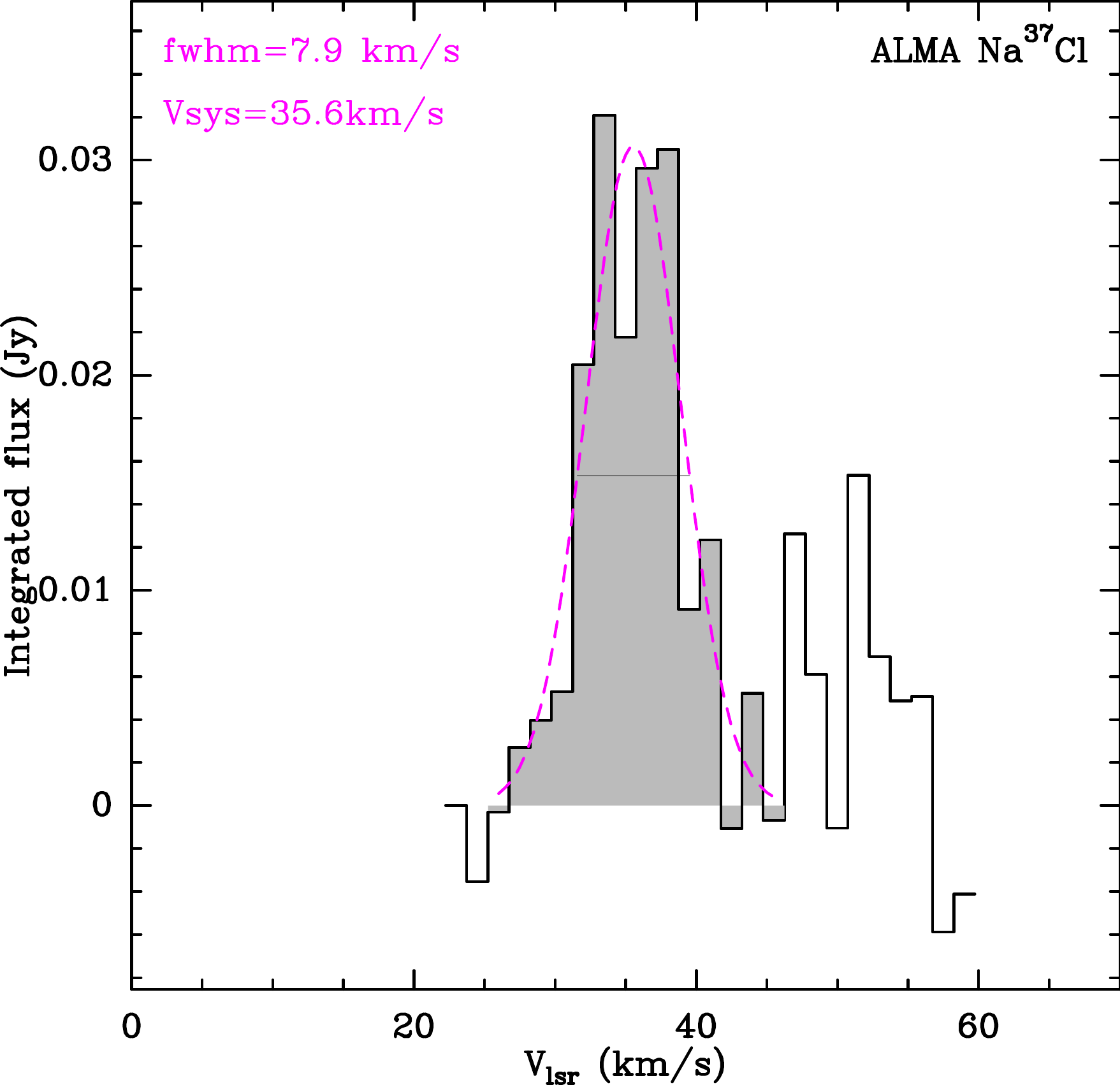}
   \caption{Integrated intensity maps (contours) on top of the
     continuum at 330\,GHz (color) (rotated clockwise by PA=21\degr,
     top) and integrated 1d-spectra (bottom) of the three molecular
     transitions with compact emission from the close surroundings of
     \qx\ (\cs). The coordinates of \cs\ are
       R.A.=07\h42\m16\fs915 and Dec.=$-$14\degr42\arcmin50\farc06
       (J2000). The position of \cs\ has been adopted as the
       origin of positional offsets in these and all subsequent
       figures illustrating image data. Gaussian fits to the line
     profiles (dashed lines) are shown together with the line
     centroids and full width at half maximum (bottom).}
   \label{f-compact}
   \end{figure*}

We have detected compact SiS\,($v$=1, $J$=17-16), $^{30}$SiO\,($v$=1,
$J$=7-6), and \nacl\,($v$=0, $J$=26-25) emission arising \emph{entirely}
from \cs\ (Fig.\,\ref{f-compact}).  Vibrationally excited
SiS and SiO lines such as those observed here (with
\eu$\sim$1200-1800\,K) are known to be produced in the warm inner
($\la$10$^{15}$cm) layers of the winds of evolved mass-losing stars,
where dust grain formation is probably not yet complete and where the
stellar wind may have not reached its terminal expansion velocity
\citep[see][for a recent review paper on AGB CSEs]{hof18}. 
This supports the hypothesis that \cs\ harbours \qx.
The region where the \sisv, \tsiov, and \nacl\,$v$=0 line emission is
produced is barely resolved in our $\sim$0\farcs2-0\farc3-resolution
ALMA maps. We have fitted an elliptical gaussian to the integrated
intensity maps of the \sisv\ transition, which is the line with the
highest signal-to-noise ratio, leading to a deconvolved angular
diameter of the emitting source of $\sim$0\farc08.
At a distance of $d$=1500\,pc, this implies that the emission is
confined to a small emitting volume of radius $\sim$60\,AU
($\sim$24\rs) comparable to that of the continuum-emitting core at the
center, i.e., \cs\ (\S\,\ref{res-cont}).

The spectral profiles of these transitions, integrating the line
surface brightness over the compact emitting area at \cs, are centered
at \vlsr$\sim$35\,\kms\ and have full widths at half maximum in the
range FWHM$\sim$8-12\,\kms\ (Fig.\,\ref{f-compact}, middle
panel). These narrow profiles are consistent with low expansion
velocities of \vexp$\sim$5-7\,\kms, adopting a Gaussian-like profile
and assuming isotropic mass-loss in these
inner-wind regions.
(In case of a non-spherical mass distribution, for example if there is an
equatorial density enhancement resulting in a torus or disk-like dominant
emission component, the expansion velocity could be larger after
taking into account line-of-sight projection effects).

The size and expansion velocity measured imply that the \sisv, \tsiov,
and \nacl\ emission lines sample wind layers that have been ejected
within the last $\sim$50\,yr.  The lines' centroid (\vsys=35\,\kms)
then represent the systemic velocity of the mass-losing star (\qx)
averaged during the last 50\,years.

The profile of the \tsio\,($v$=1, $J$=7-6) transition reported here is
remarkably different from that of any of the known SiO maser lines
detected in \ohs, which are characterized by two or three isolated
$\sim$2-3\,kms-wide, intense ($\approx$1-10\,Jy) emission peaks
\citep[e.g.,][]{san02,dod18}. The different broader profile and the
weakness of the \tsio\,($v$=1, $J$=7-6) line, with a peak intensity of
only $\sim$17\,mJy, points to thermal (i.e., non-maser) emission in
this case.

\subsection{A compact bipolar outflow emerging from \cs}
\label{res-sio}

In addition to disclosing the locus of \qx\ (inside \cs), our
ALMA data have also uncovered the existence of a compact
($\sim$1\arcsec$\times$4\arcsec)  bipolar outflow that
is emerging from the stellar vicinity. 
This outflow is exclusively traced in our data by
the SiO molecule, which is a well-known shock tracer.

\begin{figure*}[htbp]
   \centering 
   \includegraphics[width=0.85\hsize]{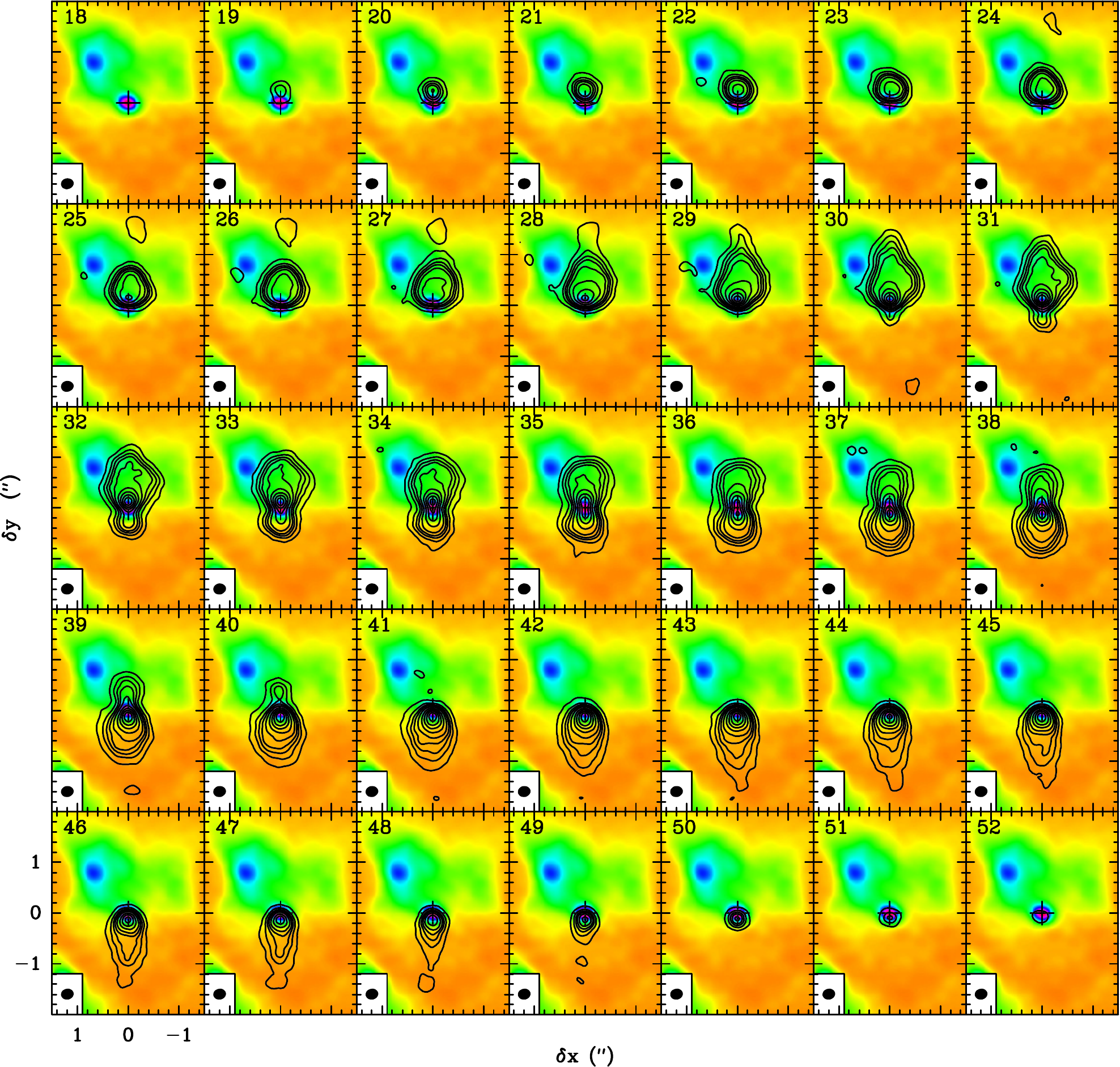}   
   \includegraphics*[bb=60 0 540 600,clip,width=0.33\hsize]{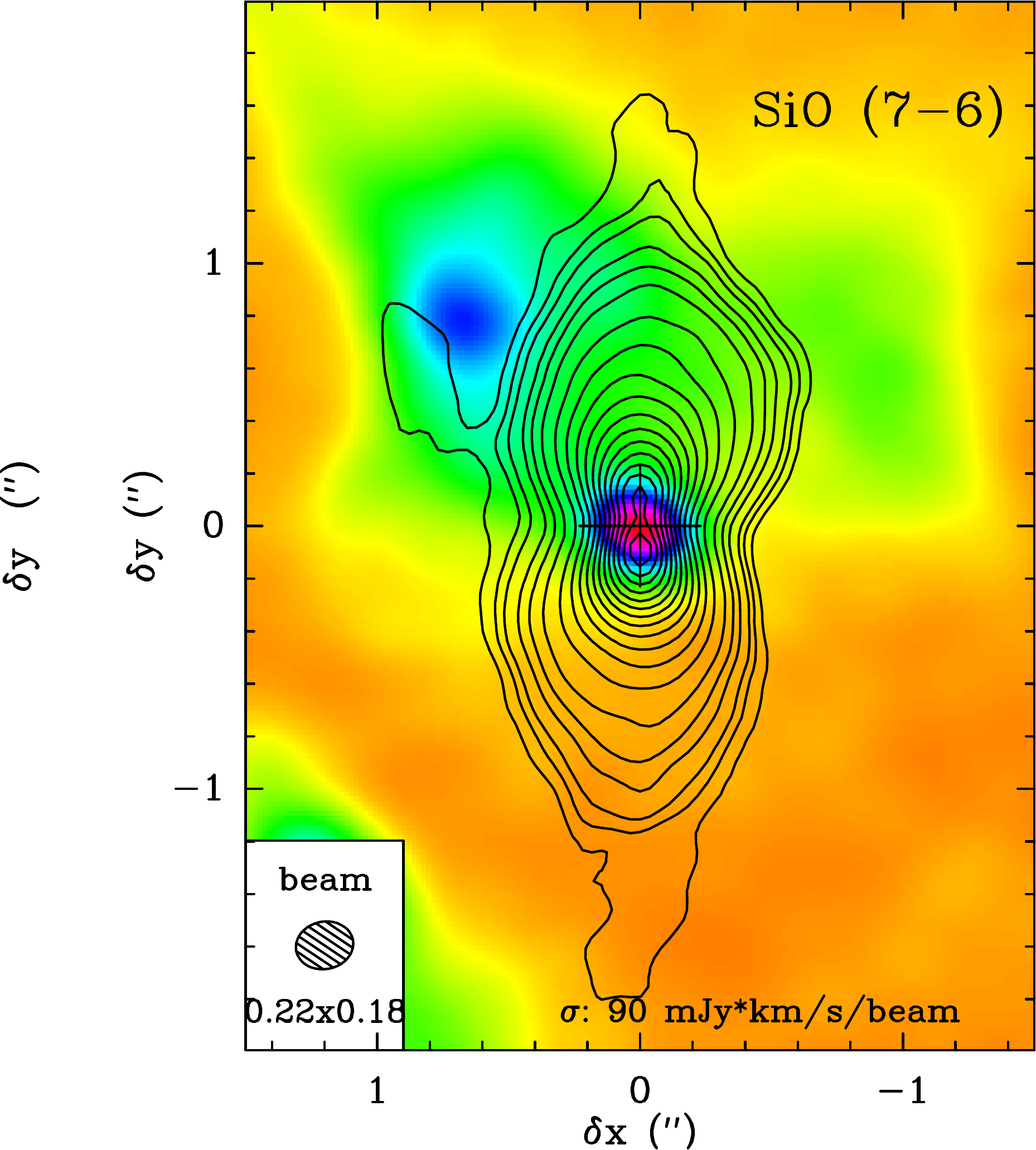}
   \includegraphics[width=0.425\hsize]{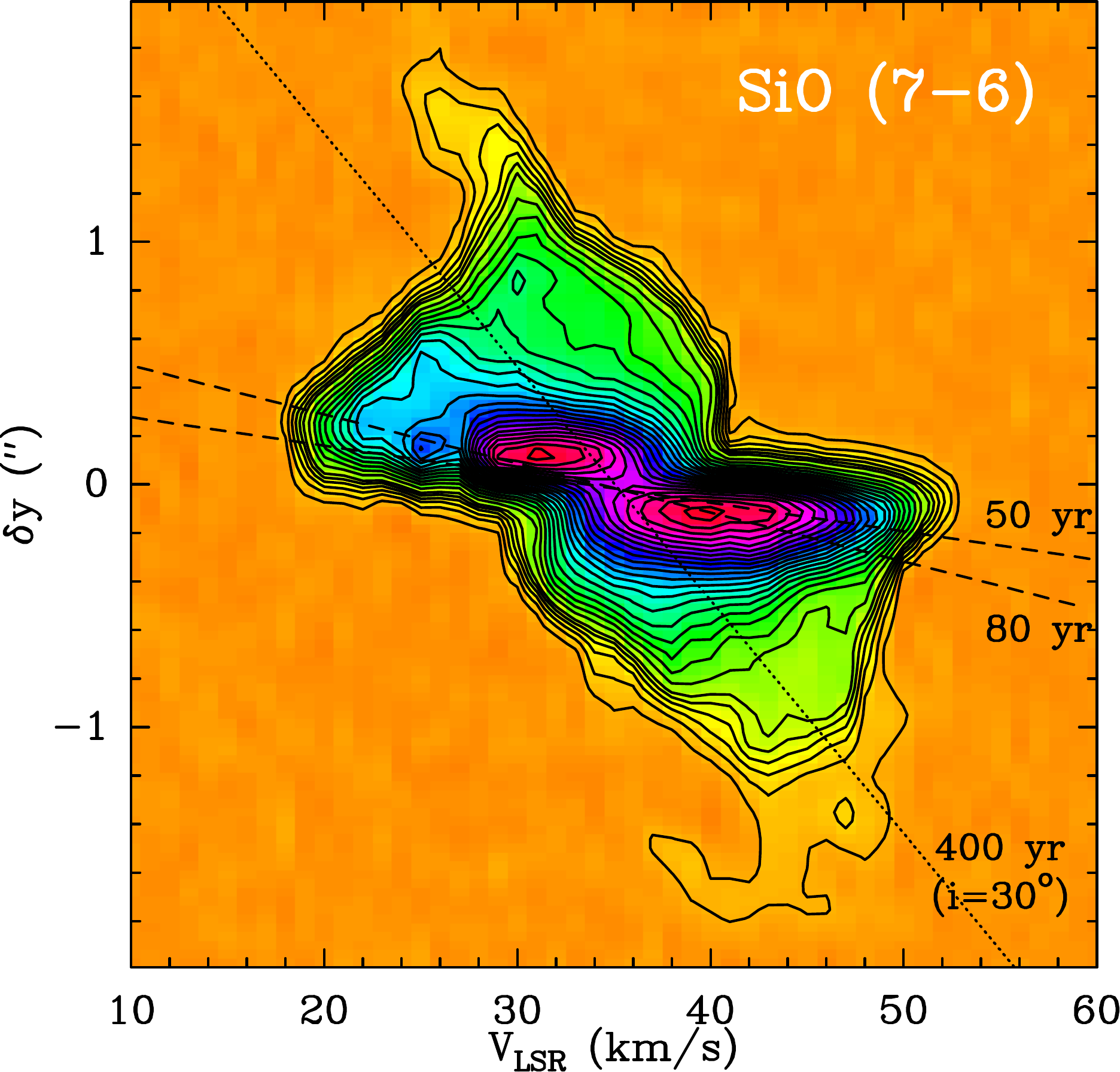}
        \caption{\siot\ compact bipolar outflow emerging from clump
          S. {\bf Top:} Velocity-channel maps rotated clockwise by
          PA=21\degr; contours are 1$\sigma$, 2$\sigma$, 3$\sigma$,
          5$\sigma$, 7$\sigma$, 10$\sigma$... by 5$\sigma$
          ($\sigma$=10\,mJy/beam). The rms of the line-free
          dv=1\,\kms-channel maps is 2.7\,mJy/beam. The background
          image is the 330\,GHz-continuum map. {\bf Bottom-left:}
          Order-zero moment map over the \vlsr=[18-53]\,\kms\ velocity
          range; contours as in top panel but with $\sigma$=90
          mJy/beam\,\kms. {\bf Bottom-right:} Axial position-velocity
          diagrams along PA=21\degr; contours are 1$\sigma$ to
          6$\sigma$ by 1$\sigma$, and from 8$\sigma$... by 2$\sigma$
          ($\sigma$=9\,mJy/beam). The lines represent different
          velocity gradients implying different kinematic ages at
          various regions of the SiO-outflow [dashed-black: the
            SiO-knots, dotted-black: flame-shaped lobes (average).]}
         \label{f-sio}
   \end{figure*}

\begin{figure*}[htbp] 
   \centering 
   \includegraphics[width=0.85\hsize]{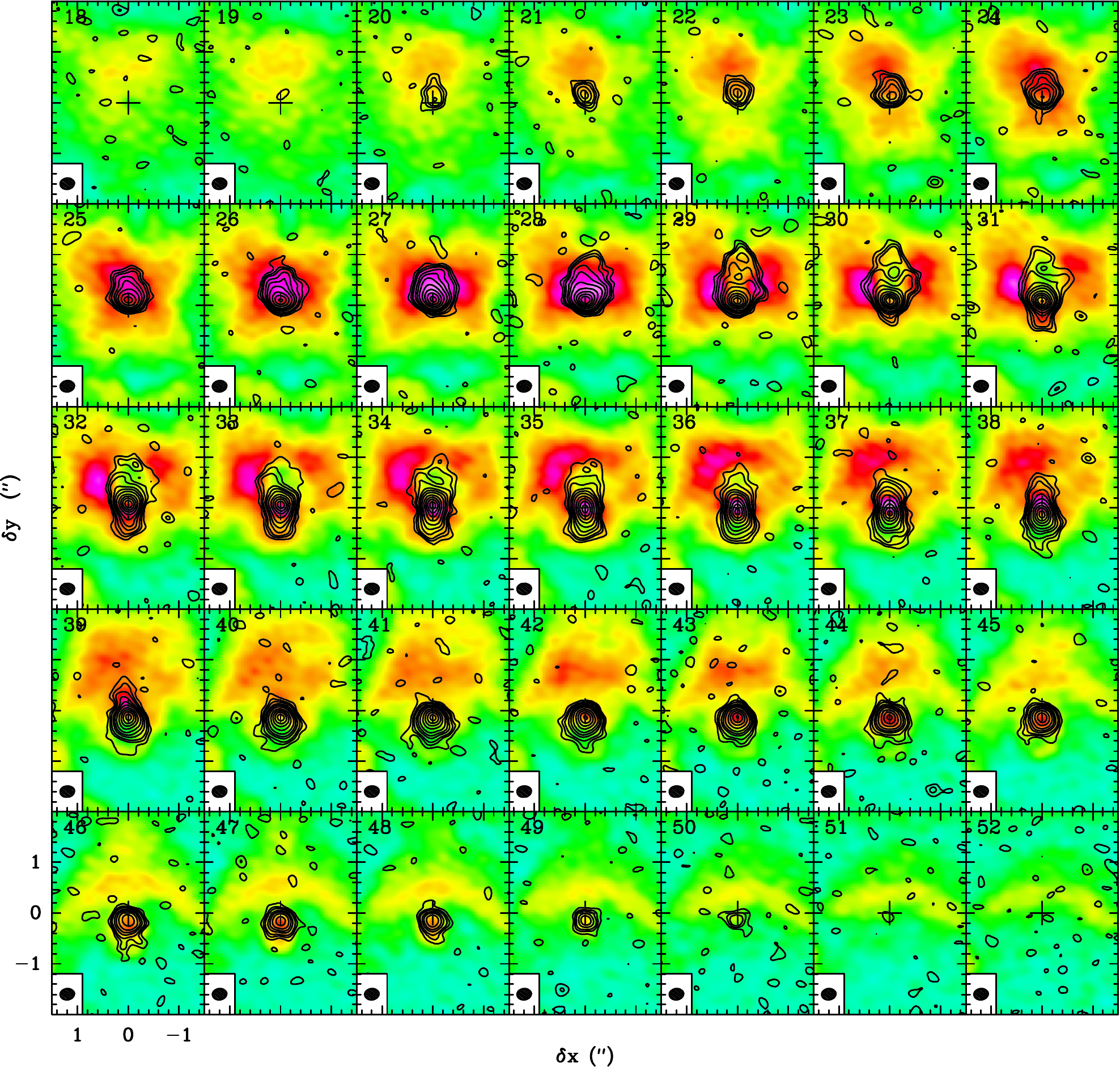}
   \includegraphics*[bb=60 0 540 600,clip,width=0.33\hsize]{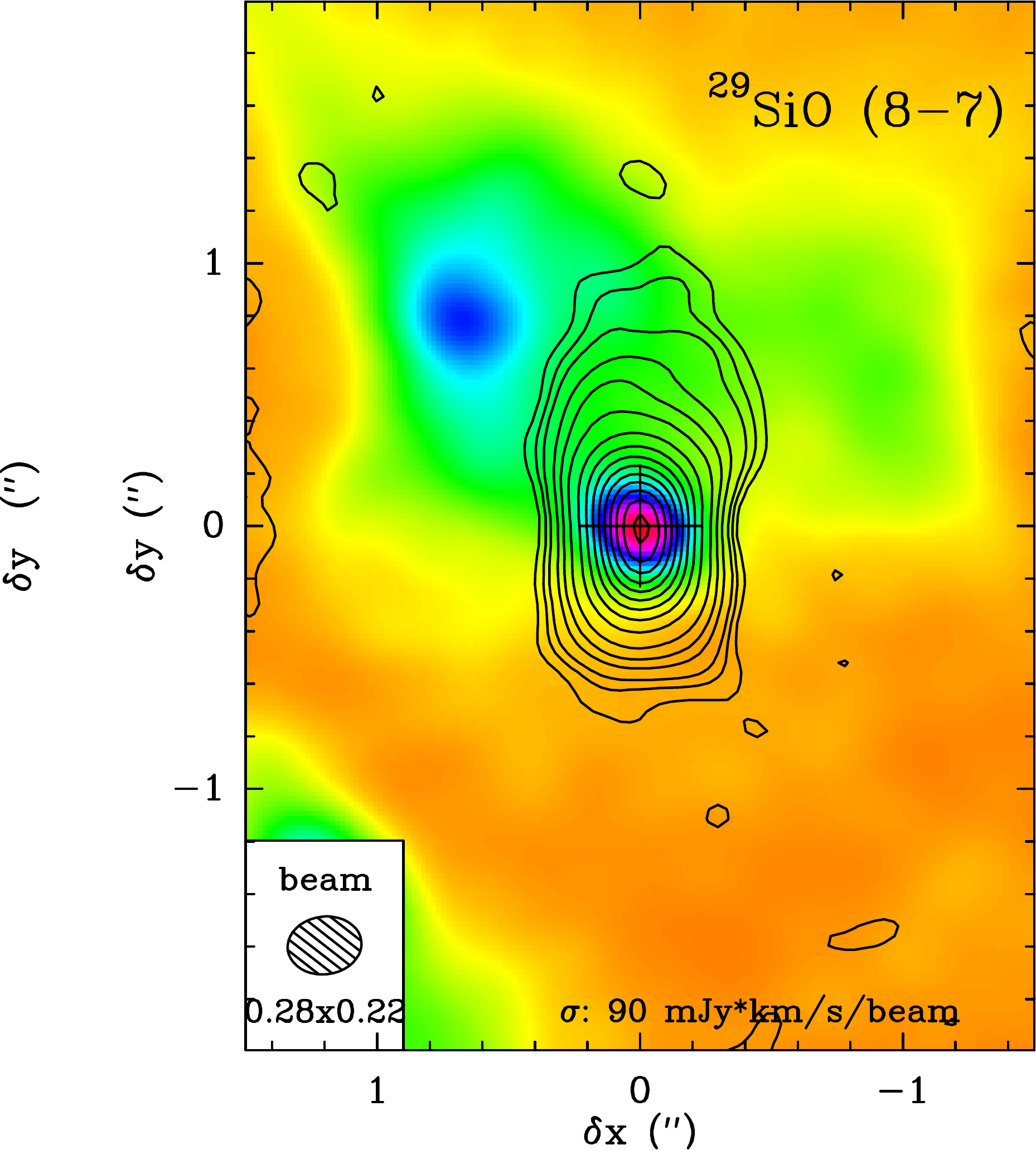} 
   \includegraphics[width=0.425\hsize]{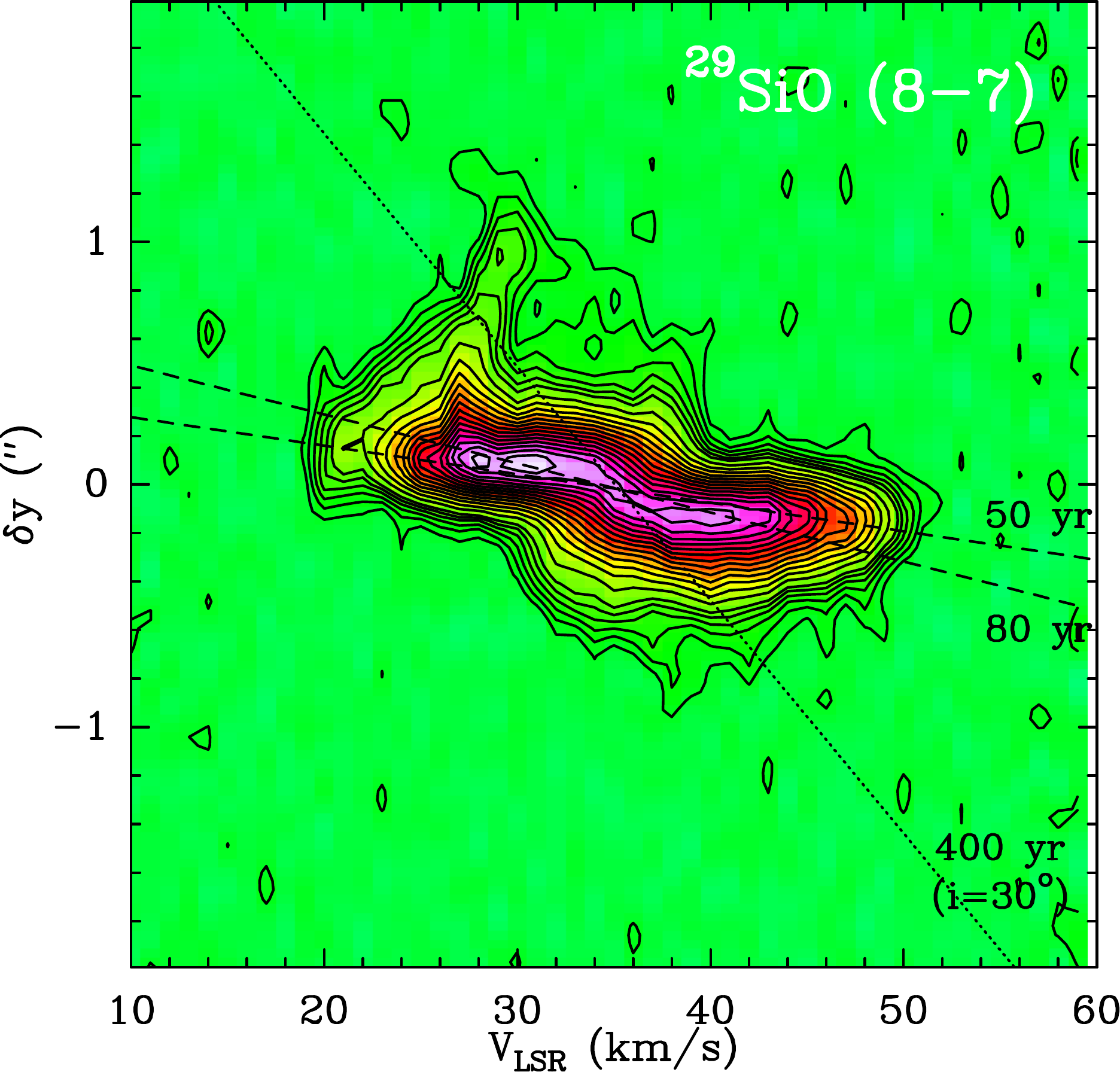}
   \caption{As in Fig.\ref{f-sio} but for the transition \vsiot. The
   velocity-channel maps (top) are overplotted on the \trecet\ maps;
   \vsiot\ map contours are 1$\sigma$, 2$\sigma$, 3$\sigma$, 5$\sigma$,
   7$\sigma$, 10$\sigma$... by 5$\sigma$ ($\sigma$=9\,mJy/beam). In
   the bottom panels levels as in Fig.\,\ref{f-sio}.
   }
         \label{f-29sio}
   \end{figure*}

Our \siot\ and \vsiot\ velocity-channel maps
(Figs.\,\ref{f-sio} and \ref{f-29sio}) show very similar brightness
distributions, both transitions displaying a notable bipolar
morphology with two flame-shaped lobes
at either side of \cs. The SiO lobes are oriented along
PA$\sim$16\degr-18\degr, that is, similarly but not totally equal to the
large-scale CO outflow.  
The emission is centered at \vlsr$\sim$35\,\kms, with most of the
blue-shifted (down to \vlsr=18\,\kms) and red-shifted (up to
\vlsr=53\,\kms) emission coming from the north and south SiO lobe,
respectively. This is
consistent with a dominant expansive kinematics centered at \cs.

The flame-shaped SiO lobes have a relative surface brightness dip at
their respective centers suggestive of a dense-walled structure (this is best seen
in the north lobe in the \vlsr$\sim$29-30\,\kms-channel maps of
\vsiot, Fig.\,\ref{f-29sio}).  Our \siot\ maps show a
pinched-waist emission distribution that is also hinted, but is less
apparent, in the maps of the \vsiot\ line. 
From the maps of both transitions, SiO and \vsio, we derive a similar
size for the pinched-waist, which has a deconvolved half-intensity
diameter of $\diameter_{\rm 1/2}$$\sim$0\farc18$\sim$270\,AU (as
measured in the \vlsr=35\,\kms-channel map).

Unlike the large-scale CO outflow, the SiO emission distribution is
rather (but not perfectly) symmetric about the nebular axis and the
equator.
An {\tt S}-shape distribution
is hinted, with both the east side of the north lobe and the west
side of the south lobe being moderately brighter than its mirror image
about the symmetry axis of the nebula.
Along the nebula axis, the SiO outflow reaches out to distances of
$\sim$1\farcs8$\sim$2700\,AU from the center, 
with the north lobe being slightly brighter at the tips. 
The dimensions of the outflow deduced from the maps of the weaker,
optically thinner \vsiot\ transition are moderately smaller along the
nebula axis as a result of the lower signal-to-noise ratio of these
maps.

The surface brightness distribution of SiO and \vsio\ peaks at two
diametrically opposed compact regions located at offsets
$\delta$y$\sim$$\pm$0\farcs11 along the nebula axis (hereafter,
referred to as SiO knots).  This is most clearly seen in the
position-velocity (PV) diagrams along the major axis of the bipolar
outflow (Fig.\,\ref{f-sio} and \ref{f-29sio}, bottom panels).
The distribution of the SiO emission increases abruptly from the
center (offset 0\arcsec, \vlsr$\sim$35\,\kms) to the bright SiO knots,
where the full width at zero-level intensity (FWZI) of the emission
reaches a maximum value of $\sim$23\,\kms.

This feature in the axial-PV diagram (i.e., abrupt velocity rise from
the center to a compact region with the largest velocity-spread) is a
well-known signature of bow-shocks. The latter are produced, for
example, when a collimated fast wind or bullet collides with dense
ambient material \citep[e.g.,][]{har87,lee03,bal13} and have been
observed before in other pPNe, for example, in the molecular outflow
of IRAS\,22026+5306 \citep{sah06}, in the ionized central regions of
M\,1-92 \citep{den08}, and in the compact, shock-excited axial
blobs or knots of Hen\,3-1475 \citep[e.g.,][and references
  therein]{rie06}. Analytical bow-shock models \citep{har87} as well
as numerical computations \citep[e.g.,][]{den08} show that the FWZI of
the emission profile produced by a radiating bow-shock equals the
shock velocity, \vs; this would imply that \vs$\sim$23\,\kms\ for both
the north and south bright SiO knots of \ohs.  Using the simple
analytic formula derived by \cite{har87}, which relates the centroid
of the bow-shock profile (\vlsr$\sim$29.5 and $\sim$41\,\kms\ at
$\delta$y=+0\farcs11 and $\delta$y=$-$0\farcs11, respectively) with
the shock-velocity, we deduce that the bow-shocks are viewed at an
angle $i$$\sim$30\degr\ with respect to the plane of the sky, again similar but not the same as the large-scale outflow.

The axial PV diagram of SiO and \vsio\ is somewhat structured and it
is not possible to derive a unique velocity gradient from the center
to the tips of the SiO lobes. The large velocities observed in the
very inner regions of the outflow, near the SiO knots, are consistent
with a large axial velocity gradient ranging between \vgrad$\sim$85
and 50\,\kms\,arcsec$^{-1}$ (black dashed lines in Figs.\,\ref{f-sio}
and \ref{f-29sio}). This gradient implies short kinematical ages of
\tdyn$\sim$[84-142]$\times$$\tan{(i)}$ years, that is,
\tdyn$\sim$50-80\,yr adopting $i$=30\degr\ (or \tdyn$\sim$60-100\,yr,
adopting the commonly used value of the inclination for the
large-scale CO outflow, $i$=35\degr).

The lower velocities at the intermediate-to-outer regions of the
flame-shaped SiO lobes imply larger kinematic ages of
\tdyn$\sim$690$\times$$\tan{(i)}$\,yr (about 400-500\,yr, for
$i$=30-35\degr) and larger at the lobe tips. In these regions,
however, the interpretation of the gradient is not straightforward
since
it probably
involves 
outflow deceleration from swept-up ambient material.

\subsection{The periphery of the SiO outflow}
\label{res-minihg}

\begin{figure*}[]
   \centering 
   \includegraphics[width=0.4\hsize]{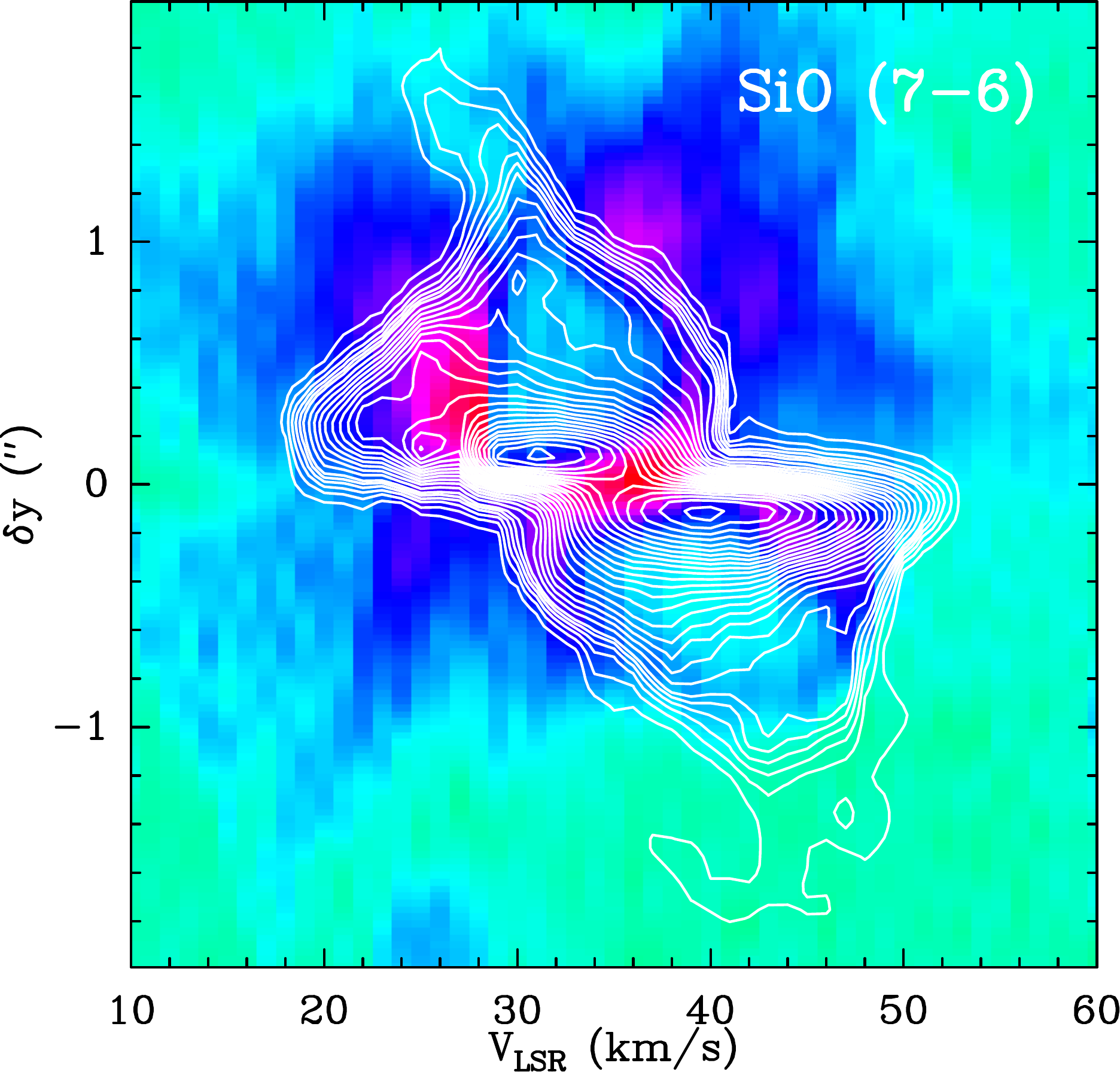}
   \includegraphics[width=0.4\hsize]{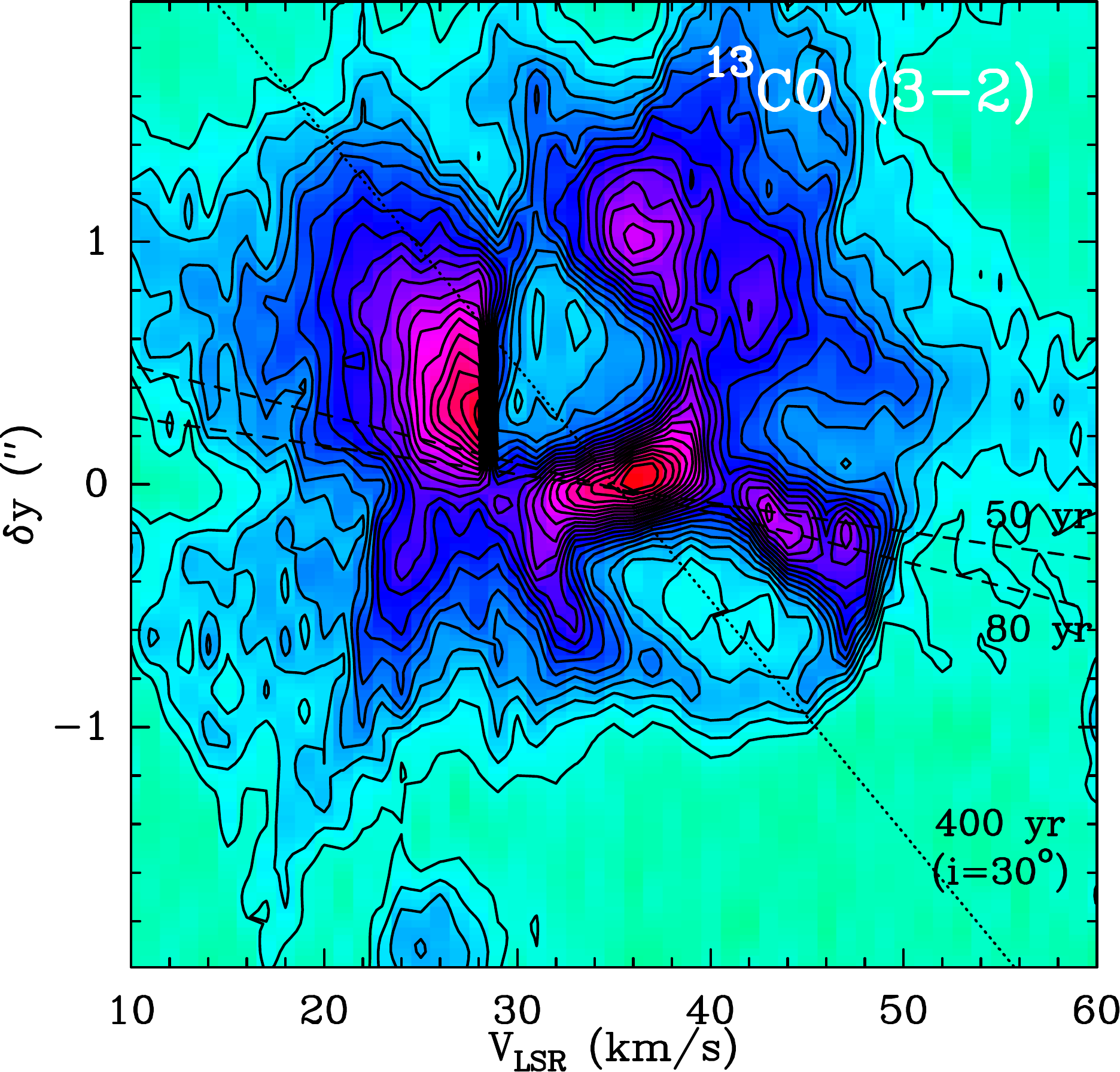}
      \caption{{\bf Left:} Axial PV diagrams of SiO\,$J$=7-6 (white
        contours) atop \trecem\,$J$=3-2 (in a color-scale).
        \siot\ contours as in Fig.\,\ref{f-sio}. {\bf Right:} Axial PV
        diagrams of \trecem\,$J$=3-2 alone. Level spacing is
        20\,mJy/beam.}
   \label{f-pv-SiO-13co}
   \end{figure*}

The close environments of the SiO outflow are remarkably
disrupted. For example, this is clearly noticed in the \trecem\,(3-2)
and CS\,(6-5) velocity-channel maps, and their respective axial PV
diagrams, in regions within $\sim$$\pm$2\arcsec\ from the center
(Figs.\,\ref{f-pv-SiO-13co} and \ref{f-cs65}). These data show a surface
brightness dip or minimum in the regions and LSR velocities where the
SiO outflow is prominent. This suggests that the SiO outflow (perhaps
jointly with an underlying powering wind not directly seen in these
data)
is carving out the ambient material as it propagates outwards.

Since the expansion center of the SiO outflow (near \cs) is located
about $\sim$0\farc6 to the south of the waist's center, the north
SiO lobe is mainly colliding with, and piercing through, the slowly
expanding equatorial waist (see \vlsr=[29:38]\,\kms-channel maps in
Fig.\,\ref{f-pv-SiO-13co}) whereas the south SiO lobe is colliding
with the, presumably less dense, ambient material residing mainly in
the base of the south large-scale lobe (and also partially in the
south part of the waist).

\begin{figure*}[htbp]
   \centering 
   \includegraphics[width=0.425\hsize]{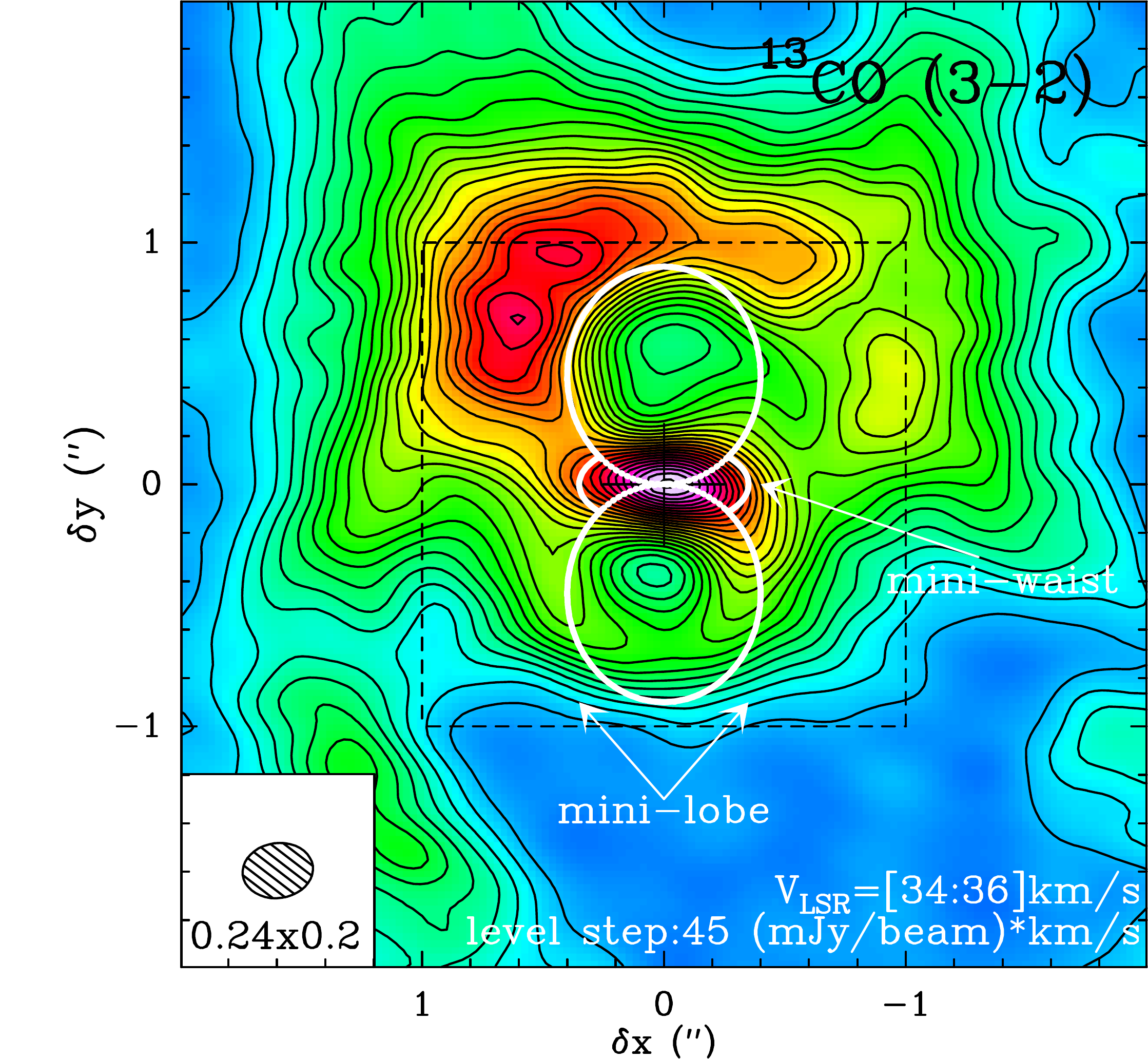} \hspace{0.25cm}
   \includegraphics*[bb=0 0 600 530,clip,width=0.45\hsize]{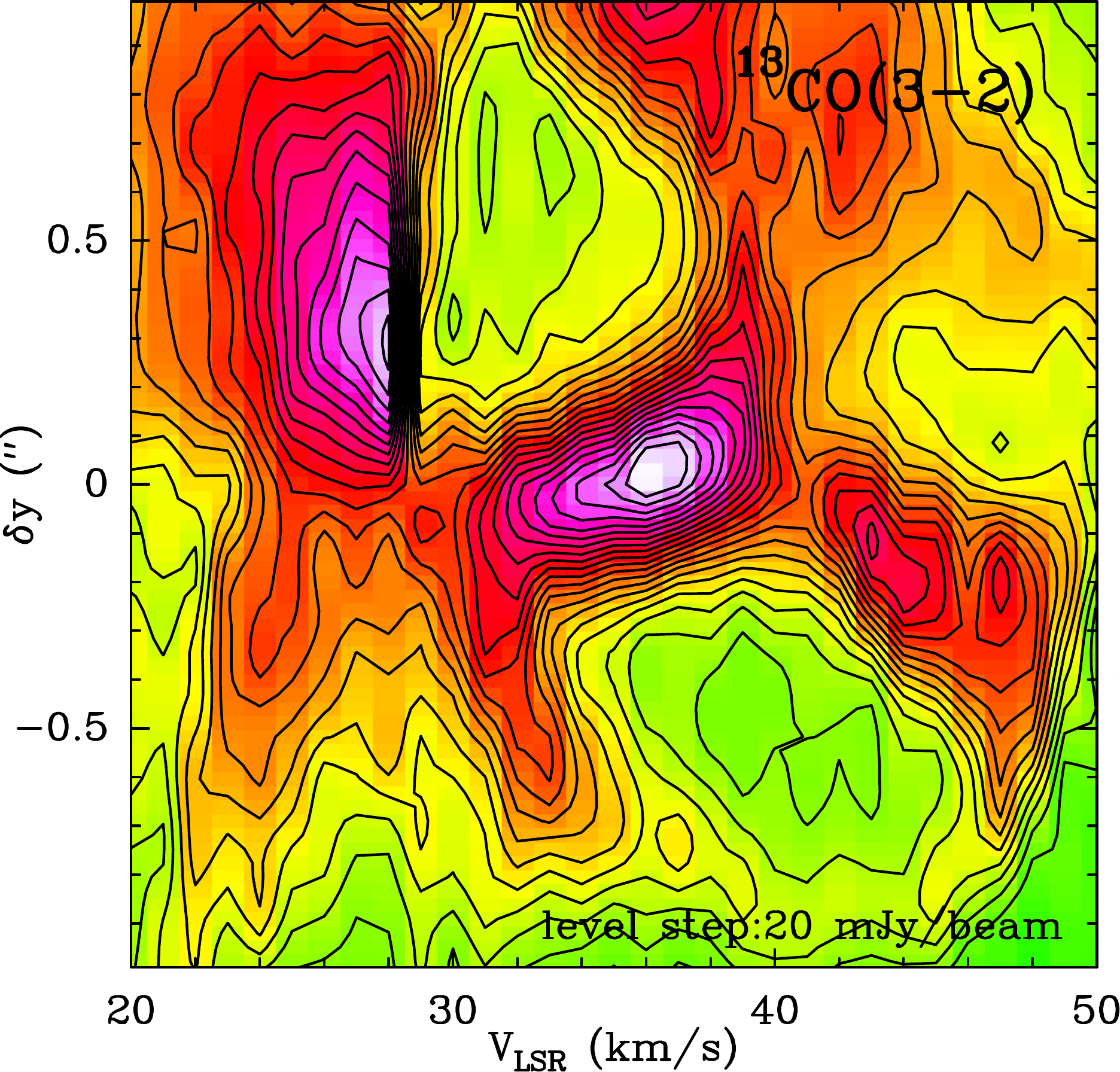} 
   \includegraphics[width=0.425\hsize]{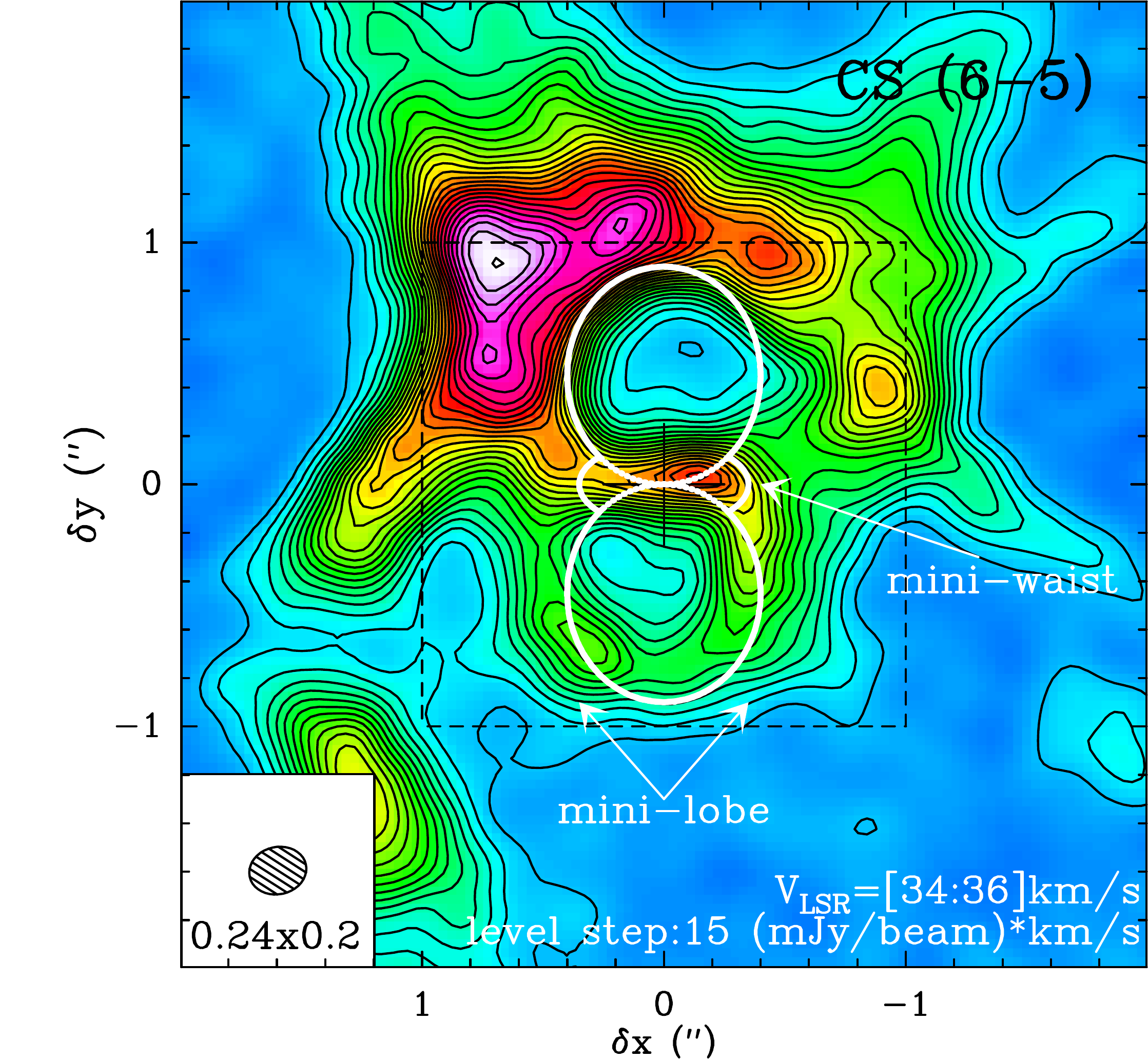}  \hspace{0.25cm}
   \includegraphics*[bb=0 0 600 530,clip,width=0.45\hsize]{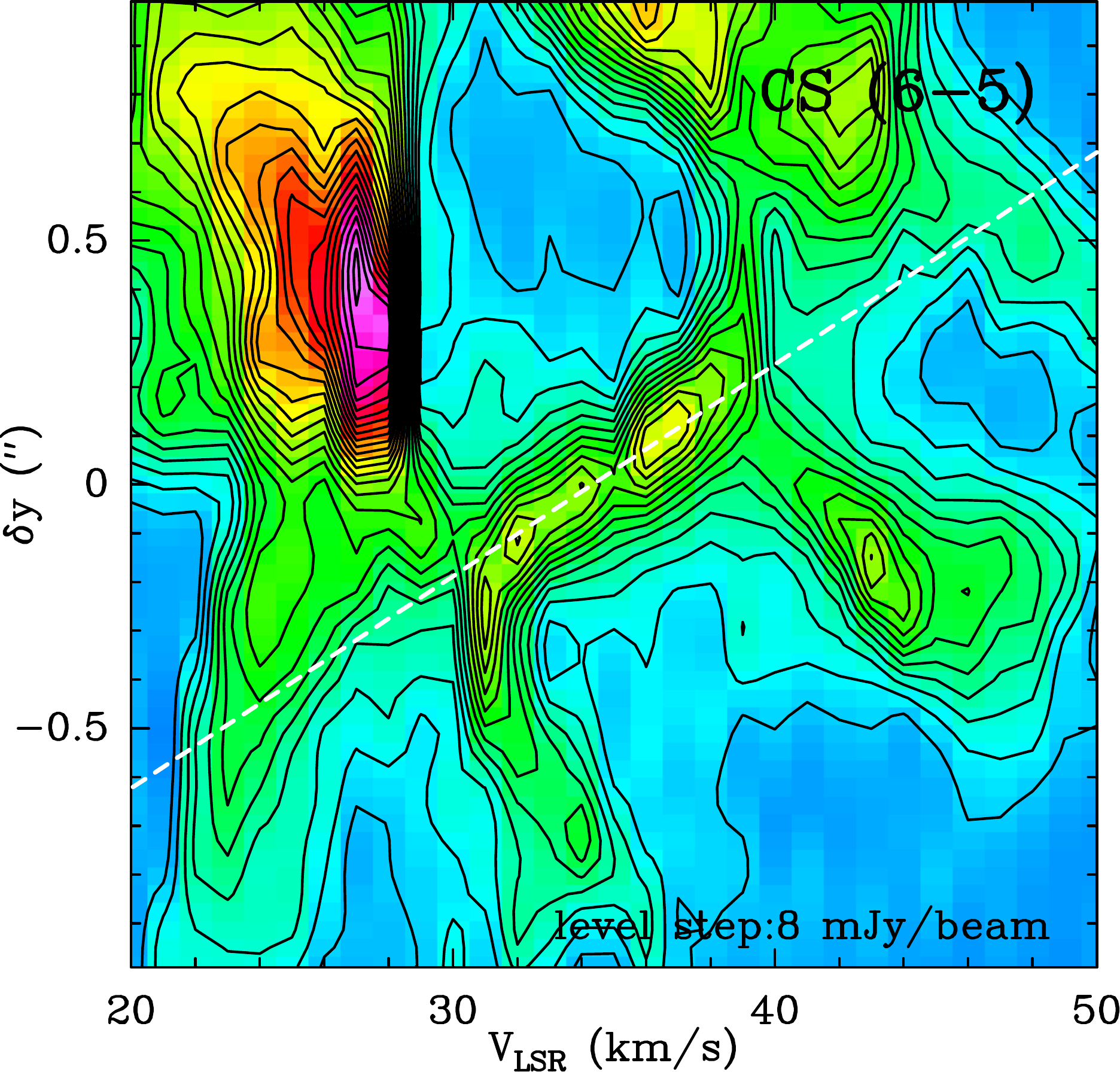} 
   \includegraphics[width=0.425\hsize]{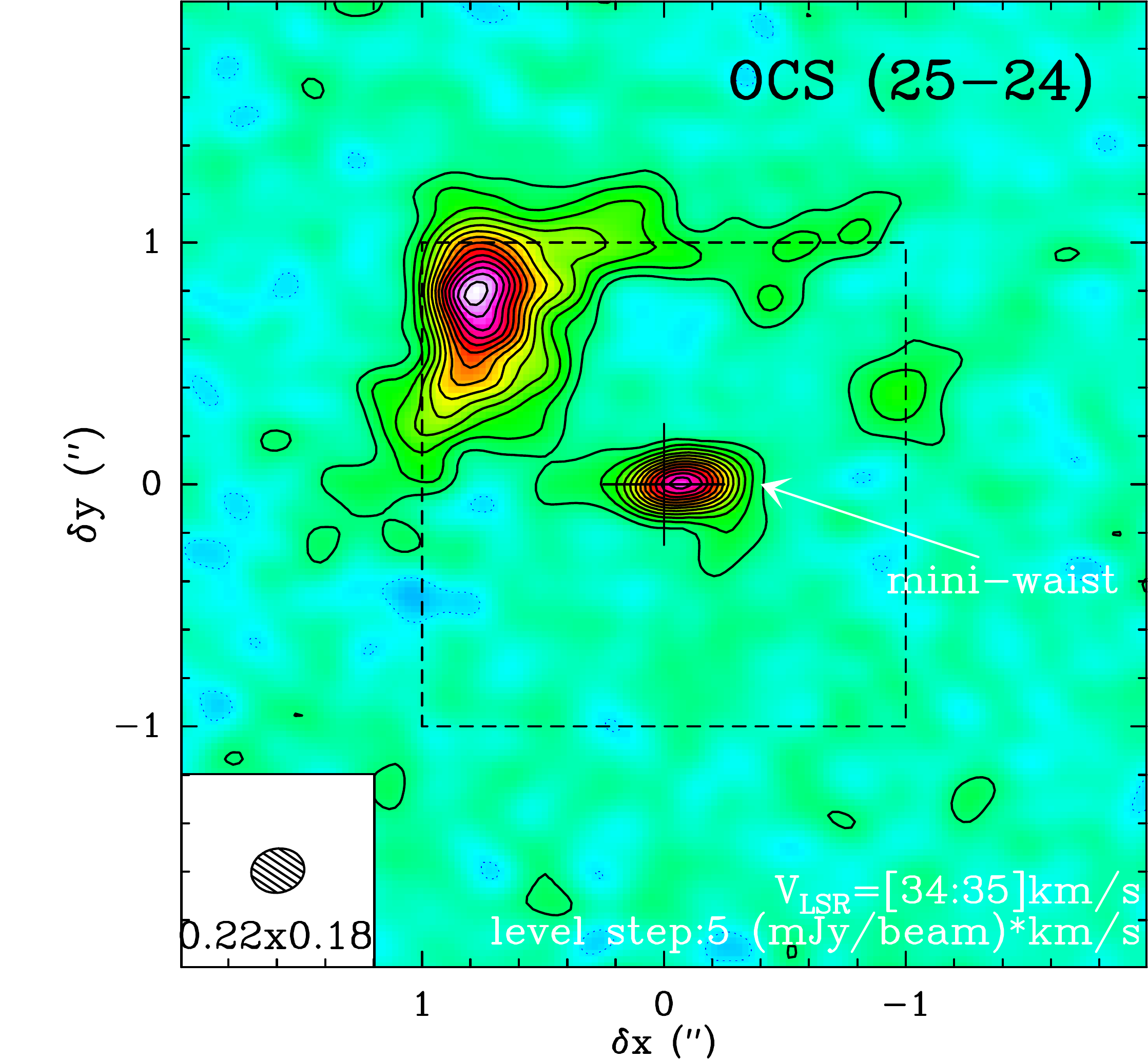}              \hspace{0.25cm}
   \includegraphics*[bb=0 0 600 530,clip,width=0.45\hsize]{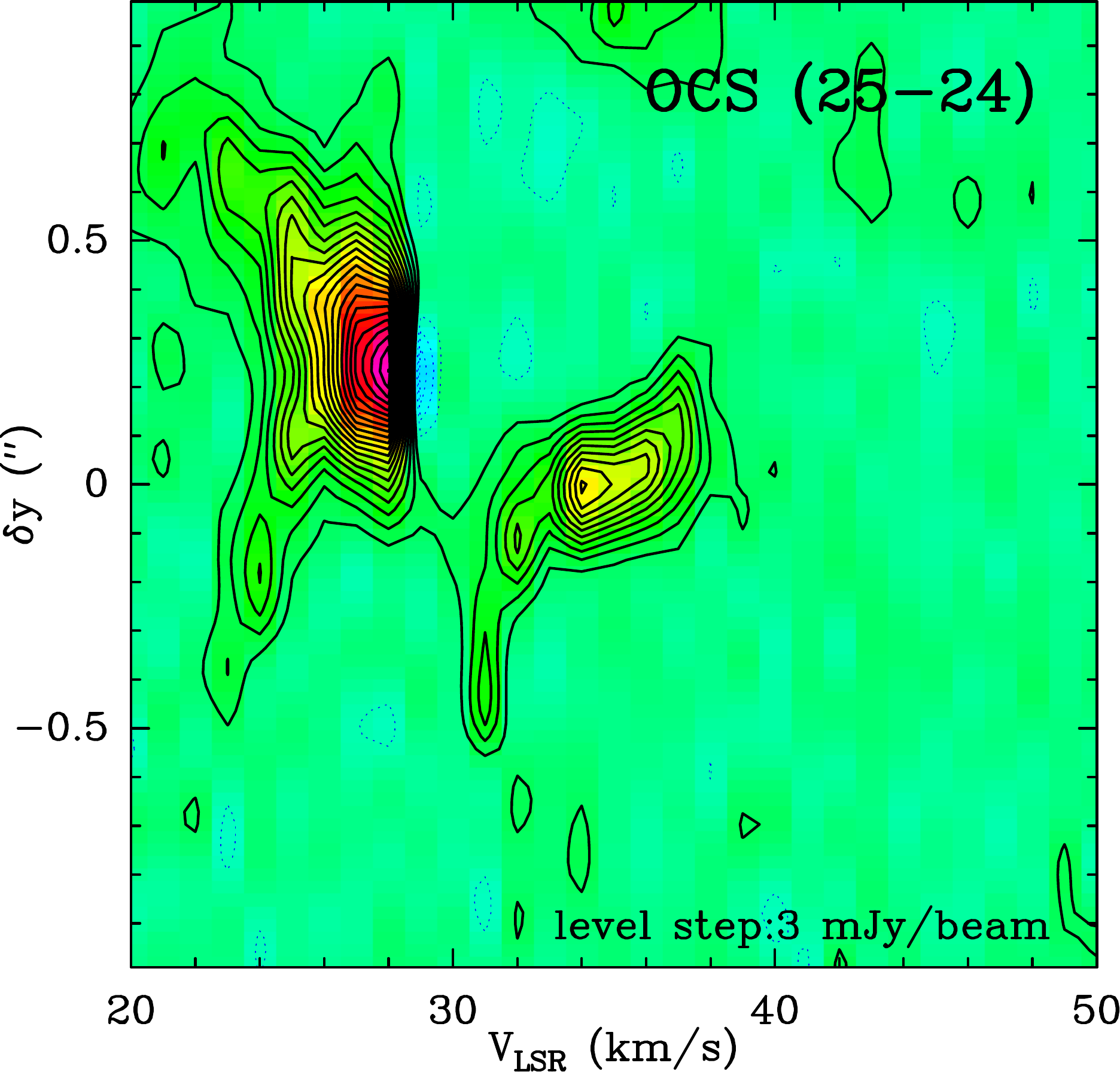}  
   \caption{{\bf Left:} Zero-order moment maps of \trecet, CS\,($J$=6-5),
     and OCS\,($J$=25-24) integrated over the central channels near
     \vlsr=35\,\kms, showing the
     $\sim$1\arcsec$\times$2\arcsec\ mini-hourglass centered at
     \cs\ (schematically depicted atop).  The OCS emission is largely
     restricted to the equatorial waist, of both the mini- and
     large-hourglass. The dashed squares indicate the smaller FoV used
     in the axial PV diagrams to the right.
     {\bf Right:} PV diagrams along PA=21\degr. The linear velocity
     gradient along the mini-waist is indicated with a dashed line.}
   \label{f-miniw}
   \end{figure*}

In agreement with this wind-collision
scenario, we identify a small-scale hourglass-shaped structure
adjacent to the SiO outflow in the ALMA maps of most molecules.
This structure, dubbed as the mini-hourglass (or mini-hg), is
nested inside the dense central regions of the large-scale nebula and
is clearly recognized, for example, in the axial PV diagrams
shown in Figs.\,\ref{f-pv-SiO-13co}-\ref{f-miniw}. 
The mini-hg, and the two polar cavities inside, are also
directly visible in the velocity-channel maps of most transitions near
the systemic velocity, \vlsr=35\,\kms\ (Fig.\,\ref{f-miniw},
top-panel). We believe the walls of this hourglass-shaped structure
are mainly made of compressed material formed as the SiO outflow
propagates throughout the dense, central regions of the nebula.

The south lobe of the mini-hourglass, or south mini-lobe, can be
traced (and is relatively well isolated from other nebular components)
along the nebula axis down to $\delta$y$\sim$$-$1\arcsec\ spanning the
range of velocities \vlsr=30-52\,\kms. The south mini-lobe follows a
velocity gradient similar to that of the outer regions of the south
flame-shaped SiO lobe (dotted line in the axial PV diagrams of
Fig.\,\ref{f-pv-SiO-13co} and \ref{f-cs65}). This suggests a similar age
for both the extended flame-shaped SiO lobes and the mini-lobes 
(\tdyn$\sim$690$\times$$\tan{(i)}$\,yr) or a similar dynamics. 
The north mini-lobe is not
so neatly delineated in our maps.
Yet, we clearly identify its central cavity (in the range
\vlsr=29-38\,\kms) surrounded by very bright emission from the dense
ambient material into which the SiO outflow is plowing.
The velocity gradient observed in the north mini-lobe is lower than in
the south, as is also the case for the flamed-shaped lobes
sampled by SiO and \vsio\ emission. This is consistent with a larger
resistance to the SiO outflow propagation toward the north (offered
by the ambient material in the dense large waist) than toward the
south, where the SiO outflow plunges into a presumably less dense
ambient medium.

The equatorial waist of the compact mini-hourglass (referred to as the
mini-waist) is partially resolved in our ALMA maps and is best
isolated in the OCS\,(25-24) emission maps (Fig.\ref{f-miniw}), where
we measure a full extent (at a $\sim$2$\sigma$ level) of
$\sim$0\farc35$\times$0\farc7, with the long axis oriented along
PA$\sim$21+90\degr.
This is consistent with a torus or disk-like structure with an outer
radius of about 500\,AU whose symmetry axis is viewed at an
  inclination angle of $i$$\sim$30\degr, that is,
orthogonal to the SiO outflow
(\S\,\ref{res-sio}).

The axial PV diagrams in Fig.\,\ref{f-miniw} show that south/north
emission from the mini-waist (within $\delta y$$\pm$0\farc18) is
blue-/red-shifted, that is, the velocity gradient 
has opposite sign to that observed along the mini-lobes, which is 
indicative of equatorial expansion.
The expansion velocity of the mini-waist is notably low and is not
constant across the waist: it increases \emph{linearly} from
the inner to the outer regions (with values as low as
|\vlsr-35|$\sim$1\,\kms\ at the center and reaching
|\vlsr-35|$\sim$5\,\kms\ at the edges).
The gradient
observed, \vgrad$\sim$\,20-30\,\kms\,arcsec$^{-1}$, leads to a
kinematical age for the equatorial mini-waist of
\tdyn$\sim$[360-240]/$\tan{(i)}$ years.
Assuming that the lobes and the waist of the mini-hourglass
of \ohs\ have the same kinematic age, as expected if both structures
resulted from the same physical process (e.g., a sudden
mass-loss ejection or mass-acceleration event), then we derive a value
of the inclination of $i$$\sim$30\degr-35\degr\ and a kinematical age
of about 500\,yr for these components.


{\sl -- The \metanol-outflow.}  We have detected two transitions from
methanol at 304.2 and 307.2\,GHz (see Table\,\ref{t-lines} and
Fig.\,\ref{f-ch3oh}). Their brightness distributions exhibit a clear
bipolar morphology. The dimensions of the region where most of
the \metanol\ emission is produced are similar to that of the compact
SiO outflow. Weaker \metanol-emission components are also detected at
larger distances from the center, up to $\sim$3-4\arcsec\ along the
nebula axis. The methanol emission profile is sharply peaked near
\vlsr$\sim$\,2, 29, 44 and 71\,\kms, with extremely weak (undetected)
emission at intermediate velocities.

The axial PV diagrams of the \metanol\ and \trecet\ lines are plotted
together in the bottom panels of Fig.\,\ref{f-ch3oh}. In these
diagrams, it is easy to see that the \metanol\ emission closely
follows the spatio-kinematics of the mini-hourglass, neighboring the SiO outflow. 
The \metanol\ emission is rather patchy, with the
brightest emission arising in the front side of the north mini-lobe
(at \vlsr$\sim$29\,\kms) and close to its tip.


  \subsection{The large-scale nebula}
  \label{res-lsnebula}

  \subsubsection{The hourglass-shaped structure}
  \label{res-largehg}

\begin{figure*}[htbp]
   \centering 
    \includegraphics[width=0.925\hsize]{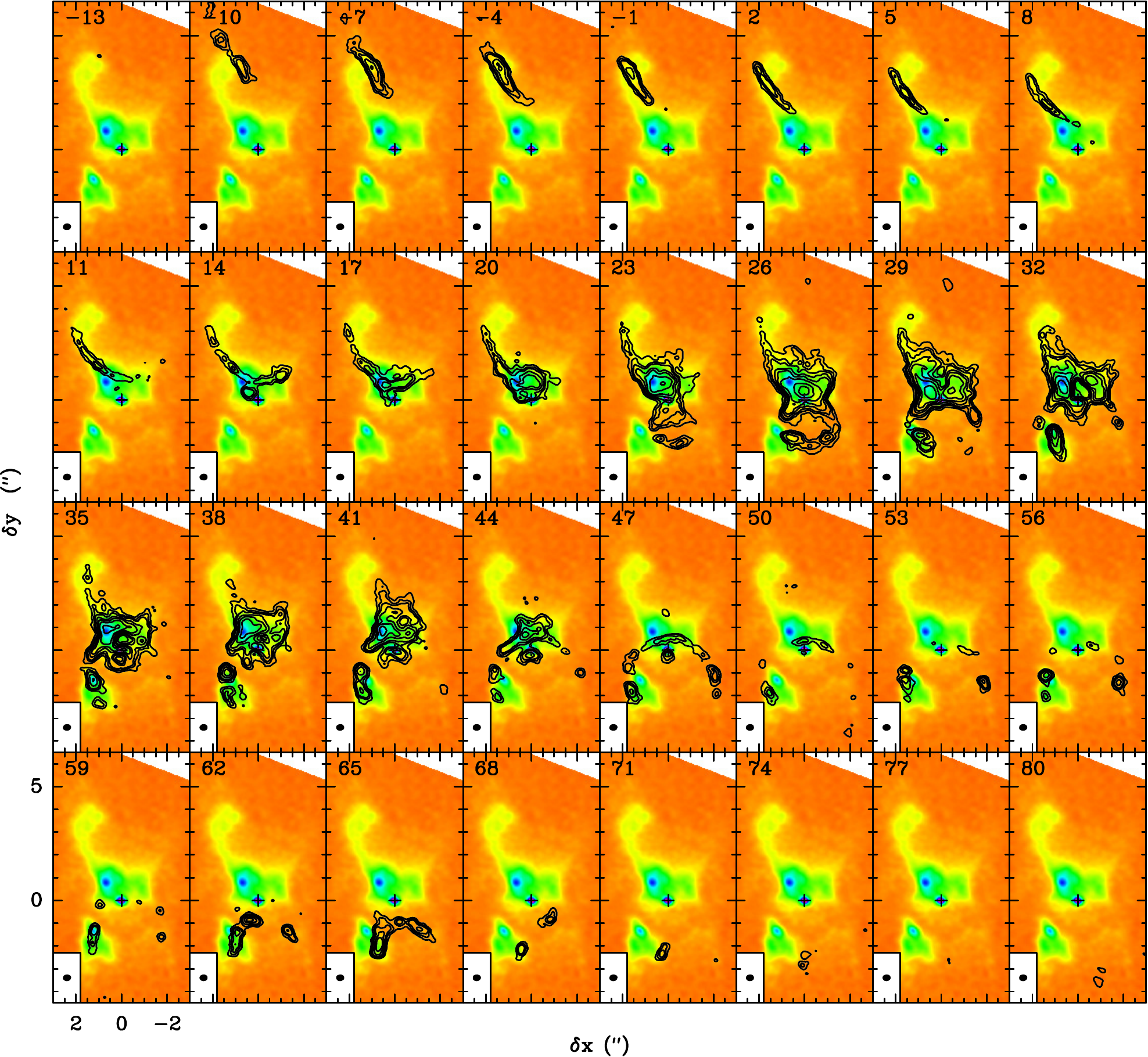}
    \includegraphics*[bb=30 0 400 600, width=0.265\hsize]{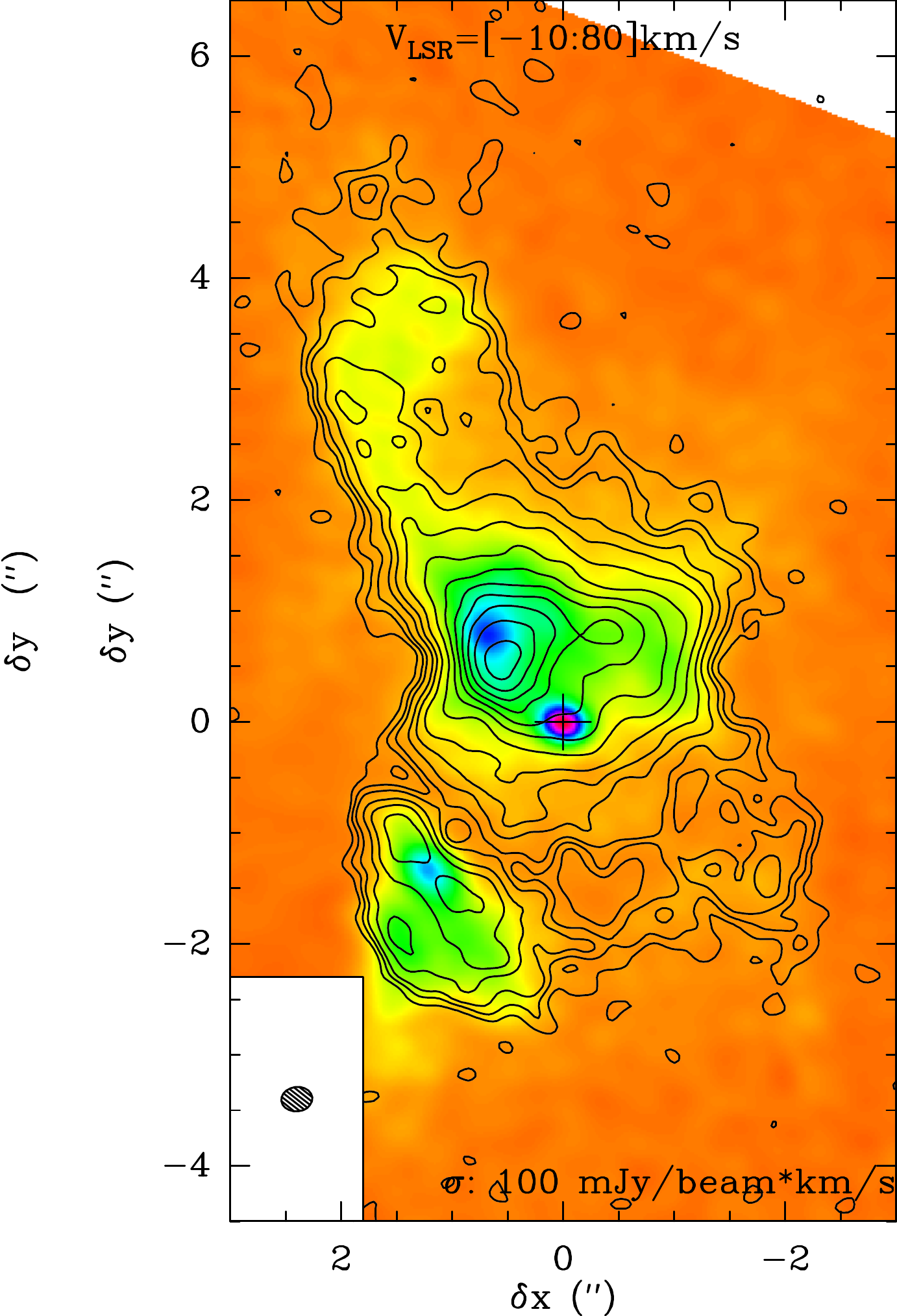}
    \includegraphics[width=0.42\hsize]{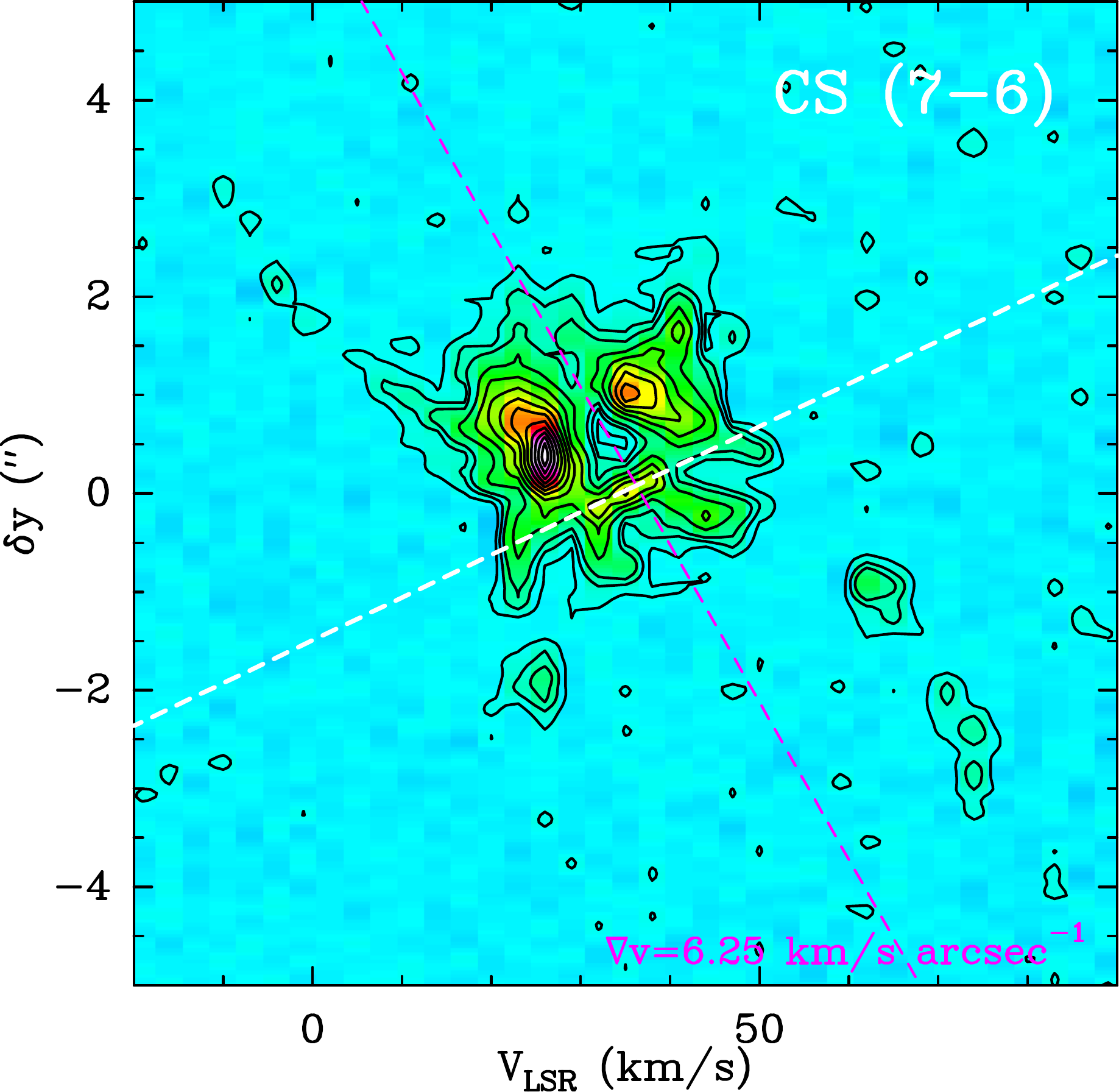}
       \caption{CS\,(7-6) velocity-channel maps (top), zero-order
         moment maps (bottom-left) and axial PV diagram (bottom-right)
         in the low-to-intermediate \vlsr\ range that samples the
         large-scale hourglass; beam HPBW=0\farc28$\times$0\farc22
         (PA=$-$83\degr). Contours are 1$\sigma$, 2$\sigma$,
         3$\sigma$, 5$\sigma$, 7$\sigma$, 10$\sigma$,... by 5$\sigma$
         (with $\sigma$=15\,mJy/beam
         in the velocity-channel maps and
         $\sigma$=100\,mJy\,\kms/beam in the integrated intensity
         map). In the PV diagram contours are 1$\sigma$,
         3$\sigma$, 5$\sigma$, 10$\sigma$... by 5$\sigma$
         ($\sigma$=6\,mJy/beam).}
   \label{f-cs76}
   \end{figure*}

  The $\sim$8\arcsec$\times$4\arcsec-sized hourglass nebula traced by
  the dust thermal continuum emission maps (\S\,\ref{res-cont} and
  Fig.\,\ref{f-cont}) is the dominant emission component of most of
  the species mapped in this work. In Fig.\,\ref{f-cs76} we present
  the velocity-channel maps of CS\,(7-6), one of the dense gas tracers
  observed by us with ALMA, where this component is clearly seen in
  the \vlsr=[$-$10:+80]\kms\ range.  (See also the \trecet\ and
  CS\,(6-5) maps already introduced showing the
  4\arcsec$\times$4\arcsec\ central regions of the hourglass,
  Fig.\,\ref{f-pv-SiO-13co} and Fig.\,\ref{f-cs65}).

 
  The large-scale hourglass nebula (or large hg) encompasses
  the equatorial waist and the base of the bipolar lobes known from
  previous works\footnote{The large-scale hourglass nebula roughly
    corresponds to clumps $\sim$I2-to-I4 as originally defined and
    labeled by \citealp{alc01}.}.
    The unprecedented angular resolution of our ALMA maps shows that
    the bipolar lobes, at their base, are essentially large cavities
    (of $\sim$3\arcsec\ in width) surrounded by $\sim$0\farc5-wide
    walls with a nearly parabolic morphology. Our maps reveal a highly
    structured and non-uniform surface brightness distribution in the
    lobe walls and in the equatorial waist. As in the continuum maps,
    the east side of the \lhg\ is much brighter than the west side.
    
  The spatio-kinematic distribution of the large hg confirms an
  overall radial expansion with a linear velocity gradient along the
  nebula long axis (PA=21\degr) of
  \vgrad$\sim$6.0-6.5\,\kms\,arcsec$^{-1}$ (Fig.\,\ref{f-cs76}); the
  same gradient is preserved along the more distant, fastest regions
  of the large-scale lobes (\S\,\ref{res-hv}). This value of the
  gradient is consistent (within uncertainties) with that
  reported from previous studies (\S\,\ref{intro}).

Although the large equatorial waist is well resolved spatially and
spectrally in our ALMA data, an accurate description of its
morphology, dimensions, and kinematics is difficult. This is because
the nebular equator is remarkably disrupted by the interaction with
the SiO outflow, which has plowed into it from the south, and also because
the waist is the region from where the highly-structured bipolar lobes
emerge, and therefore the emission from different nebular structures
and substructures, with complex kinematics, overlap.
With these limitations in mind, the large-scale equatorial waist can
be roughly described as an expanding cylindrical structure orthogonal
to the bipolar lobes.

\begin{figure*}[htbp]
   \centering 
   \includegraphics[width=0.33\hsize]{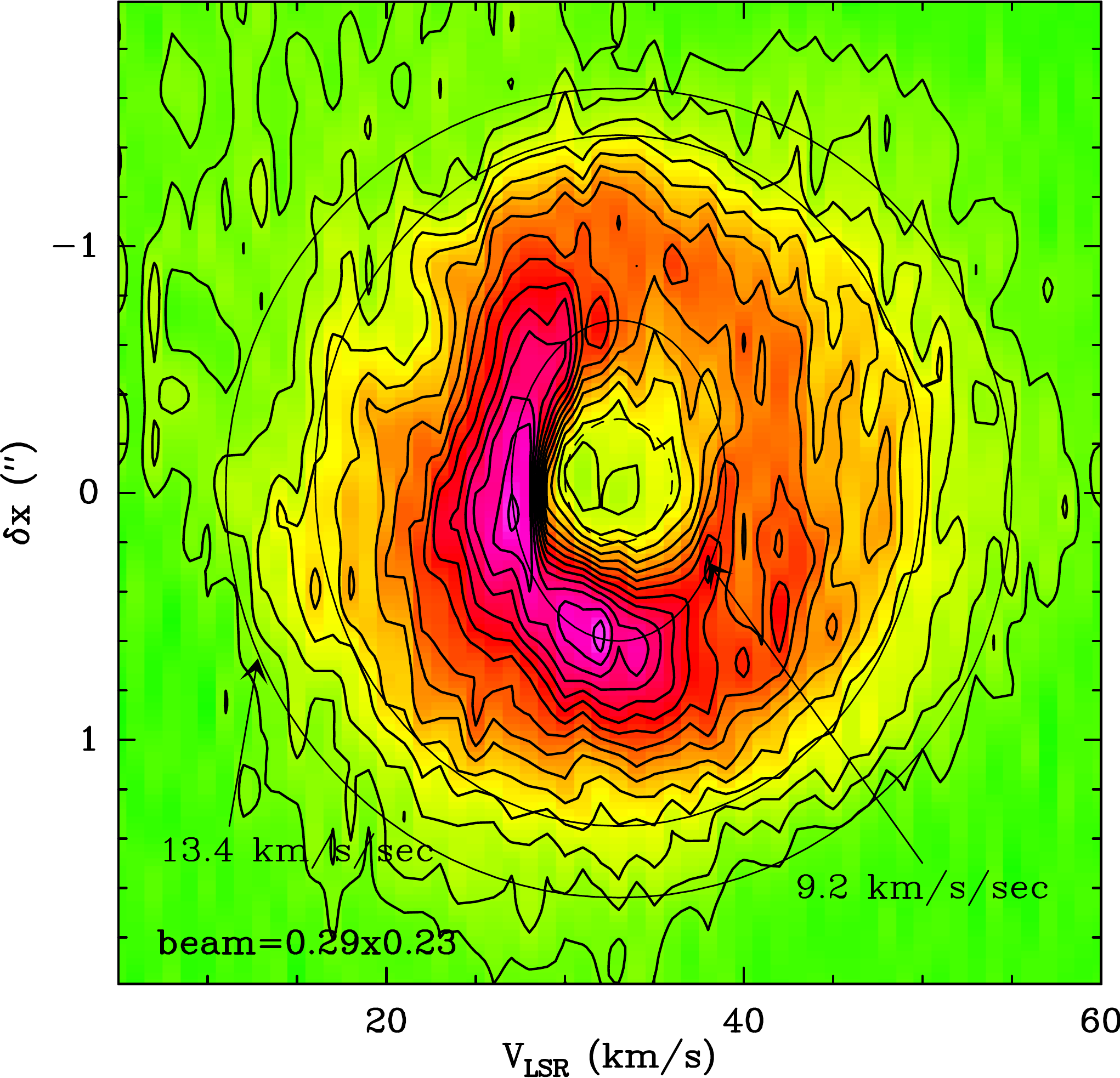}
   \includegraphics[width=0.33\hsize]{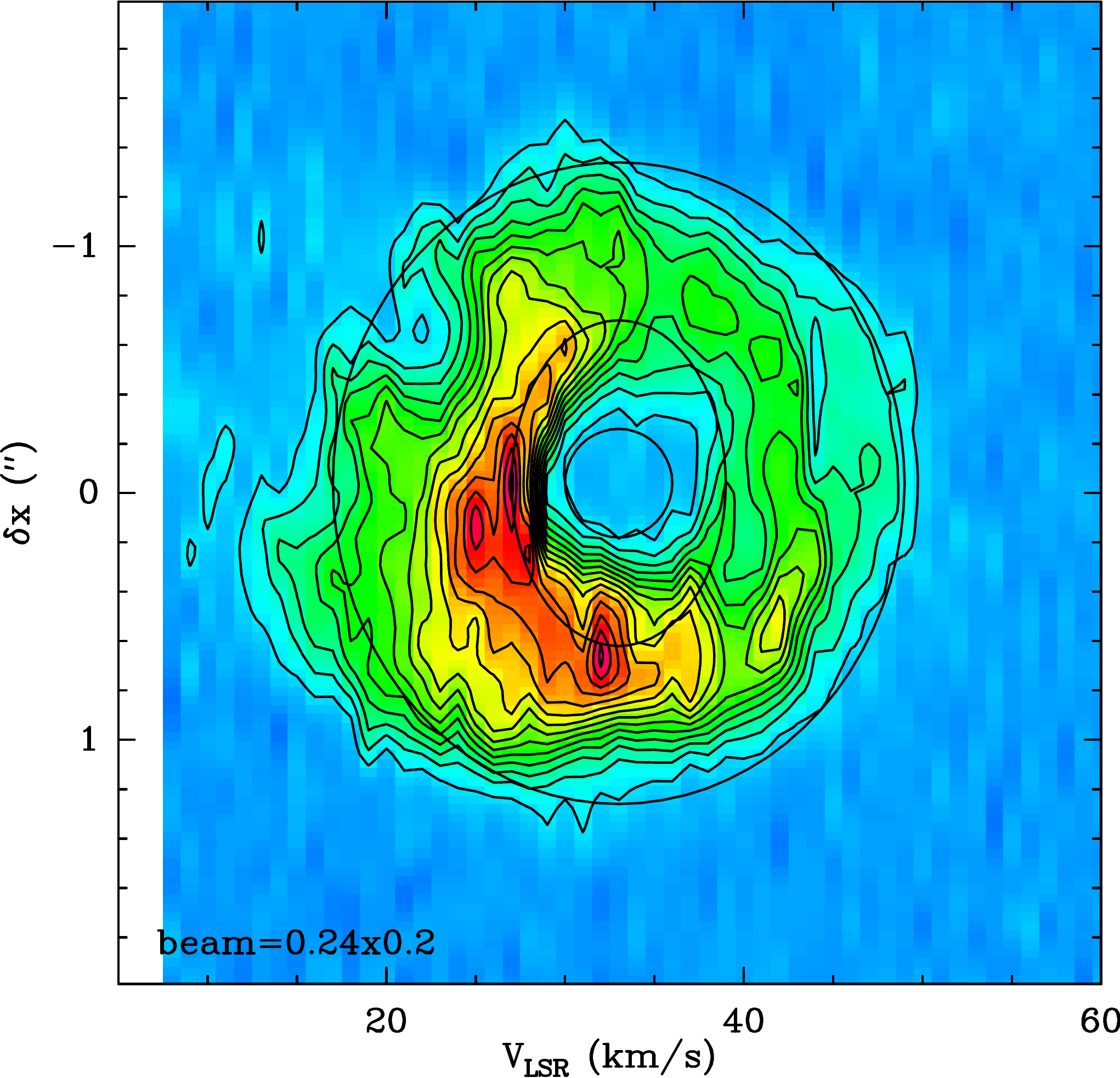}
   \includegraphics[width=0.33\hsize]{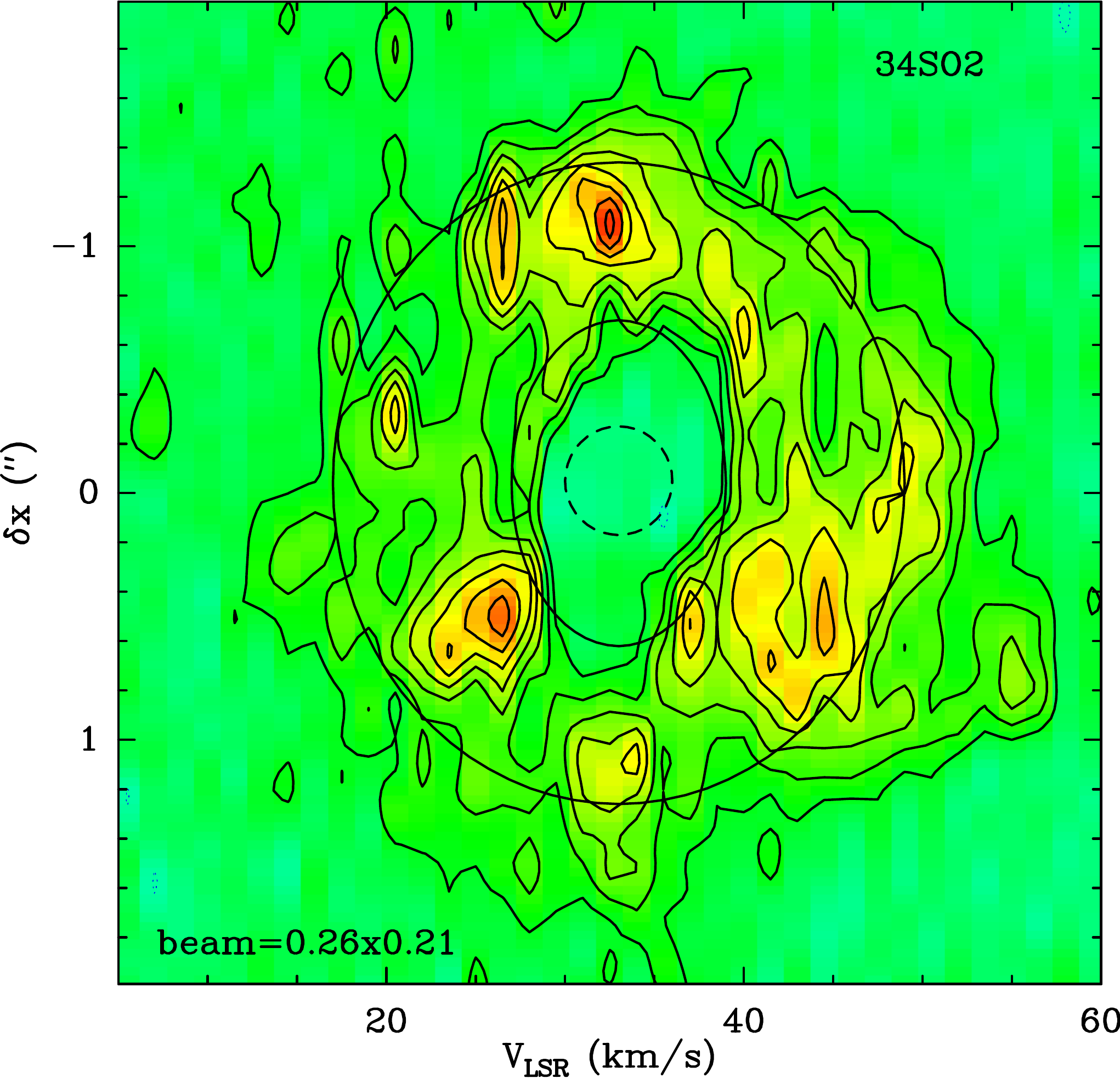}
   \caption{PV diagram along the nebula equator (PA=21+90\degr)
     through the waist center ($\delta y$=0\farc6) of three transitions:
     \trecet, CS\,(6-5) and \tsodost\ (from left to right). The large ellipses
     indicate the radial (\los) expansion velocity at 
     representative radii; the dashed ellipse (3\kms$\times$0\farc25)
     shows the central cavity. All ellipses are centered at 33\,\kms,
     $\delta$x=$-$0\farc05.}
      \label{f-largew}
   \end{figure*}

The centroid of the waist is located at offset $\sim$($-$0\farc05,
0\farc6). The \trecet\ PV cut along PA=21+90\degr\ (the equator)
passing through the waist center is shown Fig.\,\ref{f-largew}. The
\trecet\ line and the rest of transitions that sample the waist
consistently point to a systemic velocity of \vlsr$\sim$32-33\,\kms,
i.e., slightly different from that of \cs\ (\vlsr$\sim$35\,\kms). The
outer radius of the waist is $r_{\rm w}$$\sim$1\farc8 at a
$\sim$2$\sigma$-level in the \trecet\ maps (and slightly smaller, $r_{\rm w}$$\sim$1\farc4,
in the maps of other less abundant molecules, e.g.\,CS). The waist extends
about $\sim$$\pm$0\farc6 about its equator, but this is uncertain because there is not a sharp boundary
separating the waist from the base of the lobes (probably, both are
part of a unique hourglass-shaped structure).  We identify a small
cavity, of radius $\sim$0\farc2-0\farc25, at the center of the
waist. From here, the emission increases quite sharply with the
radius, reaching a peak at about $r$$\sim$0\farc6-0\farc7 and fading
gradually at larger distances.

Toward the center of the waist, the radial velocity
increases from |\vlsr-\vsys|$\sim$3\,\kms\ at the inner edge (immediately beyond
the cavity) to about |\vlsr-\vsys|$\sim$25\,\kms\ at the outermost regions
(Fig.\,\ref{f-largew}).
The waist layers at $r$$\sim$0\farc65 where the \trecet\ emission
peaks, are characterized by a radial velocity of
|\vlsr-\vsys|$\sim$6-7\,\kms. Adopting a cylindrical shape for the
waist with its revolution axis inclined with respect to the sky-plane
by $i$=35\degr, this implies a deprojected velocity gradient of
about 10$\times$$\cos{(i)}$$\sim$8.2\,\kms\,arcsec$^{-1}$, leading to
a kinematical age of \tdyn$\sim$715\,yr/$\cos{(i)}$$\sim$870\,yr,
similar to that of the large-scale outflow.

\begin{figure*}[htbp] 

   \centering 
   \includegraphics[width=0.31\hsize]{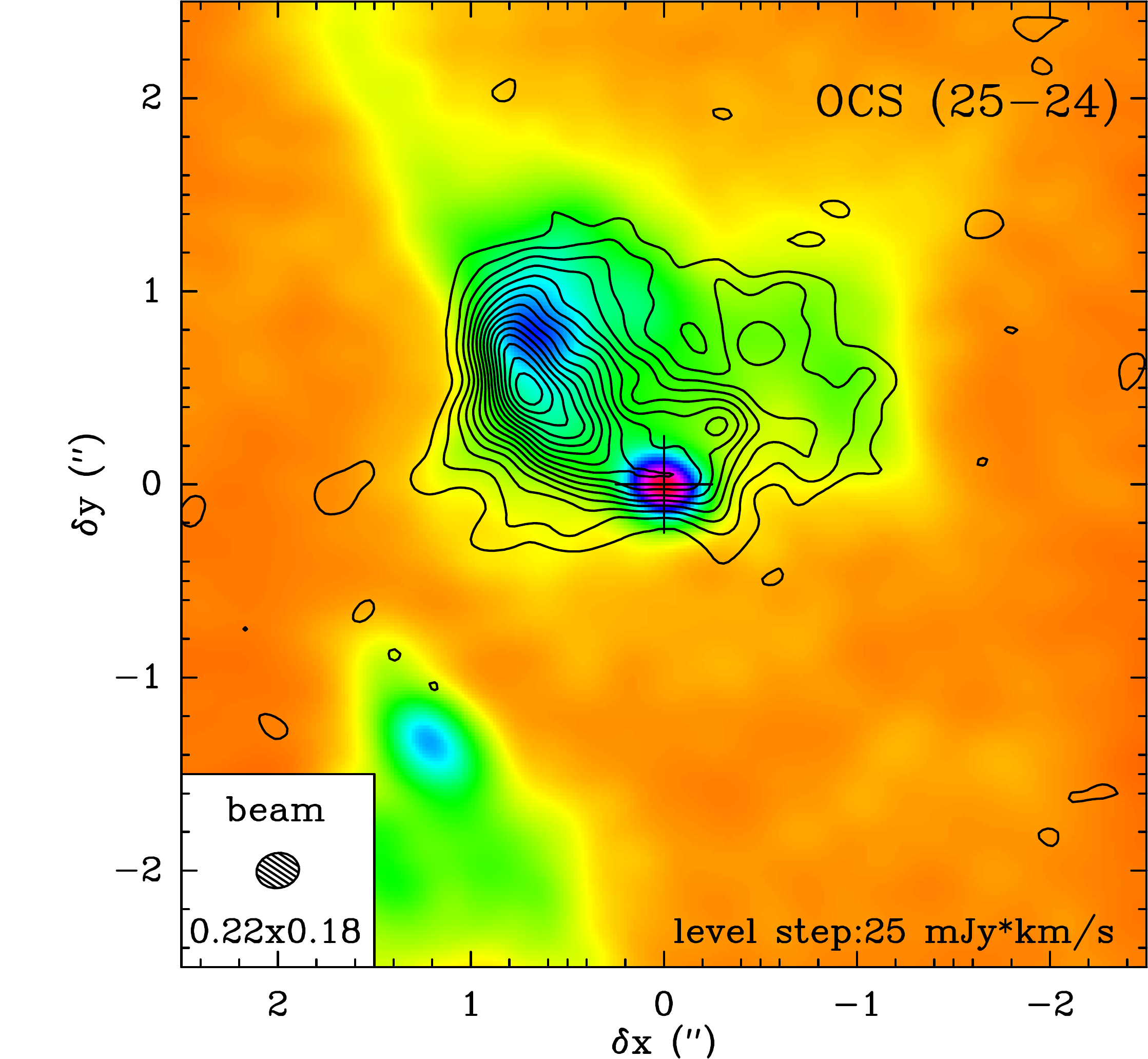}
   \includegraphics[width=0.31\hsize]{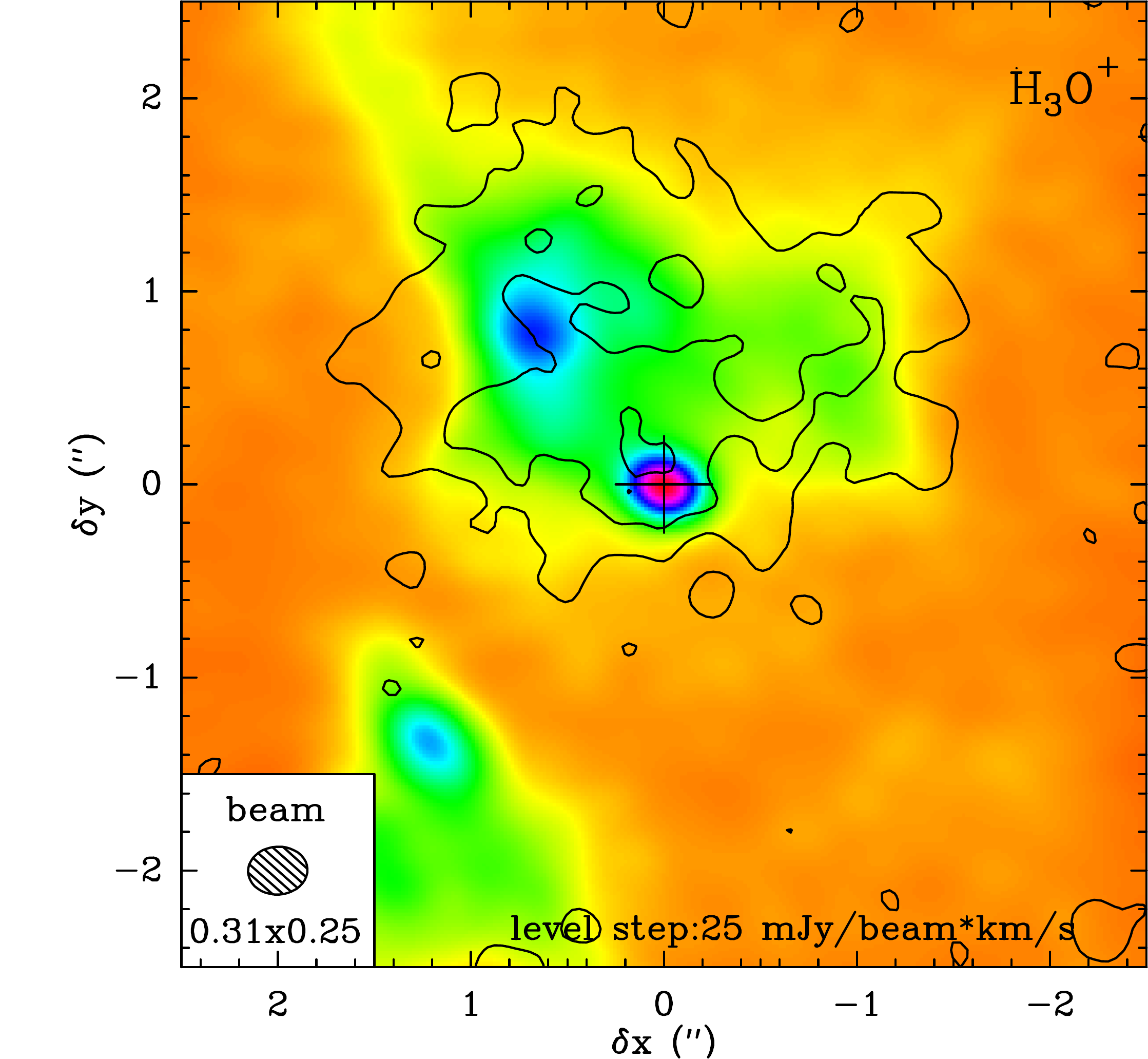} 
   \includegraphics[width=0.31\hsize]{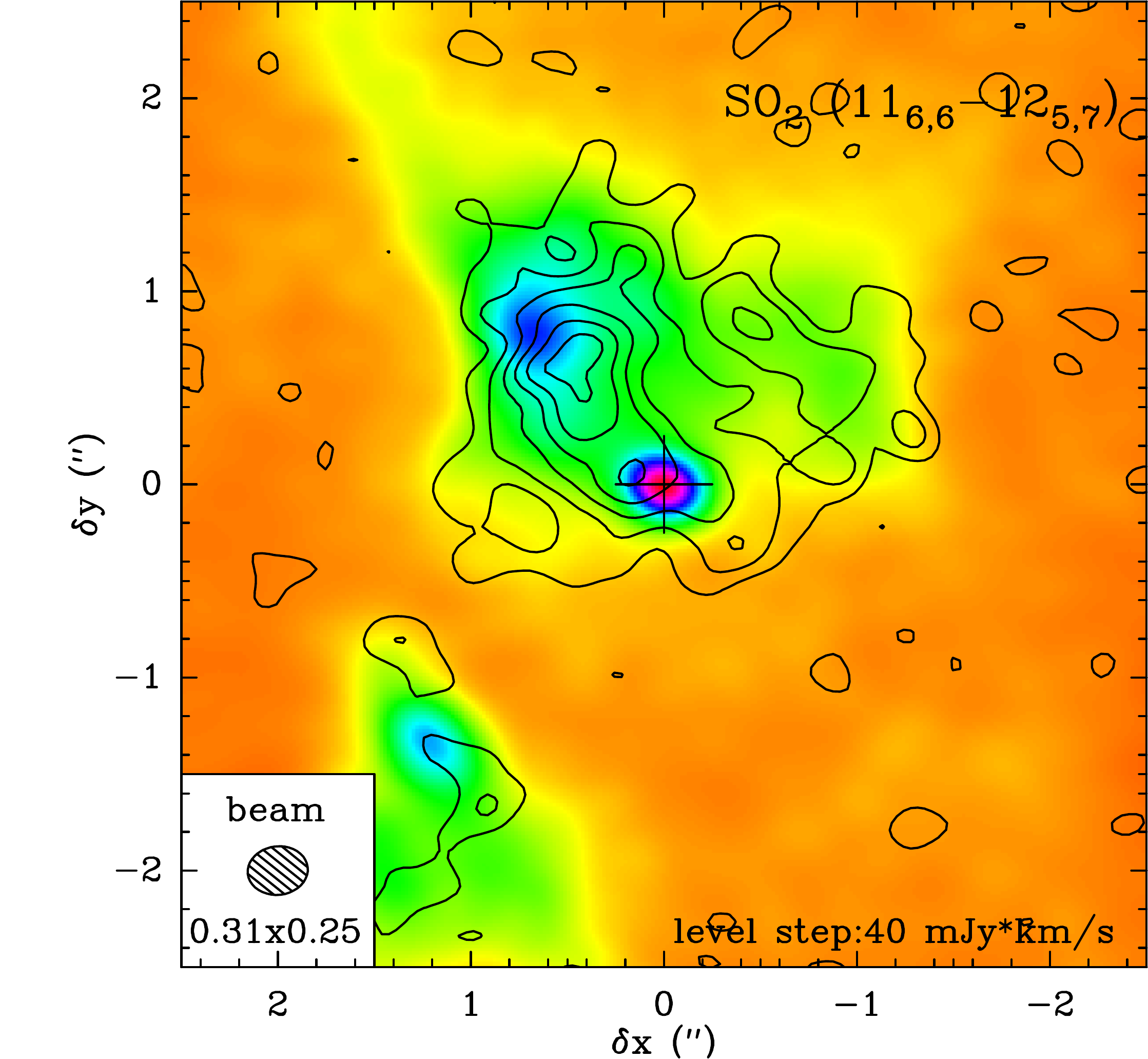}
      \vspace{0.35cm}
   \includegraphics[width=0.31\hsize]{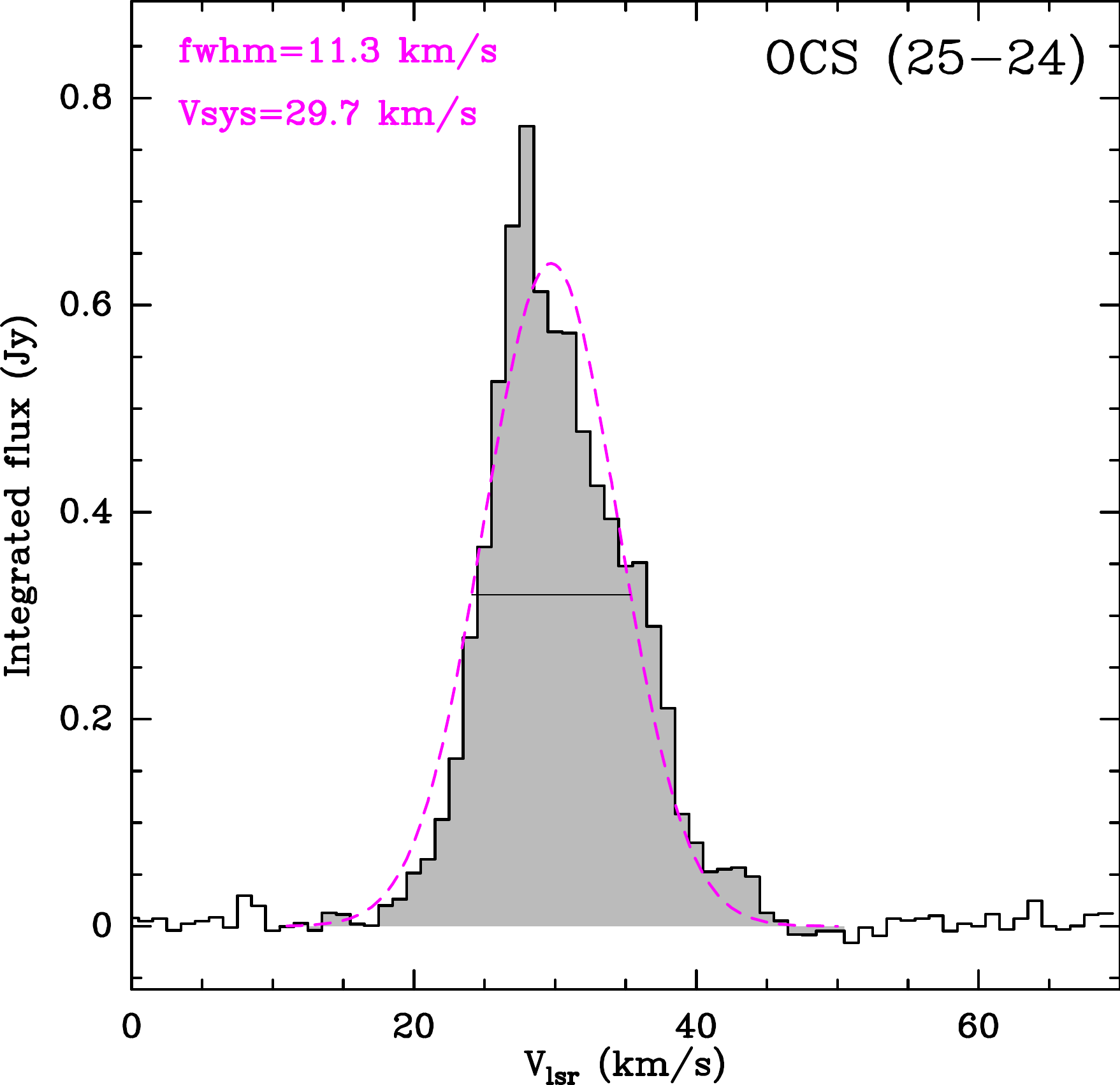}   
   \includegraphics[width=0.31\hsize]{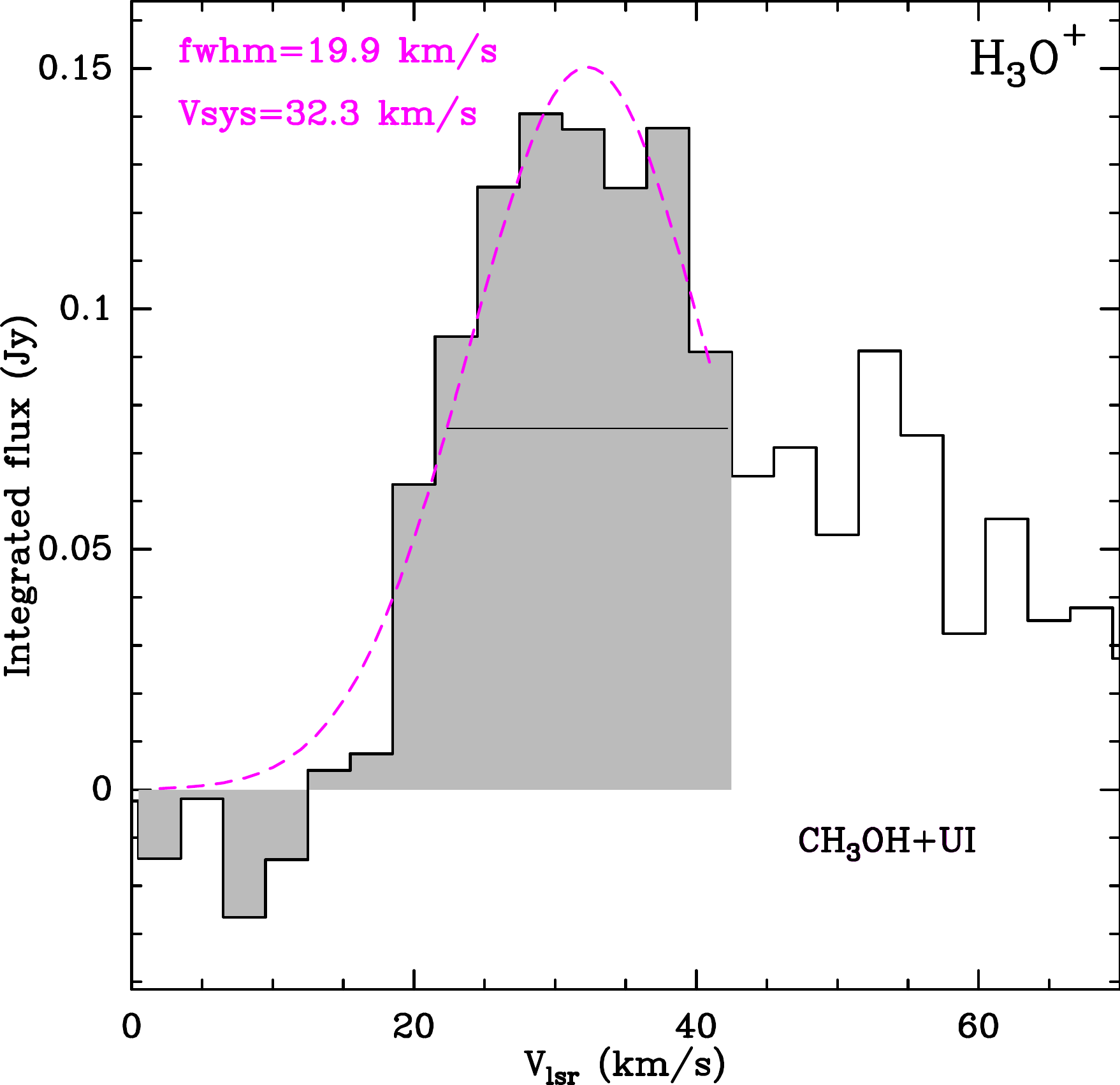}
   \includegraphics[width=0.31\hsize]{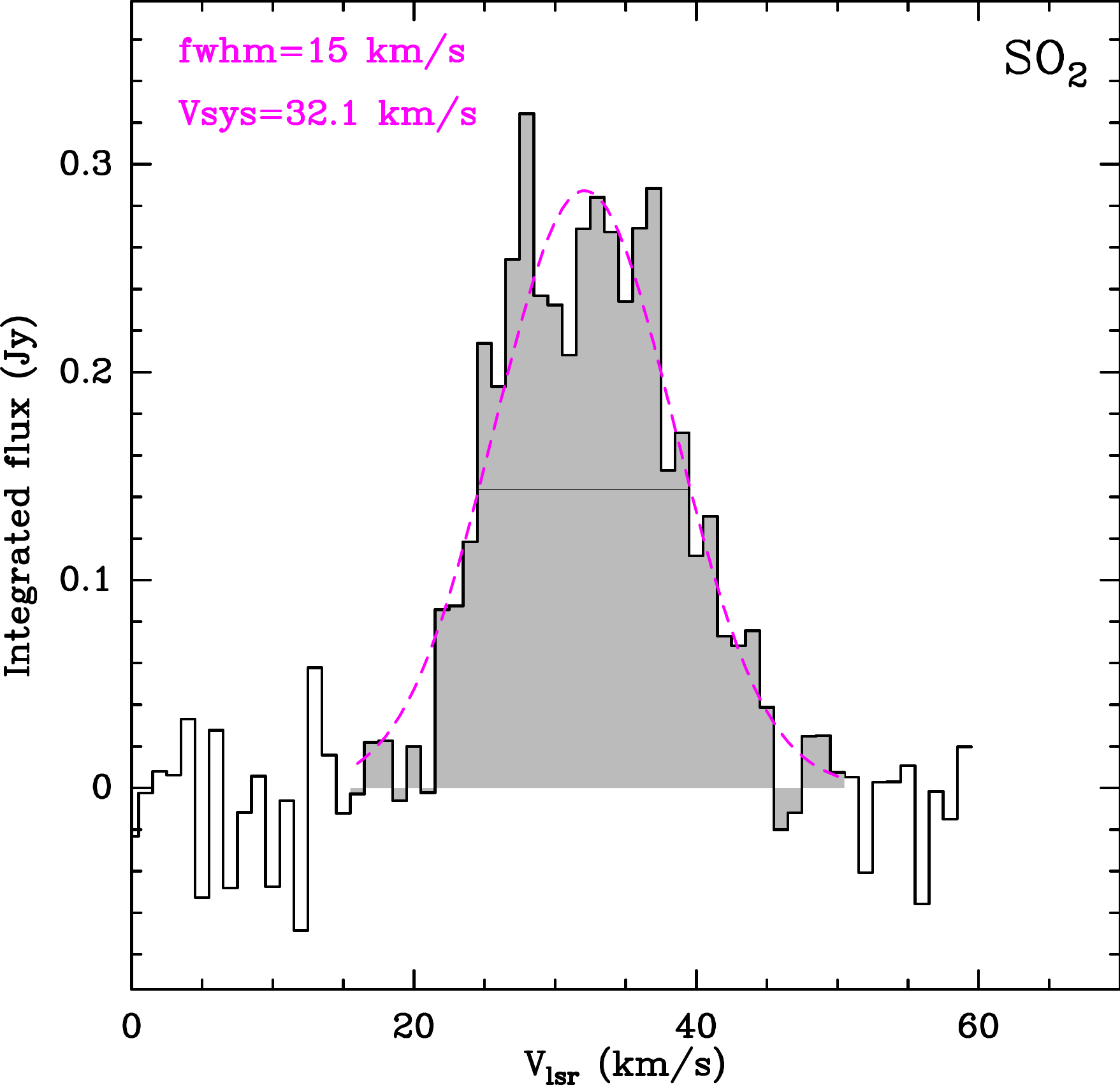}
   \includegraphics[width=0.31\hsize]{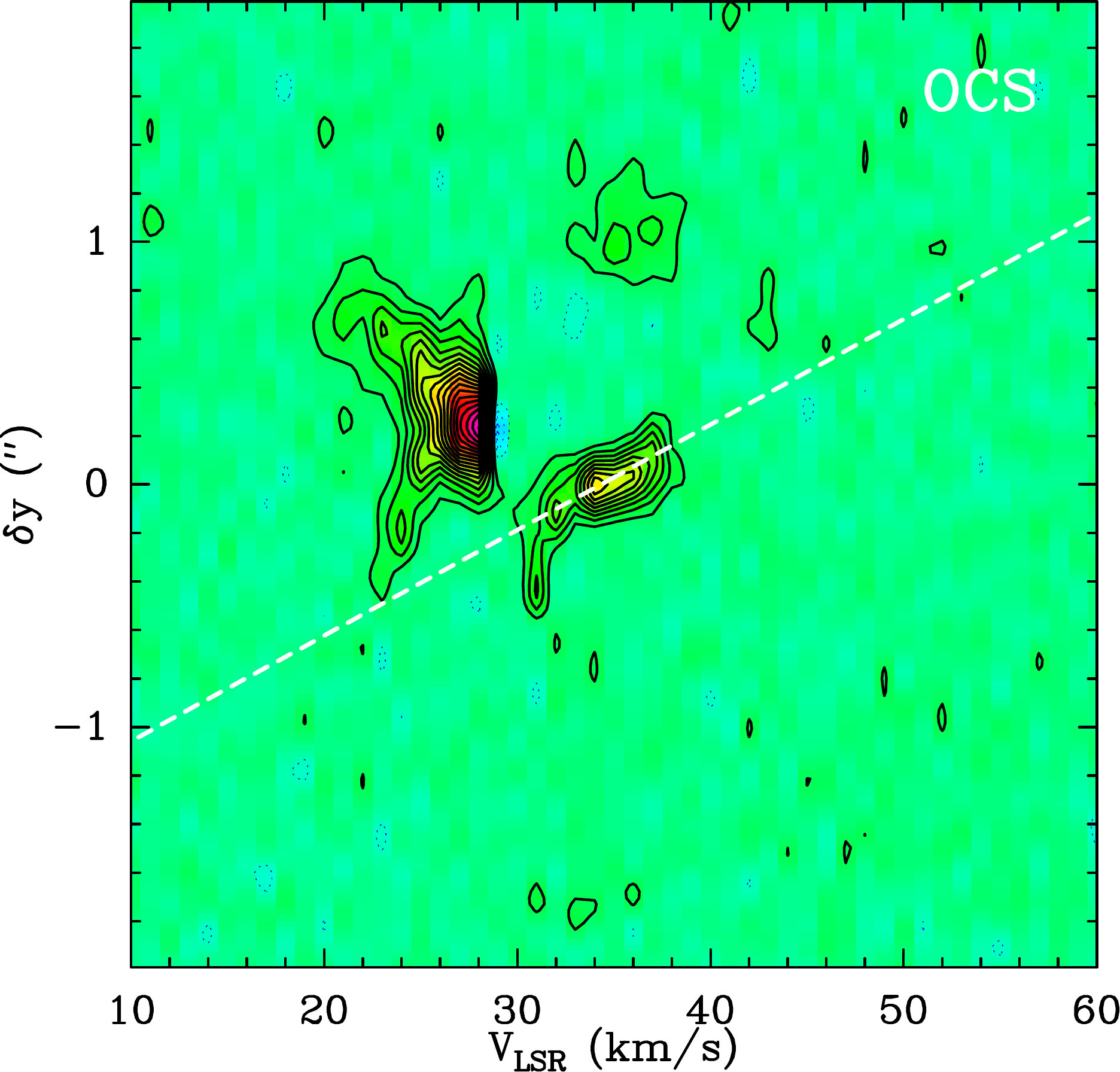}
   \includegraphics[width=0.31\hsize]{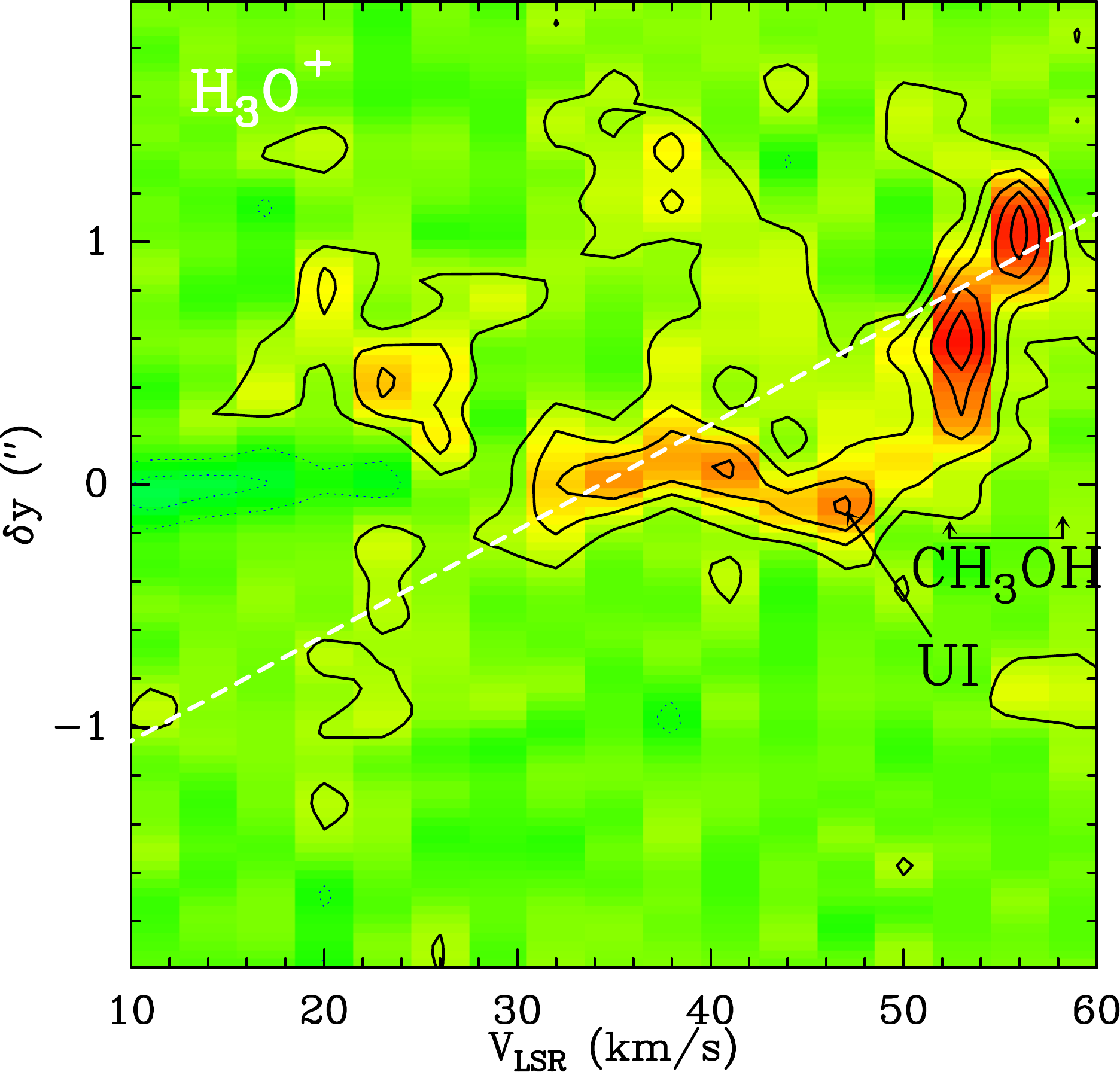}
      \includegraphics[width=0.31\hsize]{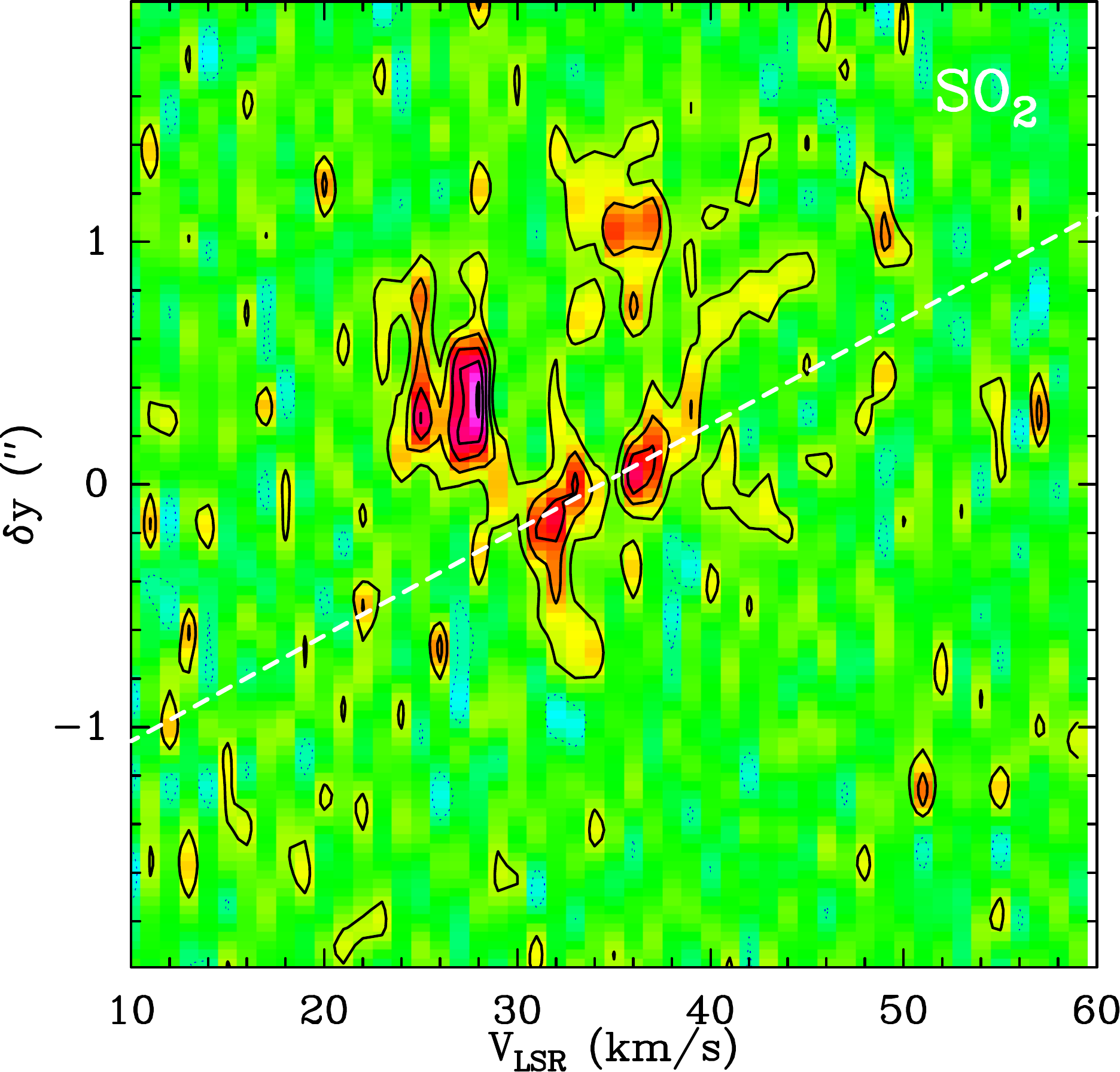}  
   \caption{Integrated intensity maps (top), integrated 1d-spectral
     profiles (middle), and axial PV diagrams (bottom) of molecular
     transitions whose emission is largely restricted to the
     equatorial waist (\S\,\ref{res-largehg}). The integrated
     intensity maps are superimposed to the 330\,GHz-continuum map.
     The velocity range of the integrated intensity maps of OCS and
     \sodos\ is \vlsr=[18:42]\,\kms, and for \htresomas\ is
     \vlsr=[14:50]\,\kms.  Level step in PV diagrams is 3.5\,mJy/beam
     (OCS), 1.5\,mJy/beam (\htresomas), and 6\,mJy/beam (\sodos).}
      \label{f-waist}
   \end{figure*}

{\sl Selective tracers of the waist.}  The emission from certain
molecular lines, namely, \ocst, \htresomast, and \sodost, is largely
restricted to the $\sim$2\arcsec$\times$1\arcsec\ central parts of the
\lhg\ (Fig.\,\ref{f-waist}).
Emission from OCS is observed over a relatively narrow velocity range
(\vlsr=[18:42]\,\kms) tracing selectively the equatorial $waist$ of
the \lhg\ and, also, of the compact mini-hg nested inside (see also Figs.\,\ref{f-miniw} and \ref{f-ocs}). This is
probably because the relatively high-excitation requirements of this
transition (\eu$\sim$190\,K) are not attained at large
distances along the fast lobes, which are predominantly cold (\S\,\ref{s-rd}).
   The SO$_2$ emission maps resemble those of
   OCS, which is probably explained by the similar upper-level energy
   of both transitions (Table\,\ref{t-lines}).
   The surface brightness distribution of
   \htresomas\ is slightly more extended, rounded, and uniform than
   that of OCS and SO$_2$.
   This is consistent with this ion being mainly a
   product of the photodissociation chain of \water\ and CO by the
   interstellar UV radiation in the outer layers of the envelope
   \citep{san15}. 
As OCS, the \htresomas\ and SO$_2$ transitions have also some
contribution to the emission produced in the mini-waist.

  \subsubsection{The high-velocity bipolar lobes}
  \label{res-hv}
  
  The high-velocity (HV) bipolar lobes are, after the large hg, the
  second major component of the molecular outflow of
  \ohs\ (\S\,\ref{intro}). In our ALMA maps, the \docem\,(3-2)
  emission from the north and south lobe reaches out to radial
  distances of $\sim$17\arcsec\ and $\sim$30\arcsec, respectively,
  along the PA$\sim$21\degr\ axis (Fig.\,\ref{f-12co}). The lobes are
  also traced by a number of molecules other than
  \docem\ (Table\,\ref{t-sketch}). These species, with emission lines
  substantially fainter than \docet, trace only certain parts of the
  lobes (normally, the CO-brightest regions).
  
\begin{figure*}[htbp]
    \centering 
    \includegraphics[width=0.43\hsize]{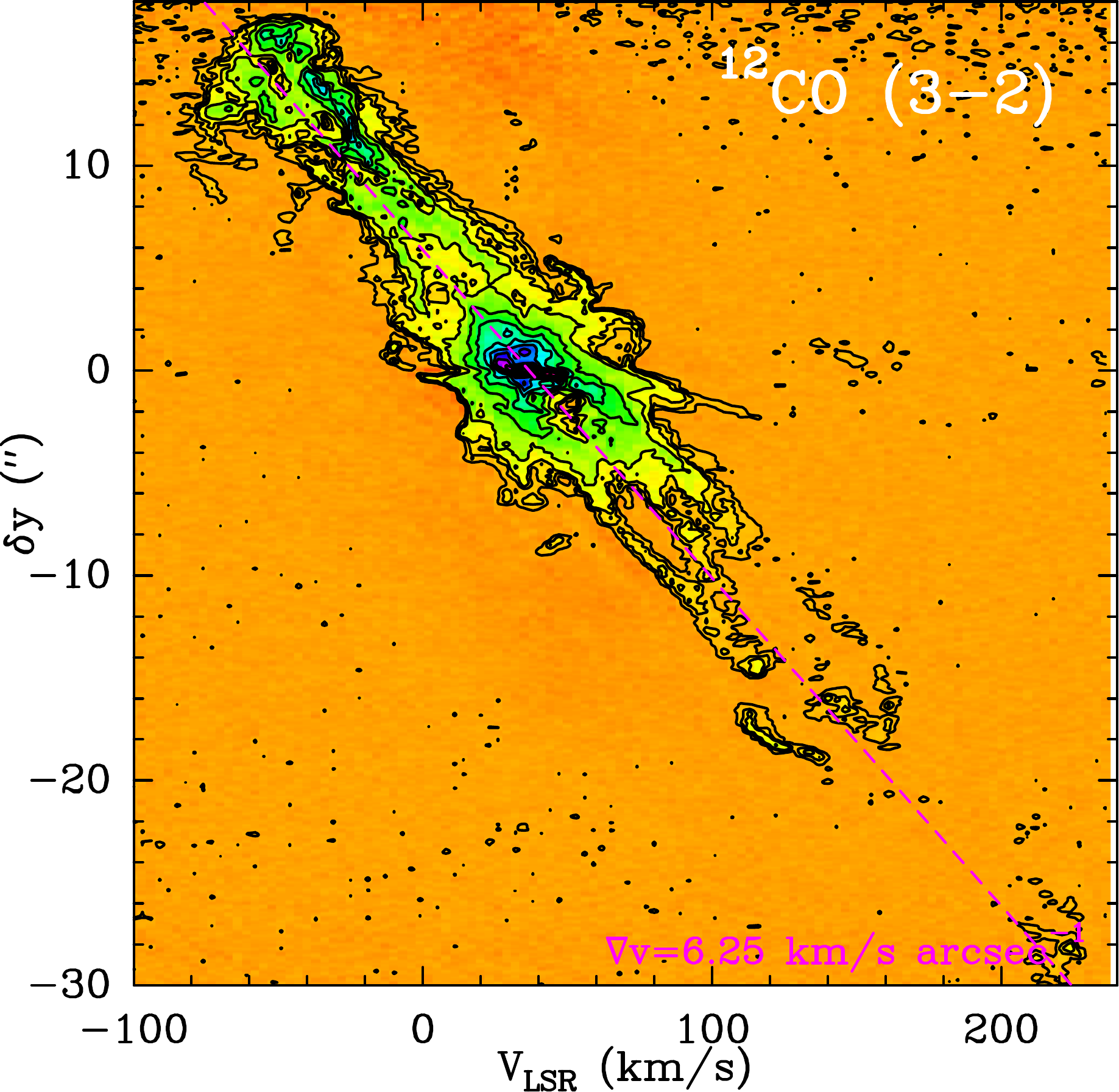}
    \includegraphics[width=0.44\hsize]{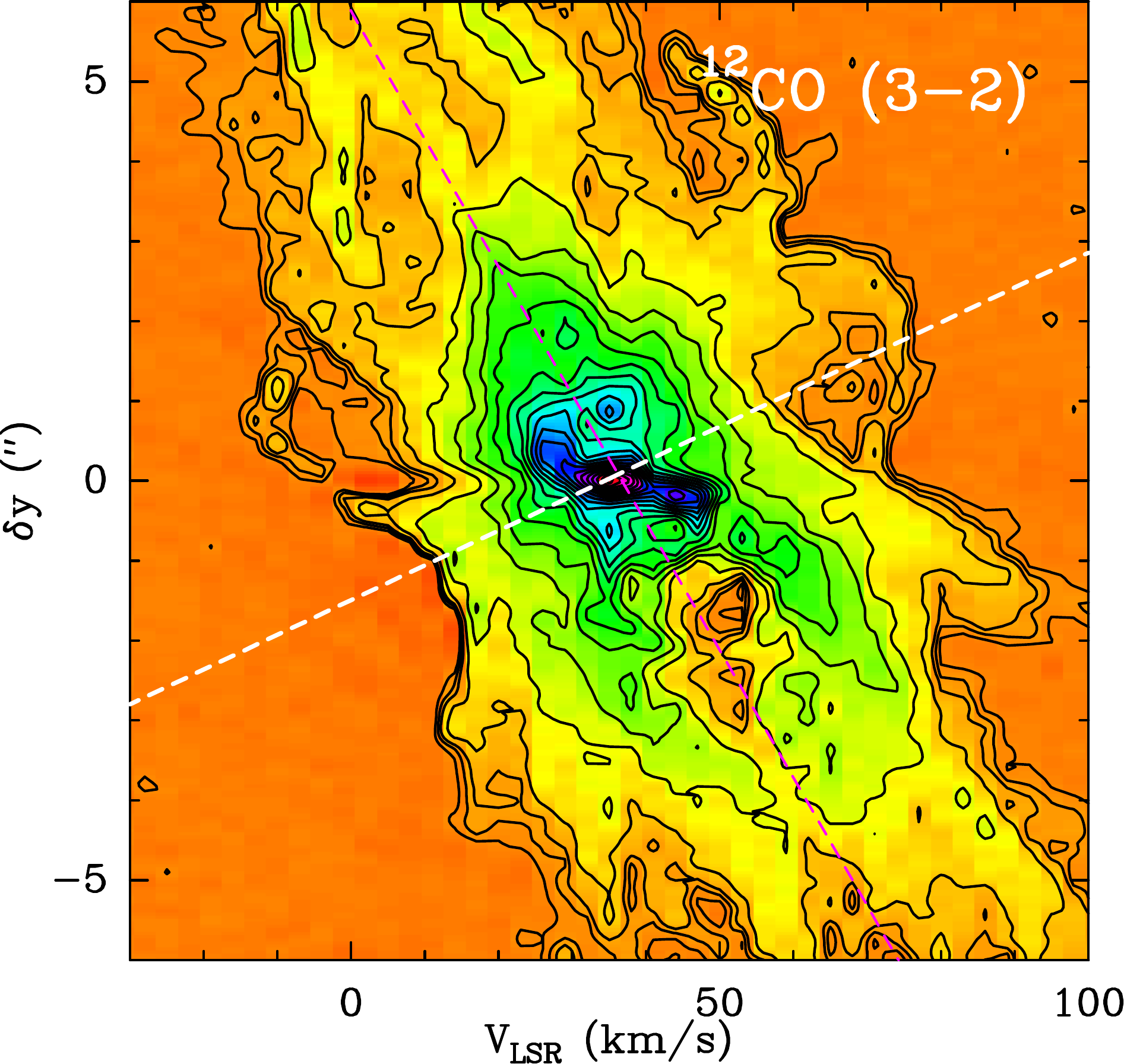}
    \includegraphics[width=0.43\hsize]{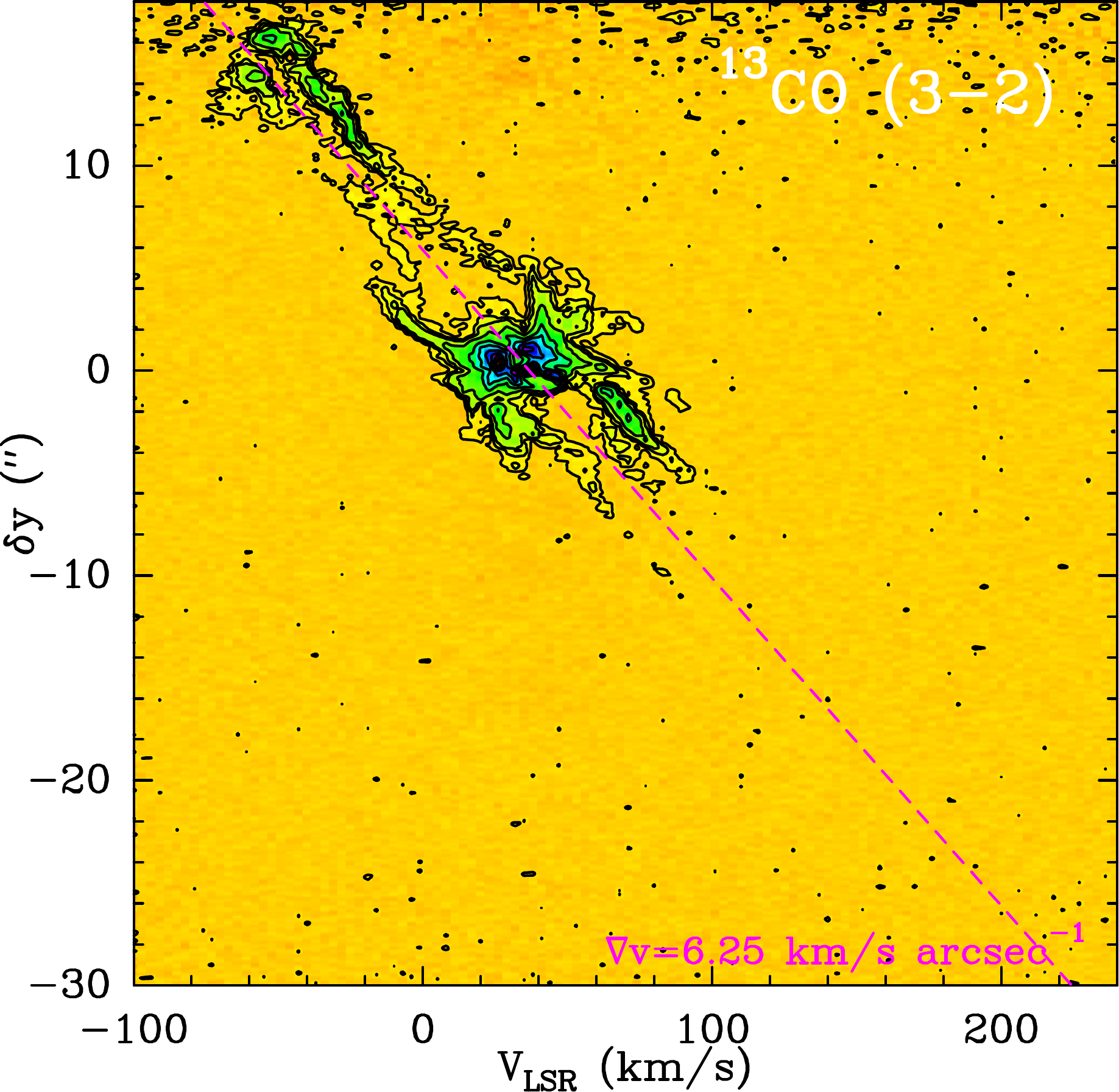}
    \includegraphics[width=0.44\hsize]{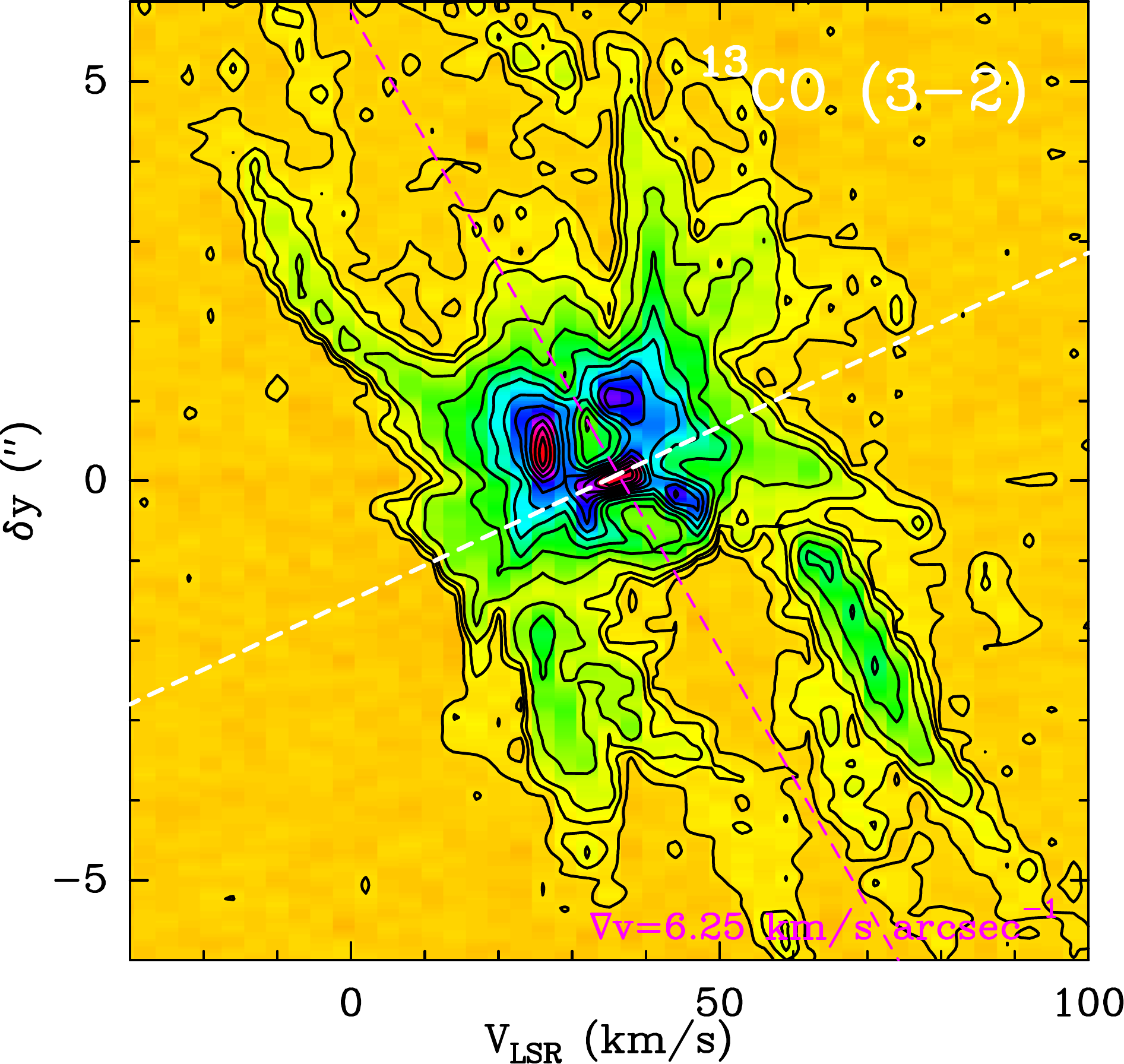}
    \caption{Position-velocity diagram along PA=21\degr\ of \docet\, (top, HPBW=0\farc31$\times$0\farc24)
      and \trecet\ (bottom, HPBW=0\farc33$\times$0\farc26). 
      show the full spatial and \vlsr\ range where CO emission is
      detected, whereas the right panel offers a closer view of the
      central regions (i.e.\, the large-hg).  Level spacing is:
      \trecem) 1$\sigma$, 3$\sigma$, 5$\sigma$, 10$\sigma$... by
      10$\sigma$ ($\sigma$=7 and 5\,mJy/beam, in the left and right
      panels, respectively); \docem) 1$\sigma$, 3$\sigma$, 5$\sigma$,
      10$\sigma$... by 10$\sigma$ ($\sigma$=6 and 4\,mJy/beam, in the
      left and right panels, respectively). The dashed lines indicate
      the overall velocity gradient along the symmetry axis of the
      large-scale outflow (purple) and along the equator of the
      mini-hg (white).
    }
         \label{f-co-pvs}
   \end{figure*}

  The HV bipolar lobes are highly structured, and a number of nested, often
  incomplete, substructures appear to be present; some of these are
  counterparts of major
  features identified in the optical and near-IR images of
  \ohs\ (skirt, spine, fingers, filaments, etc -- see
  \citealp{buj02,mea03,bal17}).
  All these substructures are characterized by a dominant expansive
  kinematics described by a radial velocity gradient,
  \vgrad$\sim$6.0-6.5\,\kms\,arcsec$^{-1}$, that is sustained from the
  base (i.e., low-latitude regions of the large hg) to the tips of the
  lobes where the highest \los-velocities are observed, 
  (|\vlsr-\vsys|$\sim$125 and $\sim$200\,\kms\ for the north and south
  lobes, respectively -- Fig.\,\ref{f-co-pvs}). The kinematic age of
  the HV-bipolar lobes implied by this velocity gradient is
  \tdyn$\sim$1150$\times$$\tan{(i)}$, that is, \tdyn$\sim$800\,yr for
  $i$=35\degr.

   \subsection{NEW macro-structures unveiled by ALMA: the fish bowls}
   \label{res-calamardos}

\begin{figure*}[htbp]
    \centering 
           \includegraphics[width=0.95\hsize]{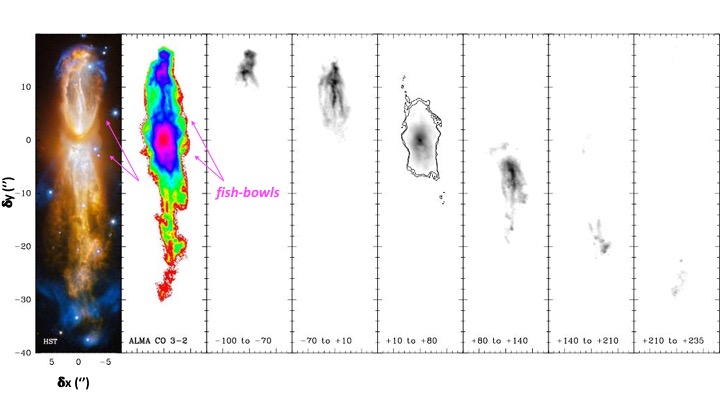}
    \caption{{\bf Left:} False color $HST$ image of the reflection
      nebulosity (red-to-yellow) and H$\alpha$-emitting lobes (blue)
      of \oh\ (Credit: ESA/Hubble \& NASA. Acknowledgement: Judy
      Schmidt. Visit {\tt
        https://www.spacetelescope.org/images/potw1705a/} for more
      details).  As in the rest of the figures, North is
      21\degr\ right of vertical. {\bf Right panels:} ALMA order-zero
      moment maps of \docem\,(3-2) integrated over the full width of
      the line profile (color-scale) and over selected LSR velocity
      ranges (grey-scale; the \vlsr\ ranges are indicated in each
      panel in units of \kms). The two bubble-like structures (or {\it
        fish bowls}) of the molecular envelope detected for the first
      time with ALMA in this work are indicated together with their
      counterpart in the $HST$ image.}
         \label{f-12co}
   \end{figure*}

\begin{figure*}[htbp]
    \centering 
    \includegraphics[width=1.0\hsize]{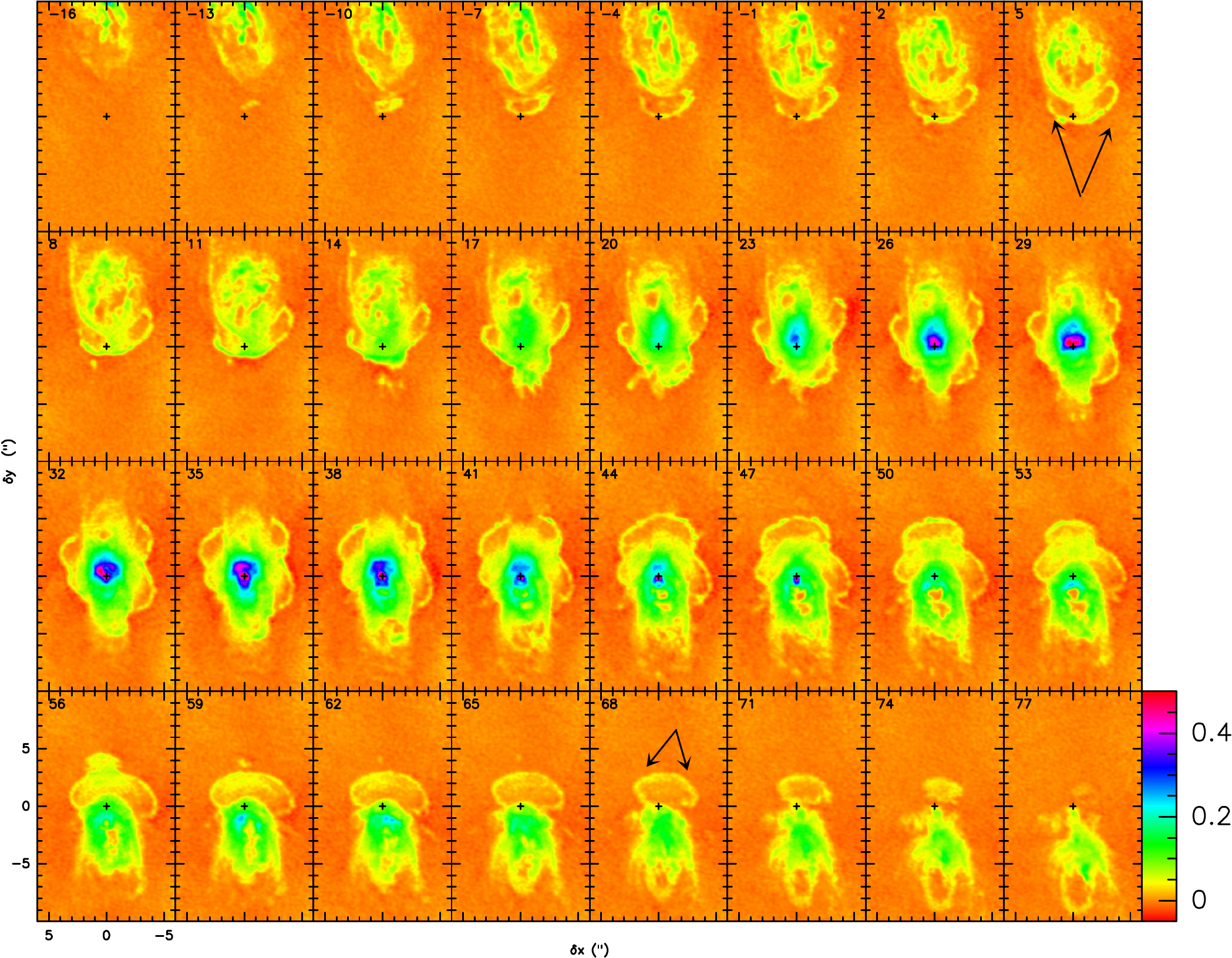}
    \includegraphics[width=0.33\hsize]{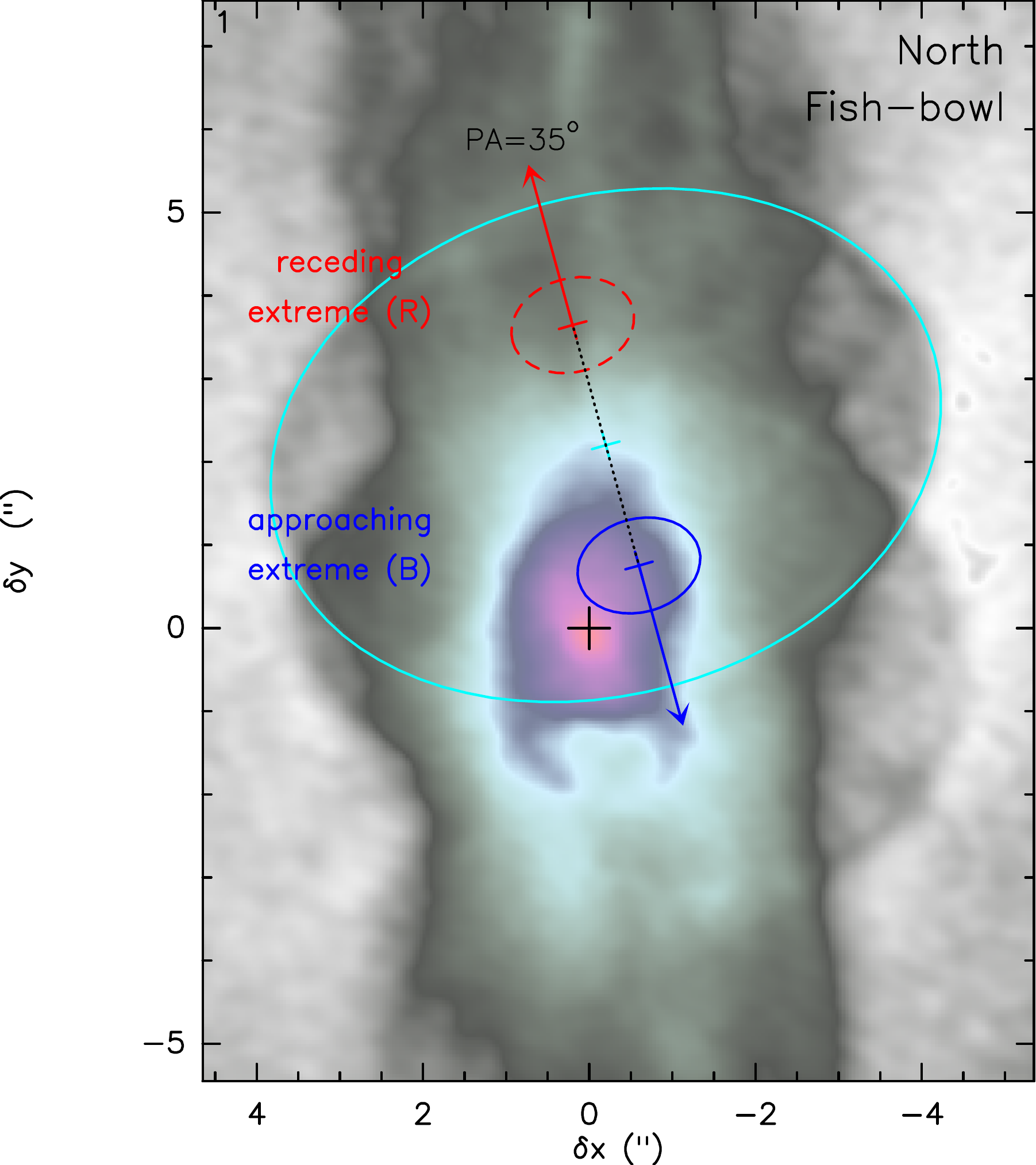}
    \includegraphics[width=0.33\hsize]{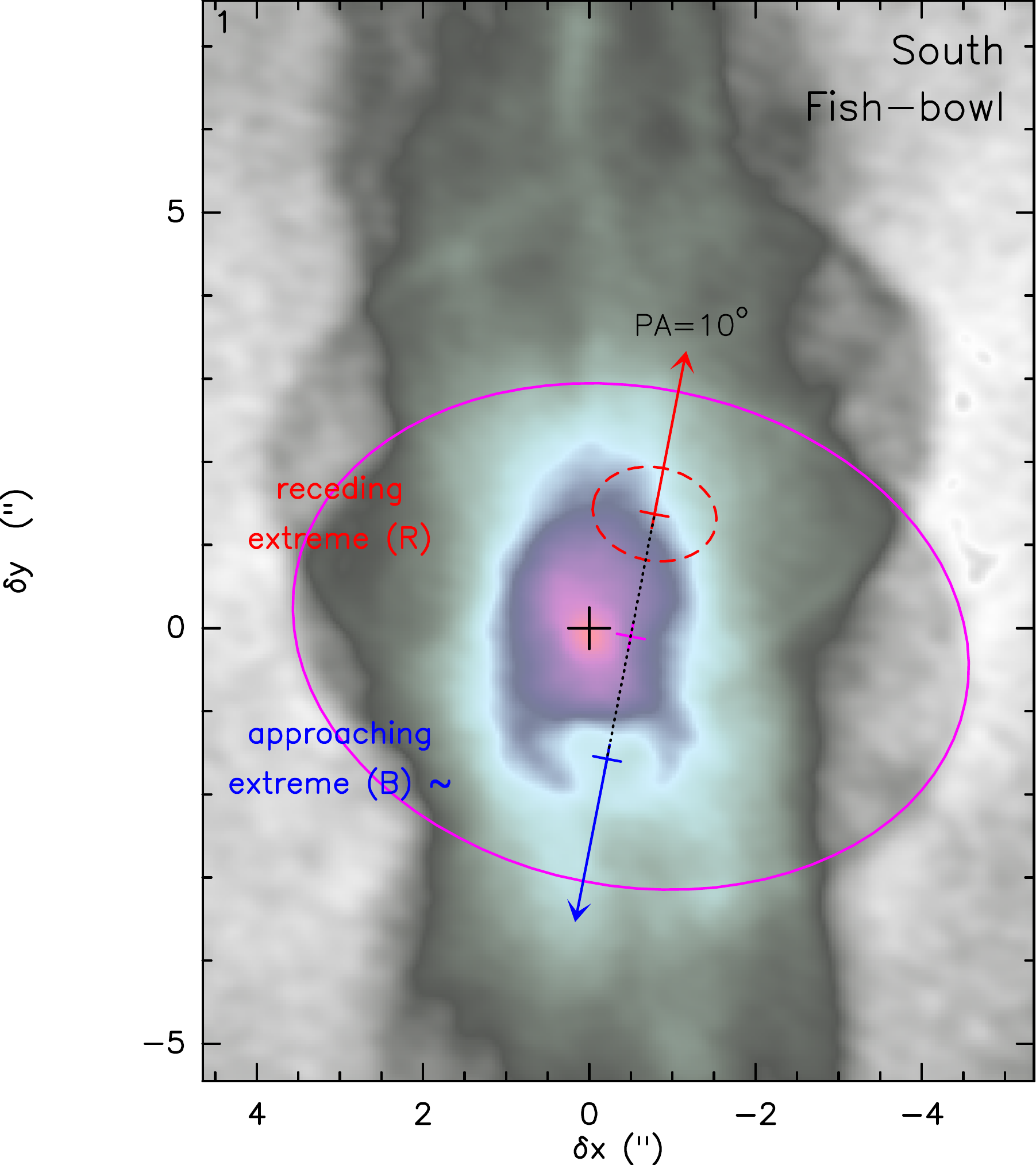}
        \includegraphics[width=0.33\hsize]{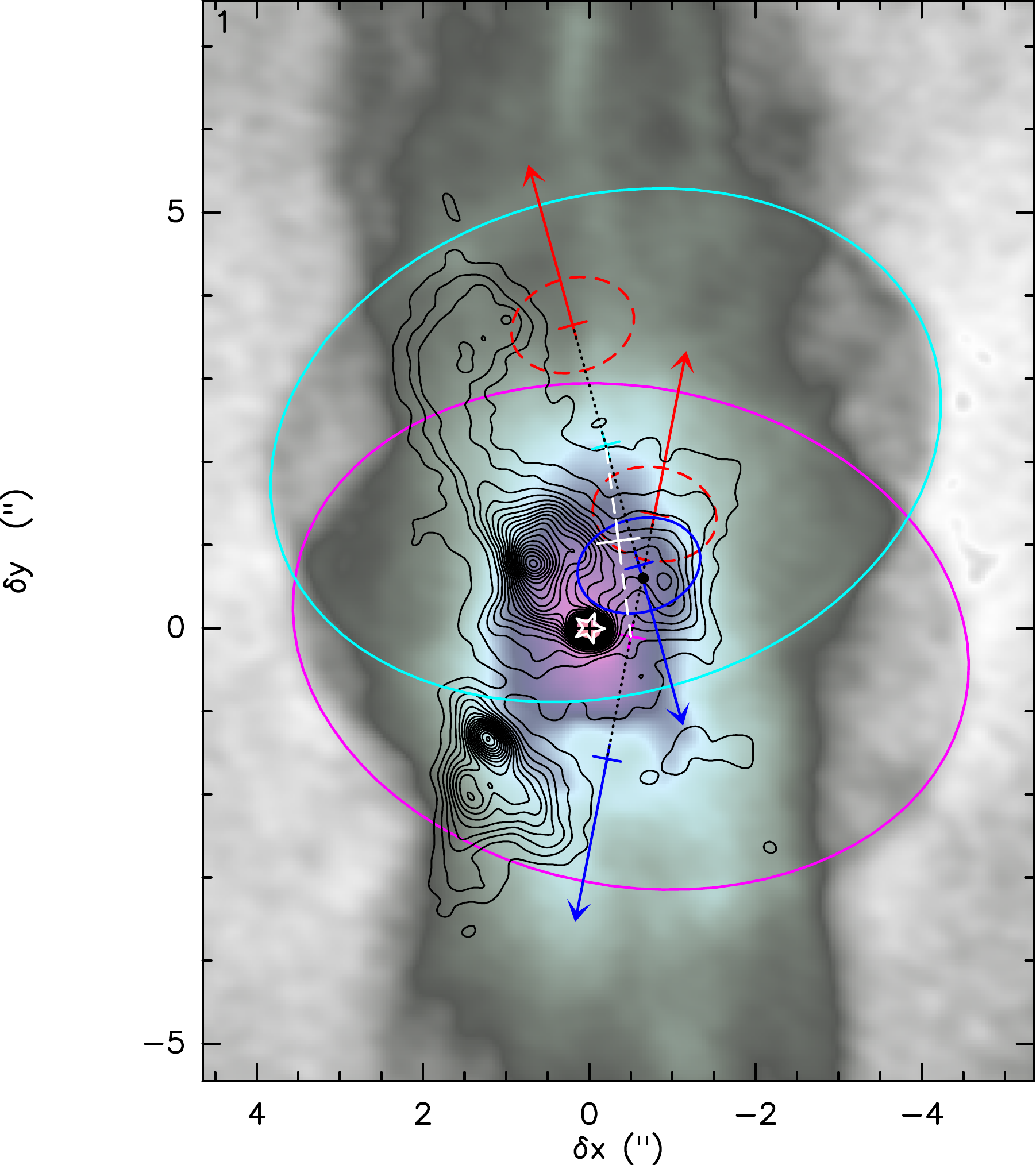}
    \caption{{\bf Top:} Velocity-channel maps of \docem\,(3-2) in the
      velocity range where two faint, elongated bubble-like features
      (dubbed {\it the fish bowls}) are discovered in the central
      regions of \ohs\ (\S\,\ref{res-calamardos}). {\bf Bottom:}
      Zero-order moment maps of \docet\ integrated over the
      \vlsr=[$-$13:+77]\,\kms\ range.  The big ellipses represent the
      projection on the plane of the sky of the outer surface of the
      ellipsoids that represent the North (left) and South (middle)
      fish bowls. The small red and blue ellipses represent
      schematically the receeding and approaching vertex of the
      ellipsoids projected in the plane of the sky. In the right panel
      the continuum emission map is atop the \docem\,(3-2) integrated
      map. The dashed line connects the centroids of the North and
      South fish bowls.}
         \label{f-calamardos}
   \end{figure*}

   The unprecedented sensitivity of the ALMA maps reported here has
   enabled us to detect for the first time the molecular counterpart
   of the two faint, rounded structures seen in the $HST$/NIR images
   toward the central regions of
   \oh\ (Fig.\,\ref{f-12co}). Curiously, no previous published studies
   of these dusty structures exist to date (to our knowledge).  These
   components are best identified in the \docet\ velocity-channel maps
   shown in Fig.\,\ref{f-calamardos} where they appear as elliptical,
   ring-like emission features around the bright central regions of
   the nebula. We refer to these structures as the fish bowls.

   The northern fish bowl (\nbw) can be relatively well isolated from
   other (lobe and waist) line-emitting regions in the velocity range
   \vlsr=[$-$13:+59]\,\kms. It comes into sight as a compact
   elongated emission feature at offset ($-$0\farc6, 0\farc75) in the
   \vlsr=$-$13\,\kms\ channel. As \vlsr\ increases, this feature
   progressively grows and develops an elliptical ring-like morphology
   that reaches a maximum size of
   $\sim$8\arcsec$\times$6\arcsec\ (with the long axis oriented along
   PA$\sim$125\degr) in channels near the centroid of the
   \nbw-emission profile, \vlsr$\sim$23-26\,\kms.
   As \vlsr\ continues increasing,
   the ring-shaped emitting region steadily shrinks and dims,
   turning into a faint compact emission clump, located at offset
   (0\farc2, 3\farc6), in channel \vlsr$\sim$59\,\kms.  
   The  ring-like feature appears clearly incomplete in several central
   channels, in particular, a vast part of the northern rim emission
   is missing at \vlsr$\leq$41\,\kms.

   The southern fish bowl (\sbw)
   comes out as a faint and compact elongated emission region at
   \vlsr=+77\,\kms\ (at offset $-$0\farc8, 1\farc4) that progressively
   becomes larger, brighter, and acquires a ring-like morphology as
   the velocity decreases. This feature reaches maximum dimensions of
   $\sim$8\arcsec$\times$6\arcsec\ (with the long axis oriented along
   PA$\sim$100\degr) in channels around \vlsr=47\,\kms. As the
   velocity continues decreasing, the size of the ring-like emission
   feature reduces. The \sbw-emission can be discerned down to
   \vlsr=17\,\kms, but only from the west side. At lower LSR
   velocities, the line-wing emission from the \sbw\ (expected to be
   faint and compact, if any) cannot be isolated or disentangled from the
   intense emission produced in the south lobe near its base. As its
   northern analog, the \sbw\ appears incomplete or broken,
   particularly, its the southern rim.

   The simplest configuration that is consistent with the observed
   spatio-kinematic distribution of the \docet\ emission from the
   fish bowls is a pair of hollow, thin-walled ellipsoids radially expanding.
   The projection of the outer surface of these ellipsoids in the
   plane of the sky is represented by the ellipses overplotted on the
   \docet\ integrated intensity maps (Figure\,\ref{f-calamardos},
   bottom panels).

\begin{figure*}[htbp]
   \centering 
   \includegraphics*[bb=150 100 600 500,width=0.33\hsize]{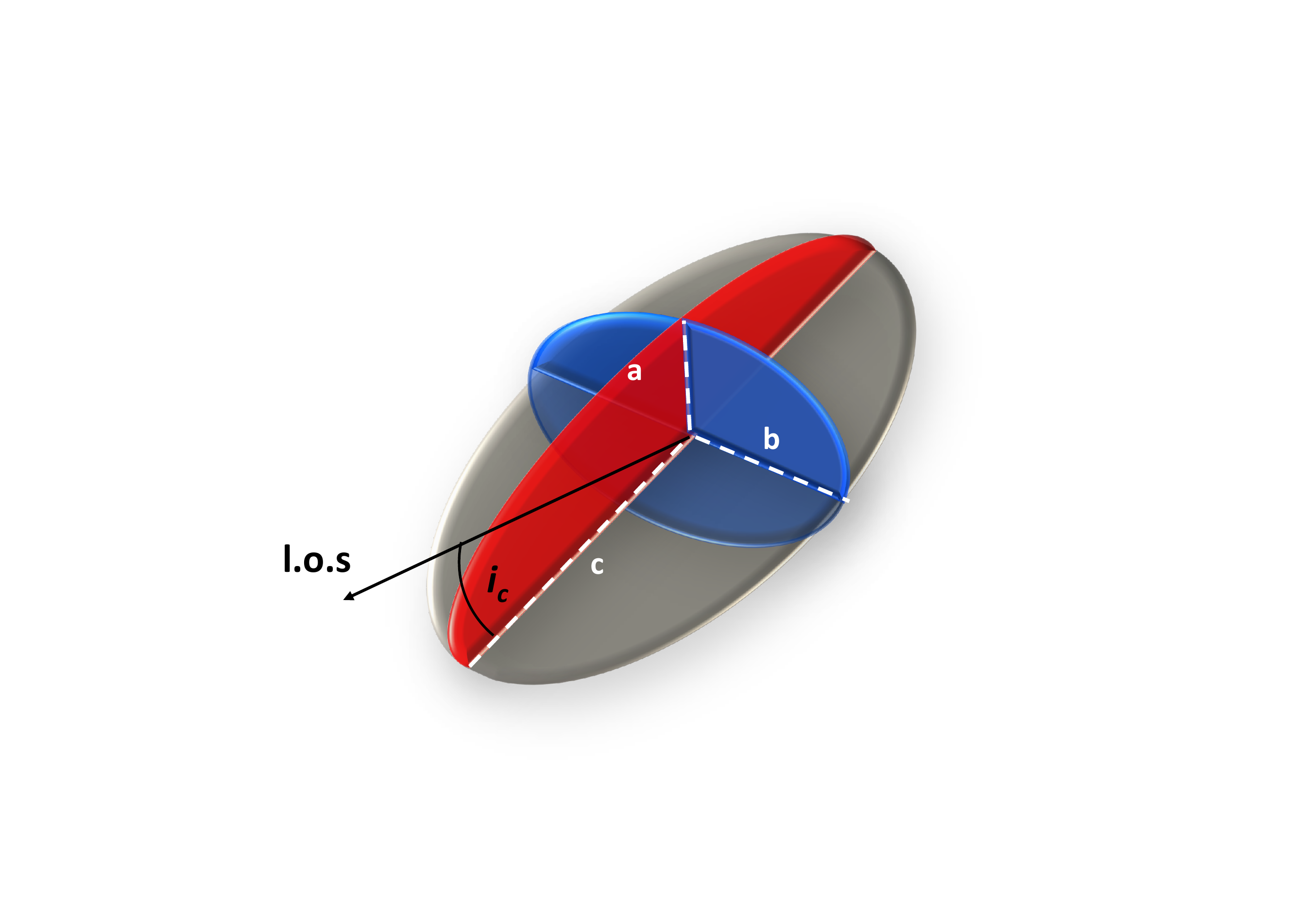}
   \includegraphics[width=0.36\hsize]{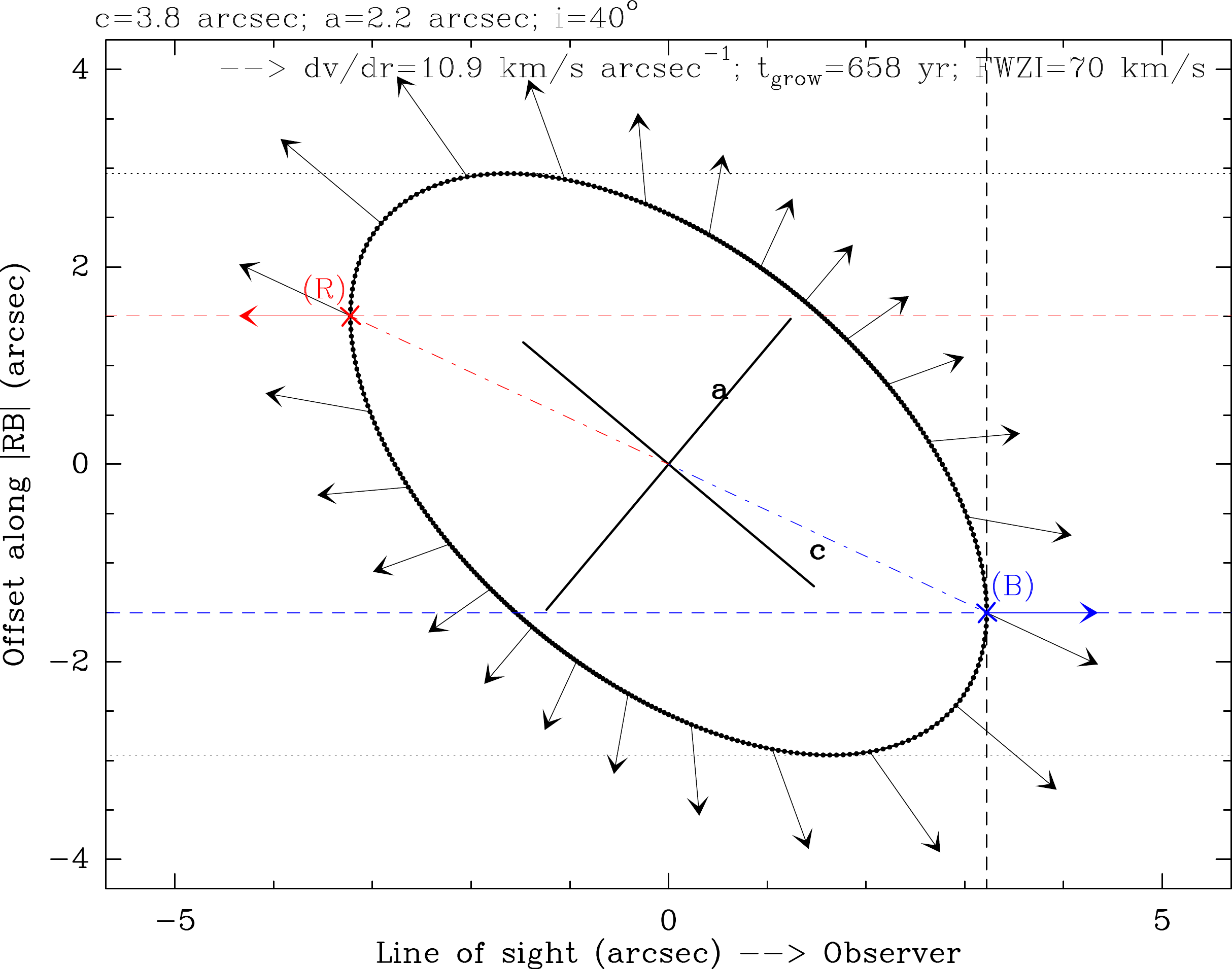}
   \includegraphics[width=0.30\hsize]{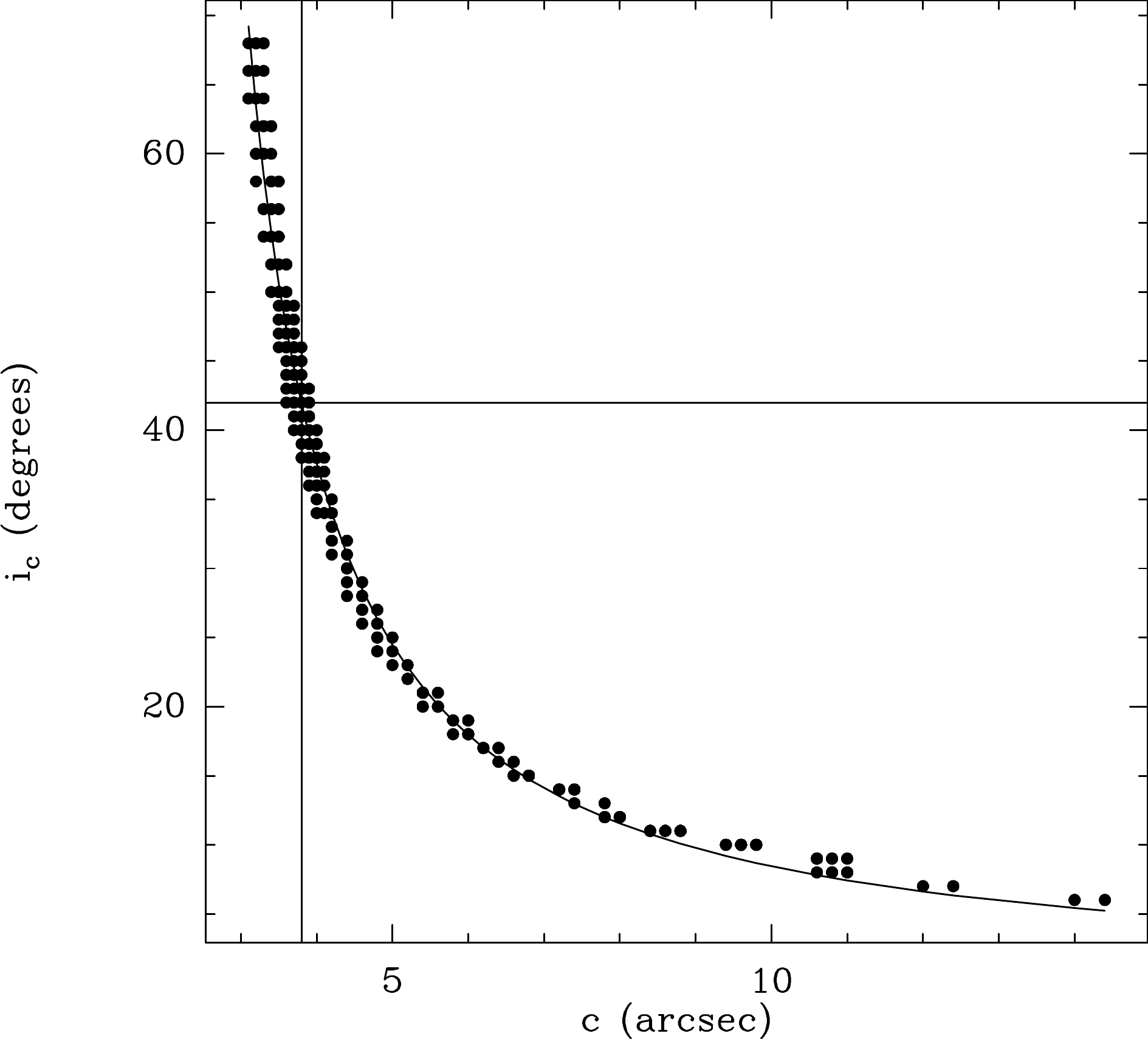}
   \caption{{\bf Left:} 3D view of a triaxial ellipsoid used to help
     visualizing the generic geometry and symmetry axes of the
     fish bowls (Fig.\,\ref{f-calamardos}) as described in
     Section\,\ref{res-calamardos}.  {\bf Middle:} Schematic view of
     the north fisw-bowl (\nbw) cut by a plane through the \los\ and
     the |$\overline{RB}$| line (i.e.\ the red $ac$-plane 
     as represented in the top panel) and adopting $c$=$b$=3\farc8.
     {\bf Right:} Relation between the size of the $c$-axis of the
     fish bowls and its \los-inclination consistent with the
     morphology and radial velocity gradient observed in our ALMA CO
     maps). The vertical and horizontal line marks the values of $c$
     and $i_{\rm c}$ that reproduce the observations and are
     consistent with a formation time $\sim$650$\pm$100\,yr.}
     \label{f-elipse}
   \end{figure*}

   A three-dimensional view of a generic ellipsoid,
   with semi-major axes $a$, $b$, and $c$, is given in
   Fig.\,\ref{f-elipse} (left) as well as a cut of it through the
   $ac$-plane, which contains the \los\, (Fig.\,\ref{f-elipse},
   middle). As we will see below, the sizes of $b$ and $a$ can be
   constrained from the observations, however, $c$ is uncertain since
   this axis can be tilted with respect to the \los\ by a given
   (unknown) inclination angle \ic\ (Fig.\,\ref{f-elipse}, right). 

   
   For radial expansion, the velocity-channel maps
   (Figure\,\ref{f-calamardos}) offer a sliced view of the emitting
   volume intersected by different planes perpendicular to the
   \los. Therefore, the channels near the systemic velocity
   (\vsys$\sim$23-26 and $\sim$47\,\kms\ for the \nbw\ and \sbw,
   respectively) represent the cross-sectional cut of the ellipsoids
   through their expansion center. This is indeed consistent with the
   fact that the largest ring-like emitting areas are observed near
   \vsys\ and enable us to estimate $b$$\sim$4\farc1 (for both fish
   bowls). The $b$-axis is oriented along PA=35\degr+90\degr\ and
   PA=10\degr+90\degr\ for the \nbw\ and \sbw, respectively.  
   esto es suponiendo que el eje b--> este contenido en el 
   del cielo. Sabemos que esto es asi de hecho porque la 
   conecta los appr. and receding extremes esta orientada 
   el eje menor de la elipse grande.  si el eje b tuviese 
   angulo resp. al plano del cielo, medido hacia la 
   lineas no coincidirian!  as well as an upper limit 
   $a$$<$3\arcsec.  

   The channels at the most extreme radial velocities
   (|\vlsr$-$\vsys|$\sim$35 and 30\,\kms\ for the \nbw\ and \sbw,
   respectively) sample the gas that is moving directly toward or
   away from us. The gas in these regions, referred to as
   approaching/blue (B) and receding/red (R) extremes, are expected to
   be near the vertex (or endpoint) of the ellipsoids along the
   $c$-axis, which is consistent with the compact emission observed in
   these velocity-channels (small ellipses in
   Figure\,\ref{f-calamardos}). The sky-plane projection of the
   $c$-axis then runs through the imaginary line that joins the
   receding and approaching extremes ($\overline{\rm RB}$). Because
   $\overline{\rm RB}$ has the same orientation as the minor ($a$)
   axis of the big ellipse (for both fish bowls), the $b$-axis must
   lie in the plane of the sky (or very close to it).

We measure a linear velocity gradient between the approaching and
receding extremes, which are separated by $\overline{\rm
  RB}$$\sim$3\arcsec\ in both cases. In the \nbw, the radial velocity
difference between R and B, $\sim$70\,\kms, results in a velocity
gradient along along the PA=35\degr\ axis of $\nabla
\upsilonup$$\sim$23$\times$$\tan{(i^{\rm N}_{\rm
    RB})}$\,\kms\,arcsec$^{-1}$, where $i^{\rm N}_{\rm RB}$ is the
line-of-sight inclination of the $\overline{\rm RB}$ line of the
\nbw. In the \sbw, the full width of the emission profile is slightly
smaller, $\sim$60\,\kms,
which leads to $\nabla \upsilonup$$\sim$20$\times$$\tan{(i^{\rm
    N}_{\rm RB})}$\,\kms\,arcsec$^{-1}$ along PA=10\degr. In the case
of the \sbw, the value of the gradient is rather uncertain since both
FWZI and $\overline{\rm RB}$ are poorly determined due to
contamination of the CO blue-wing by intense emission from other
regions of the nebula that overlap along the \los.

As already mentioned,  it
   is not possible to obtain an accurate estimate of $a$ and $c$ 
   because $i_{\rm c}$ is unknown (we note that, in general,
   \ic$>$$i_{\rm RB}$, as can be seen in Fig.\,\ref{f-elipse}-middle).
   An upper limit to $a$$\la$3\arcsec\ is given by the size of the
   minor axis of the big ellipses, which represents the sky-projection
   of the outer surface of the ellipsoids. However, $c$ and $i_{\rm
     c}$ are strongly degenerate: although a lower limit to
   $c$$\ga$3\arcsec\ is deduced from the distance between R and B,
   values of $c$ almost as large as desired are possible if $i_{\rm
     c}$ approaches 0\degr\ (i.e., pole on view). Possible
   combinations of $c$ and $i_{\rm c}$ consistent with our observables
   (i.e., the sky-projected dimensions of the ellipsoids and
   $\overline{\rm RB}$$\sim$3\arcsec\ in the CO maps) are shown in
   Fig.\,\ref{f-elipse} (right). As we discuss in \S\,\ref{dis-mloss},
   assuming that the fish bowls are nearly orthogonal to the
   large-scale outflow (\ic$\sim$40\degr), the size of $c$ would be
   similar to $b$ (solid lines in Fig.\,\ref{f-elipse}-right) and,
   thus, the fish bowls would simply approach oblate spheroids.
   

   \section{Analysis}
   \label{s-anal}

   
   \subsection{Density and temperature distribution}
   \label{s-rd}
   
We have used the ALMA maps of the CS\,(7-6) and CS\,(6-5) transitions
to derive the physical conditions within the
$\sim$4\arcsec$\times$12\arcsec\ central regions of \oh's molecular
outflow, where the prevailing 
structure is the so-called large-hourglass component
(\S\,\ref{res-largehg}).
This is also where most of the
emission from CS arises and where the ALMA maps of the two lines
overlap and, thus, their surface brightness distributions can be
compared.

We have used the classical population (or rotational) diagram analysis
technique to derive the CS column density (\ntotcs) and the rotational
temperature (\trot) distribution in these regions and at different
velocity channels. This method is described in detail and discussed
extensively by, e.g., \cite{gol99} and it has been successfully used
in the analysis of the mm-wavelength line survey of
\ohs\ \citep{san15,vel15} and the molecular emission from the
envelopes of many other evolved stars \citep[normally studied by means
  of single-dish observations; e.g.,][]{jus00,wes10,quin16,vel17}. In
the rotational diagram analysis technique, the natural logarithm of
the column density per statistical weight (N$_{\rm u}$/g$_{\rm u}$) is
plotted against the energy of the upper level above the ground state
(\eu) for a number of (at least two) transitions of the same
molecule. Assuming that the lines are optically thin and thermalized,
that is, that all levels are under local thermodynamic equilibrium (LTE)
conditions at a given unique temperature, N$_{\rm u}$/g$_{\rm u}$ and
\eu\ are related by the following formula:

\begin{equation}
\label{equ-rd}
\ln \frac{N_{\rm u}}{g_{\rm u}} = 
\ln \frac{3kW_{\rm ul}}{8\pi^3\nu_{\rm ul}\,S_{\rm ul}\mu^2} = 
\ln \frac{\ntotcs}{Z(\trot)} - \frac{\eu}{k\trot},  
\end{equation}

\noindent 
where $k$ is the Boltzmann constant, W$_{\rm ul}$ is the source
brightness temperature integrated over the channel velocity-width,
$\nu_{\rm ul}$ and S$_{\rm ul}$ are the frequency and line strength of
the transition, respectively, $\mu$ is the appropriate component of
the permanent dipole moment of the molecule, Z(\trot) its partition
function, and $u$ and $l$ refer to the upper and lower levels involved
in the transitions.  According to Eq.\,\ref{equ-rd}, for a given
molecule a straight-line fit to the points in the population diagram
provides \ntotcs/Z(\trot) from the $y$-axis intercept and \trot\ from
the slope of the fit.

\begin{figure*}[]
   \centering 
   \includegraphics[width=0.945\hsize]{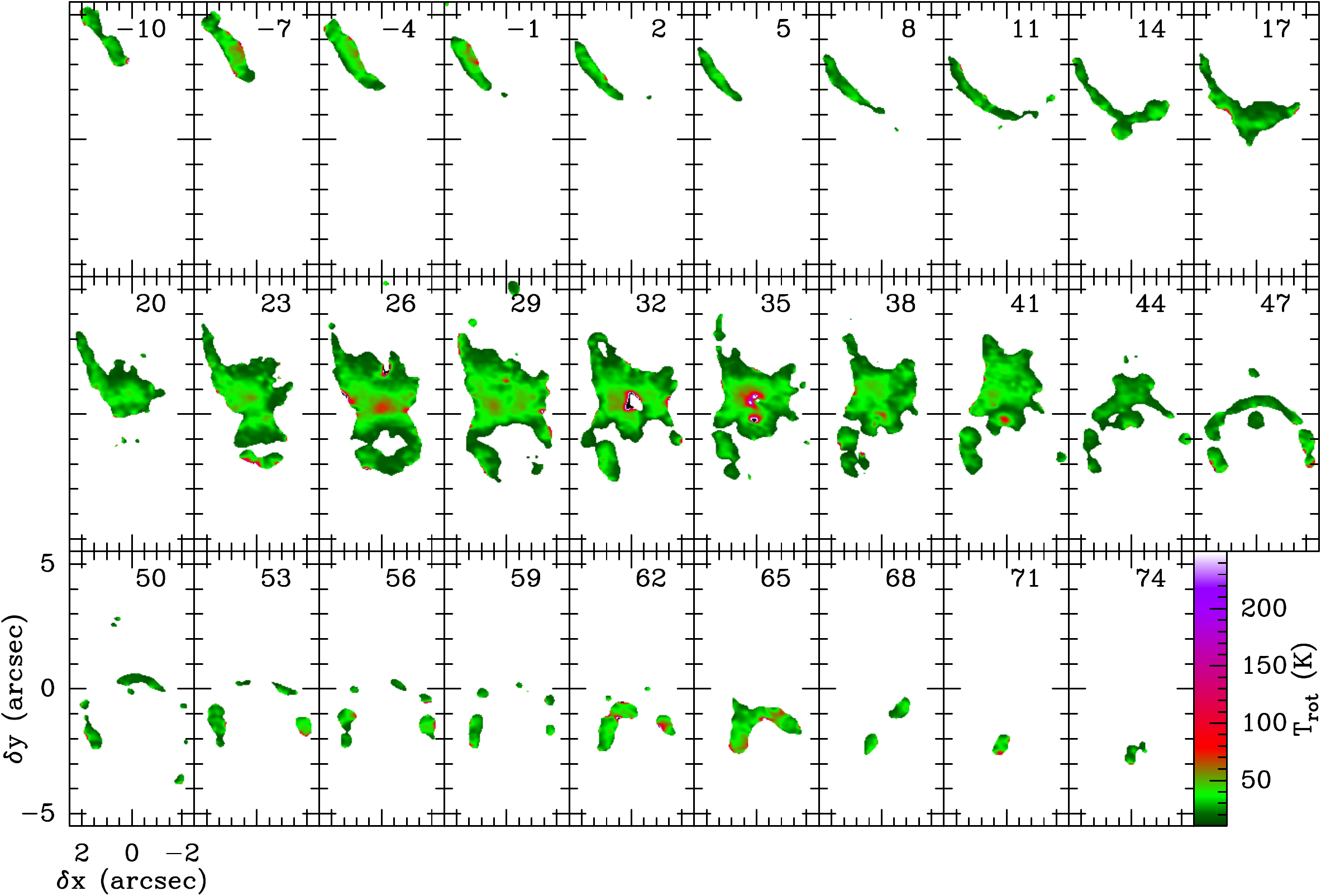} \\
   \vspace{0.45cm}
   \includegraphics[width=0.945\hsize]{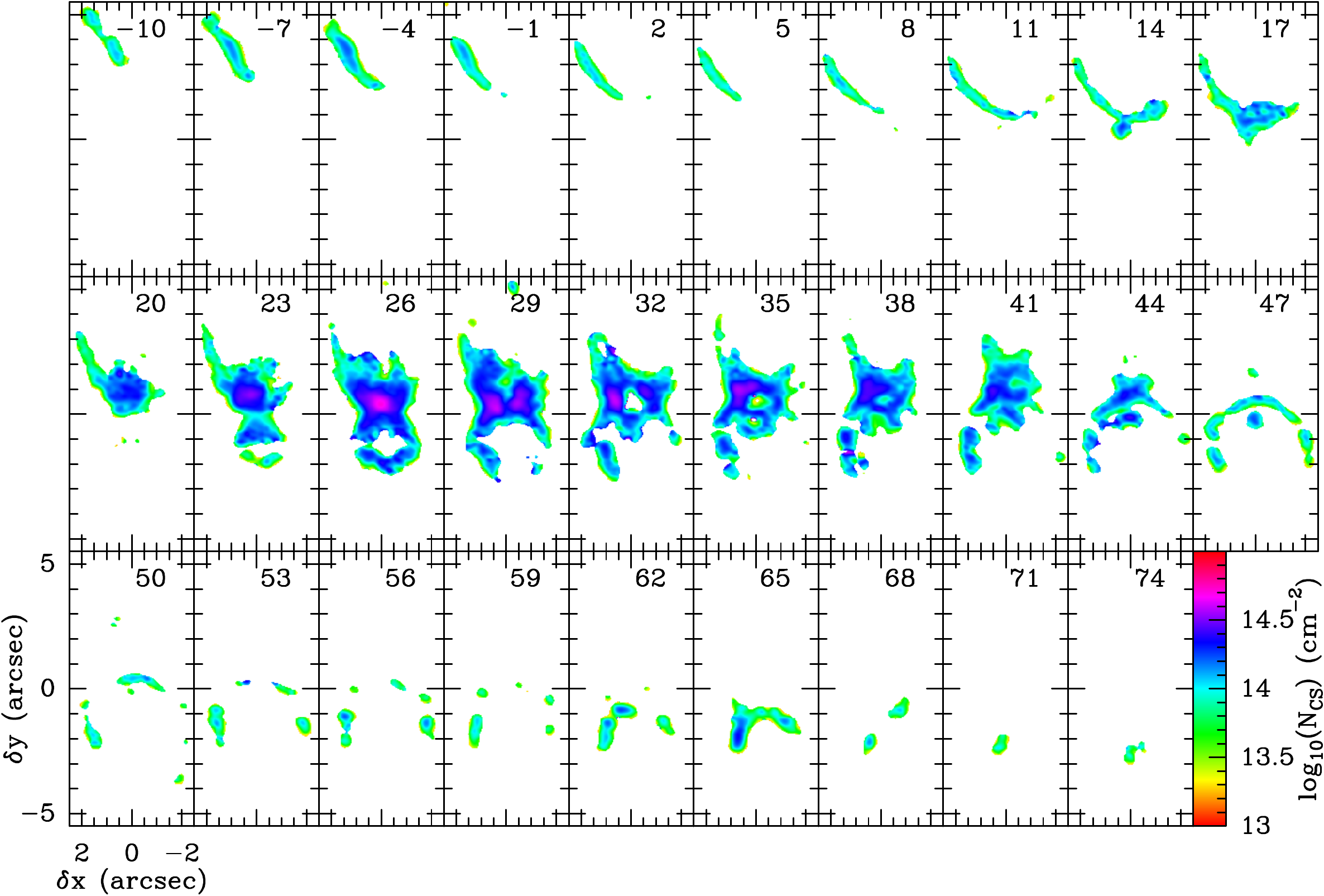}
   \caption{Velocity-channel maps of \trot\ (top) and column density
     of CS (in logarithmic scale, bottom) deduced from the
     CS\,(7-6)-to-CS\,(5-6) ratio (\S\,\ref{s-rd}).
     Beam is 0\farc31$\times$0\farc25, PA=$-$85\degr.}
   \label{f-rds}
   \end{figure*}

The surface brightness distribution of the CS\,(7-6) and
\mbox{CS\,(6-5)} emission lines integrated over the velocity width of
each channel (W$_{\rm 76}$ and W$_{\rm 65}$) are transformed to
$\ln(N_7/g_7)$ and $\ln(N_6/g_6)$ velocity-channel maps, respectively,
using the conversion factors in Eq.\,\ref{equ-rd} and after matching
the dimensions and beam size of both maps to a common value of
0\farc31$\times$0\farc25 (corresponding to the largest HPBW of the
two, the CS\,(7-6) maps). The required spectroscopic parameters and
partition function
have been obtained from the MADEX catalog \citep{madex}.
Then, a straight-line has been fitted pixel by pixel to the
$\ln(N_7/g_7)$ and $\ln(N_6/g_6)$ velocity-channel maps in order to
create the corresponding cubes of \trot\ and
\ntotcs\ (Fig.\,\ref{f-rds}).




Our velocity-channel maps of \trot\ confirm the overall low
temperatures of the molecular envelope of \ohs\ inferred from previous
works (\S\,\ref{intro}). We find values of \trot$\sim$10-30\,K in the
lobe walls and moderately larger temperatures,
\trot$\sim$40-60\,K, in most regions of the equatorial waist. There is
a trend for the innermost waist regions to show the largest
temperatures, with maximum values of \trot$\sim$70-150\,K (around
\vlsr=[26:41]\,\kms\ in Fig.\,\ref{f-rds}-top).

The CS column density in the lobe walls and outermost regions of the
waist are in the range \ntotcs$\sim$3\ex{13}-2\ex{14}\,\cm2,
increasing toward the central parts up to $\sim$5\ex{14}\cm2
(Fig.\,\ref{f-rds}-bottom). Considering both the column density and
temperature profiles derived from the CS emission maps, we deduce
moderate optical depths (at the center of our \dv=3\kms-wide channels)
for both transitions, with opacity values of $\tau$$\sim$0.05-0.8
across most regions of the nebula; optical depths close to $\sim$1 are
only expected at the dense and warm clump near the center that emits
at \vlsr$\sim$26-29\,\kms.

The column densities and rotational temperatures derived from our ALMA
data are in good agreement with the source-averaged values obtained
from a similar rotational diagram ana\-ly\-sis using multiple
transitions of CS covering a much broader range of levels (up to
\eu$\sim$400\,K) observed with the \iram\ and \hso\ single-dish
telescopes (Velilla Prieto et al., private communication).  The rotational
diagram analysis technique has been used not only for CS but for a
large number of molecules present in \ohs\ and is known to describe
remarkably well the envelope-averaged level populations, which closely
follow a straight line in that ln(N$_{\rm u}$/g$_{\rm u}$)-vs-\eu\ representation
for the majority of the species \citep{vel15,san15}. For some
molecules (including CS and a few others) the rotational diagram
analysis of single-dish data reveals two major temperature components,
one cold (\trot$\sim$15-50\,K) and one warm (\trot$\sim$100-200\,K),
which is consistent with the temperature distribution found in
this work.

Using the CS column density maps, we have generated a total (H$_2$)
number density cube taking into account the linear scale of the pixel
at a distance of $d$=1500\,pc, the overall velocity
gradient of the outflow,  and
making the simplifying assumption that the CS-to-H$_2$ fractional
abundance is constant throughout the outflow, for which we adopt a source-averaged
value of X(CS)=5\ex{-8} \citep[][Velilla Prieto et al.\, in
  preparation]{san97}. As shown in Fig.\,\ref{f-cs-dens}, we find
rather high densities inside the lobe walls
(\dens$\sim$[2-4]\ex{5}\cm3) and even higher (up to
\dens$\approx$10$^6$\,\cm3) in regions of the waist closer to the
nebula center.


Given the critical densities of the CS transitions used in this work,
\nc$\sim$[2-5]\ex{6}\,\cm3 \citep[for a 10-100\,K temperature
  range,][]{shi15}, some deviations from the LTE level population may
exist, particularly in the inner and outer edges of the lobe walls. In
these regions, the rotational temperature derived from our analysis
may then represent a lower limit to the gas kinetic temperature.
The discrepancy between \trot\ and
\tkin, however, must be small because the majority of the molecules detected in
\oh, including CO (with much smaller values of \nc), consistently point to
very similar rotational temperatures of $\sim$20-40\,K in the lobes
\citep{san15,vel15,alc01,clau00,san97,omo93,mor87}. Moreover, under
non-LTE conditions, the CS column densities derived from the rotational diagram analysis
(assuming LTE conditions) would be slightly underestimated, which
would then imply that true H$_2$ densities in the lobes
are larger than our estimates, alleviating the expected LTE departures. 
We note that LTE deviations are confirmed to be small or modest in most
regions of \oh's molecular outflow for other species, including dense
gas tracers, based on detailed molecular excitation calculations
\citep{vel15,san15}.

%

    \subsection{Molecular abundance of new species}
    \label{s-abun}
    \subsubsection{Refractory molecules in \cs}
    \label{abun-refr}
    We report the first detection of \nacl\ toward the mass-losing
    star \qx. The first identification of Na$^{35}$Cl (= NaCl) in
    O-rich evolved stars was made by \cite{mil07} toward the
    red supergiant VY CMa and the AGB star IK Tau.  Previously, NaCl
    had been detected in the circumstellar envelopes of two C-rich
    stars, the AGB star IRC+10216 \citep{cer87} and the pPN CRL\,2688
    \citep{hig03}.
The compact distribution of the \nacl\,($v$=0, $J$=26-25) emission in
\oh, which is found to be confined within the a central region (\cs) of
half-intensity radius of $\sim$60\,AU (\S\,\ref{res-clumps}), is
expected. This is because \nacl, as well as other refractory molecules
such as SiS and SiO, are thought to be formed
close to the photosphere and then to condense onto dust grains as the
envelope material flows from the central star \citep{gla96}. A similar
inner-envelope distribution, within $\sim$50\rs, is found for NaCl in
VY CMa and IK\,Tau \citep{mil07,dec16}.


We have obtained an order of magnitude estimate of the column density
and fractional abundance of \nacl\ in the inner layers of \qx's wind
(\cs) under some simplifying assumptions. We have used the
CLASS\footnote{{\tt http://www.iram.fr/IRAMFR/GILDAS}.} task {\tt
MODSOURCE} to model the observed \nacl\,($v$=0, $J$=26-25) emission
profile assuming LTE and taking as input model parameters the column
density, kinetic temperature, source size and the line width. The
source angular diameter and the line width are constrained by the
observations to $\sim$0\farc08 and FWHM$\sim$8\,\kms, respectively
(\S\,\ref{res-clumps}).  We have considered a relatively broad range
of probable values for the excitation temperature between 100 and
1000\,K in the inner layers where the emission is produced
\citep[e.g.,][and references therein]{mil07}. Adopting these
temperatures, the column densities of \nacl\ needed to reproduce the
observed \nacl\,($v$=0, $J$=26-25) profile are between $\sim$2\ex{14}
and 5\ex{14}\,\cm2, comparable to the column densities of NaCl found
in VY\,CMa and IK\,Tau \citep{mil07}.  While our crude estimate has
large errorbars, this agreement suggests that our order-of-magnitude
value is roughly correct, since the $^{35}$Cl/$^{37}$Cl isotope ratio
is known to be low, about two, in intermediate-mass AGB stars
\cite[e.g.,][and references therein]{maa16,hig03}.

The \nacl\ fractional abundance (relative to H$_2$) in \cs\ 
can be estimated if the current mass-loss rate (\mloss) and expansion
velocity (\vexp) of \qx's wind is known. 
Adopting constant values of \mloss$\sim$10$^{-5}$\,\my\ and
\vexp=7\,\kms\ for the present-day wind (see \S\,\ref{intro}), the
column density in a spherical shell with inner and outer radius of
\rin$\sim$6\,AU and \rout$\sim$60\,AU is \nh$\sim$2\ex{23}\,\cm2.  For
\rin, we have adopted the characteristic radius of the SiO masing
region in \oh, which corresponds to the dense
$\sim$10$^{9}$-10$^{10}$\cm3 layers of the extended atmosphere of
\qx\ at $\sim$2-3\rs\ \citep{san02} where molecules such as SiO, SiS,
and NaCl are expected to form abundantly.
Considering the approximate value of \nh\ obtained above, we deduce a \nacl\ fractional abundance of about \mbox{$\approx$[1-3]\ex{-9}}. 
Given the low $^{35}$Cl/$^{37}$Cl$\sim$2 ratio expected, we estimate
X(NaCl)$\approx$[2-6]\ex{-9}, which, in spite of the large errorbars
(easily of a factor $\sim$3), is again in very good agreement with the
fractional abundance of salt measured in VY\,CMa and IK\,Tau
\citep{mil07}.

Applying the same modeling procedure and assumptions, we estimate the
abundance of SiS and SiO in \cs\ using the \sisv\ and \tsiov\ thermal
line profiles observed (\S\,\ref{res-clumps}). In this case, the range
of probable excitation temperatures explored is $\sim$500-1500\,K,
given the higher-excitation requirements of these vibrationally
excited transitions (\eu$\sim$1200-1800\,K) compared to the
\nacl\,$v$=0 line (\eu=214\,K). We deduce SiS and \tsio\ column
densities of $\sim$[4-6]\ex{16}\,\cm2 and $\sim$[1-3]\ex{16}\,\cm2,
respectively, resulting in fractional abundances of
X(SiS)$\approx$[2-3]\ex{-7} and X(\tsio)$\approx$[0.5-1.5]\ex{-7}. For
a $^{28}$Si/$^{30}$Si isotopic ratio of $\sim$20-30 (roughly
consistent with the solar value, $\sim$29), as measured for some
O-rich AGB stars \citep[e.g. IK\,Tau,][and references therein]{vel15}
including \ohs\ (Velilla Prieto et al., in preparation), we deduce a
fractional abundance of SiO in \cs\ of X(SiO)$\sim$[1-5]\ex{-6}.  Our
estimates of X(SiS) and X(SiO,\tsio) are in good agreement with
measurements and theoretical expectations in the inner-envelope
regions of M-type AGB stars \citep[e.g.,][]{sch07,gon03,dec10,gob16}.
Finally, even considering the large uncertainties of our estimates,
our results indicate that SiO is clearly more abundant (by $\sim$2
orders of magnitude) in \cs\ than at large distances from the star in
the fast bipolar lobes, where a fractional abundance of
$\approx$10$^{-8}$ is calculated from previous molecular studies using
SiO $v$=0 transitions \citep{san97,mor87}. Our value of the SiO
abundance in \cs\ as well as the inferred depletion factor at large
distances is also in good agreement with recent measurements of X(SiO)
in the inner wind of $o$\,Cet (Mira) by \cite{wong16}. Using
long-baseline ALMA observations, these authors find that the SiO
abundance drops significantly from 1\ex{-6} to 1\ex{-8}-1\ex{-7} at a
radius of about 1\ex{14}\,cm $\sim$ 5\,\rs.

\subsubsection{Methanol}
\label{abun-ch3oh}

The detection of \metanol\ in evolved stars has remained elusive until
very recently. Indeed, \ohs\ is the first CSE of an AGB star in which
gas-phase methanol has been identified, after the recent discovery of
this molecule with ALMA in the post-AGB object HD\,101584
\citep{olo17}.  Only one year before, methanol was detected in
  the yellow hypergiant IRC+10420 \citep{quin16}, although this is a
  different class of object descending from a much more massive star
  ($\sim$50\,\msun) than \qx\ and HD\,101584.

\begin{figure*}[htbp]
   \centering 
   \includegraphics*[bb=60 0 560 600,width=0.35\hsize]{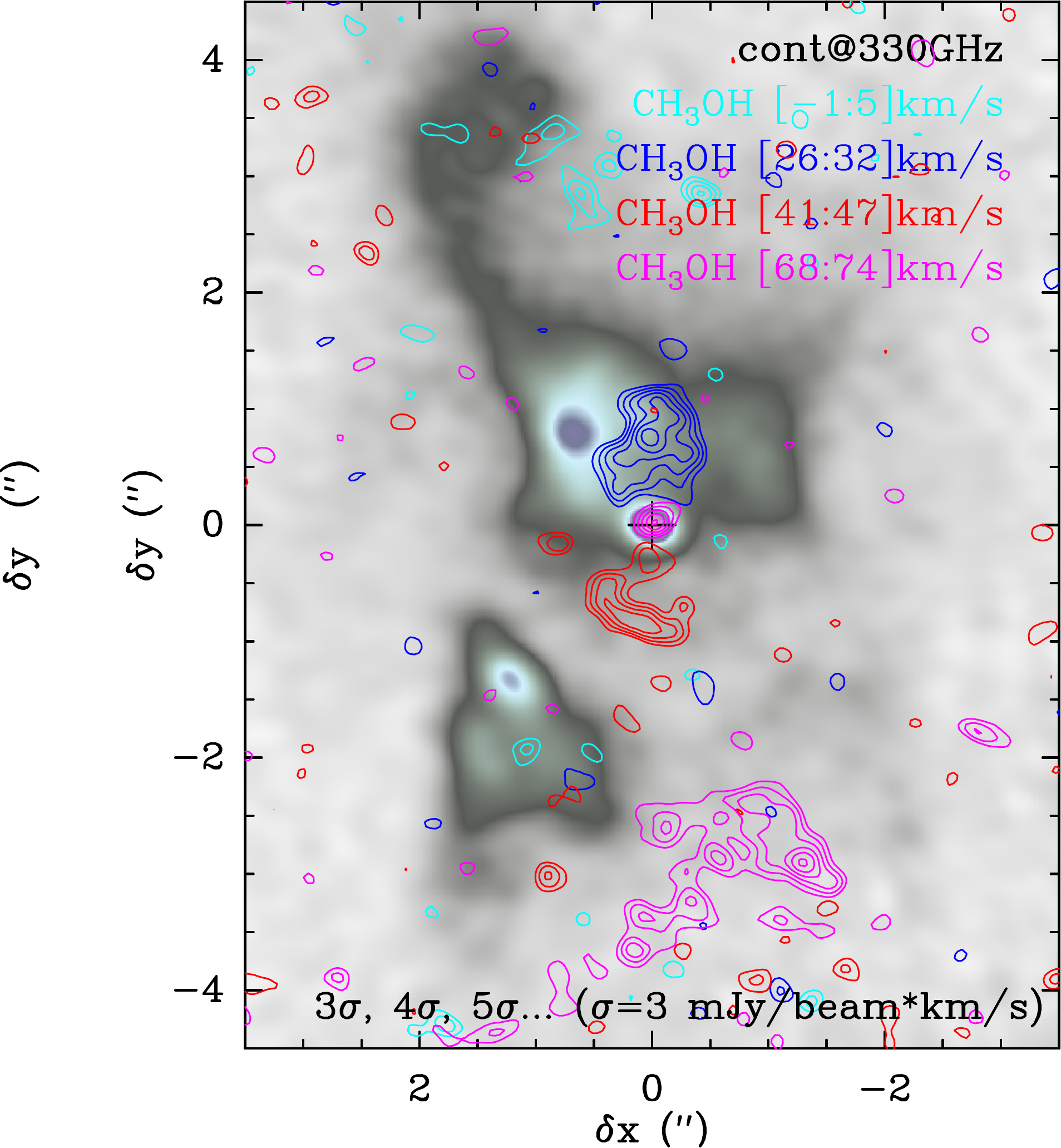}
   \includegraphics*[bb=60 0 560 600,width=0.35\hsize]{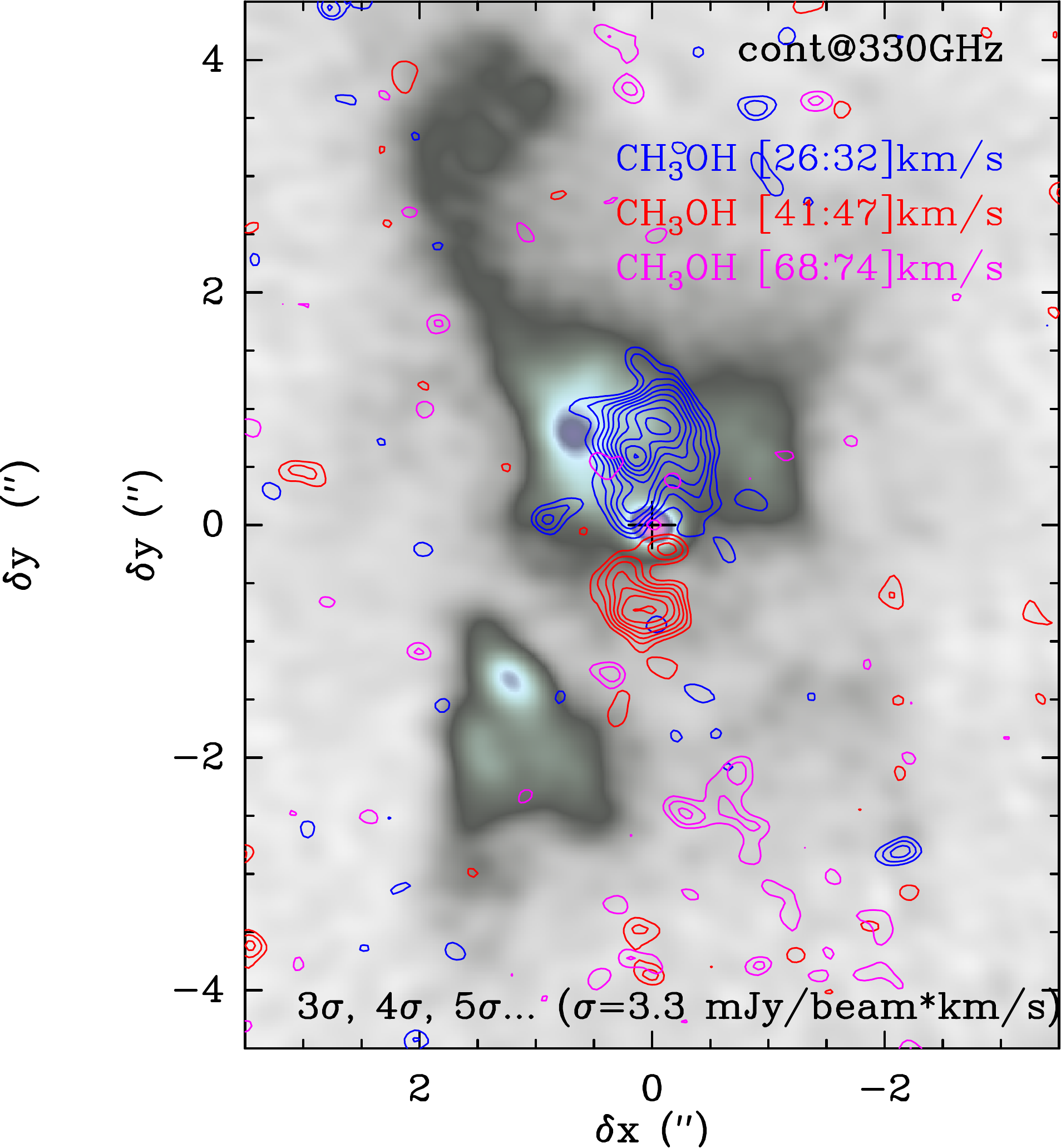} \\
   \vspace{0.25cm}
   \includegraphics[width=0.35\hsize]{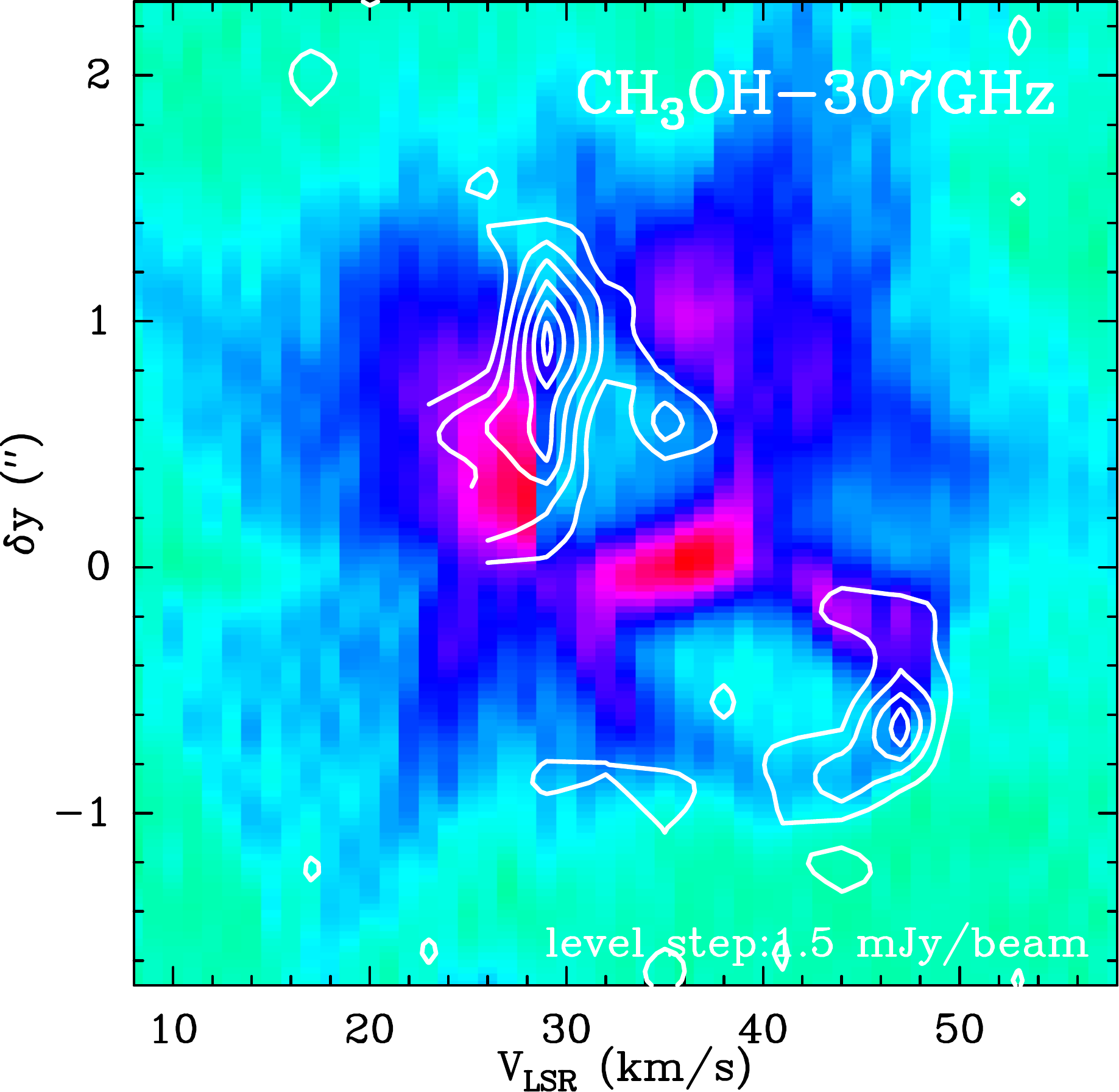}
    \includegraphics[width=0.35\hsize]{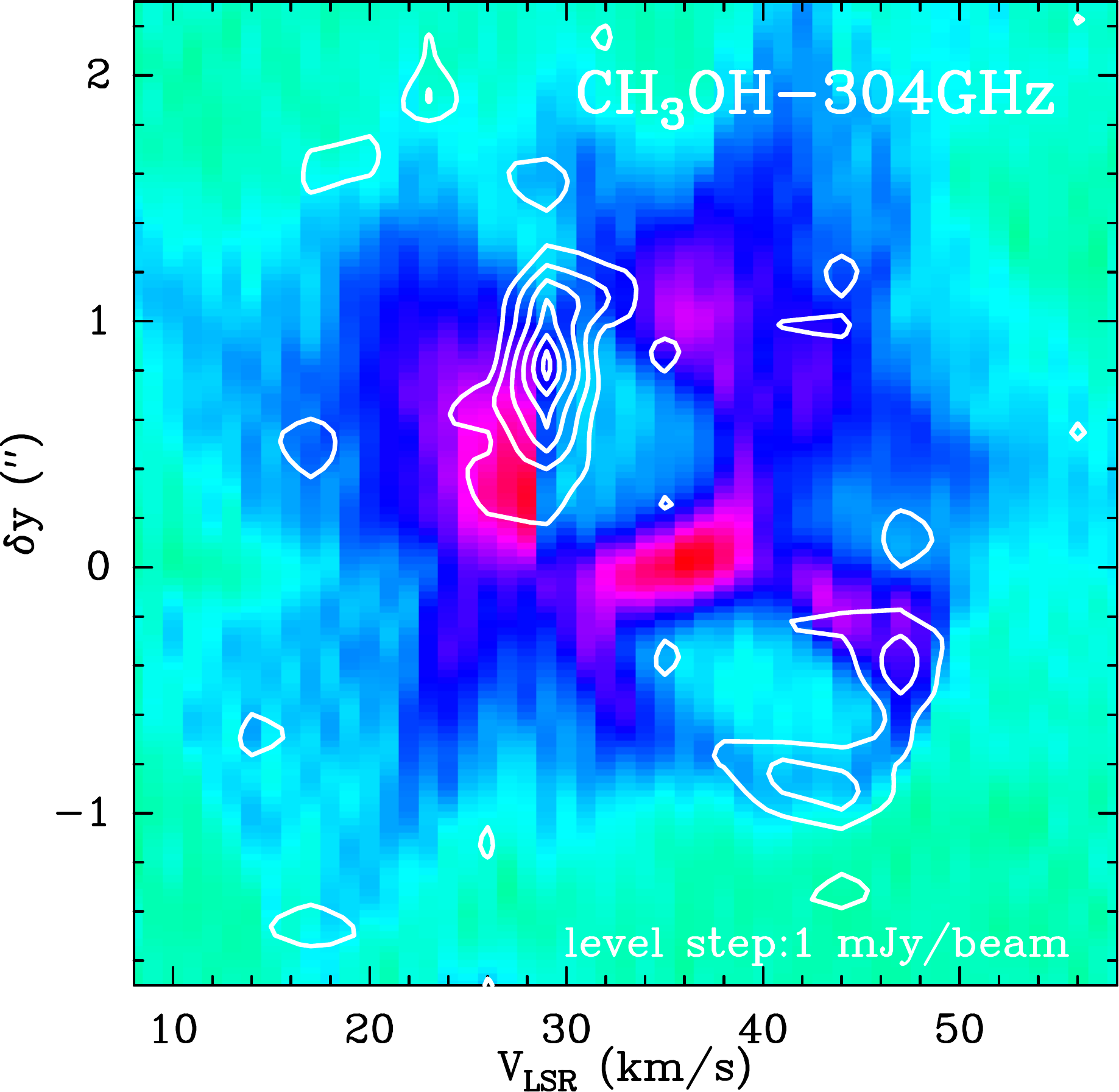}
 \caption{CH$_3$OH line emission towards \ohs, transitions at 307.166\,GHz (E$_{\rm
     up}$=38\,K) and 304.208\,GHz (E$_{\rm up}$=21.6\,K). {\bf Top:} Integrated intensity maps over the
   \vlsr\ ranges where emission is detected (indicated inside the
   boxes) overimposed on the 330\,GHz-continuum map. {\bf Bottom:}
   Axial PV diagram of \metanol\ (white contours)
   on top of \trecet\ for comparison (color-scale). }
         \label{f-ch3oh}
   \end{figure*}

In the post-AGB object HD\,101584, methanol was found to come
exclusively from two extreme velocity spots (at about
\vlsr=\vsys$\pm$140\,\kms) on either side of the object where a HV
outflow interacts with the surrounding medium. 
In a similar way, the methanol emission in \ohs\ arises solely from
the bipolar outflow, predominantly from the periphery of the compact
bipolar outflow traced by SiO and, to a lower extent, from more
distant regions along the large-scale lobes (at offsets
$\delta$y=$\pm$1\arcsec\ and $\delta$y=$\pm$3\arcsec,
Fig.\,\ref{f-ch3oh}). The main difference with respect to HD\,101584
is that the outflow expansion velocities (projected along the
line-of-sight)
where \metanol\ is detected are significantly lower in \ohs\ ($\sim$6
and $\sim$35\,\kms\ at $\delta$y=$\pm$1\arcsec\ and $\pm$3\arcsec,
respectively, see \S\,\ref{res-sio}).

   As we did for refractory species (\S\,\ref{abun-refr}), we have
   obtained an order-of-magnitude estimate of the column density and
   fractional abundance of \metanol\ in the regions of the compact
   bipolar outflow of \ohs\ where this molecule is 
   detected. The structure and physical conditions in the emitting
   volume are crudely constrained from the observations and,
   therefore, we need to make some simplifying assumptions.  As seen
   in Fig.\,\ref{f-ch3oh}, most of the
   \metanol\ emission (within \vsys$\sim$26-47\,\kms) comes from two
   $\sim$1\arcsec$\times$1\arcsec-sized regions centered at about
   $\delta$y$\pm$1\arcsec\ from \qx. Accordingly, we adopt a size for
   the whole emitting area (projected in the sky-plane) of
   $\sim$2\,arcsec$^2$ and a representative half-intensity line width
   of $\sim$20-25\,\kms\ as input values of our simple LTE (CLASS/{\tt
     MODSOURCE}) model. We have considered a rather broad range of
   probable excitation temperatures of $\sim$50-200\,K, which includes
   (but also goes beyond) the excitation temperature span in these
   \metanol-emitting $\delta$y$\pm$1\arcsec\ regions (as deduced in \S\,\ref{s-rd} -- see
   Fig.\,\ref{f-rds}). For the previous model parameters, we find that
   the two \metanol\ lines observed (at 304 and 307\,GHz) imply column
   densities of about \mbox{\ntotch$\sim$[1-8]\ex{14}\,\cm2}.
   The model reproduces the relative
   intensities of both transitions (of comparable strength in our ALMA
   data) and is consistent with these and most
   \metanol\ transitions being below (but, in some cases, close to)
   the detection limit of our $\sim$79-350\,GHz spectral survey of
   \ohs\ with the \iram\ antenna (Velilla Prieto et al., in
   preparation).

\begin{figure*}[]
   \centering 
   \includegraphics[width=0.95\hsize]{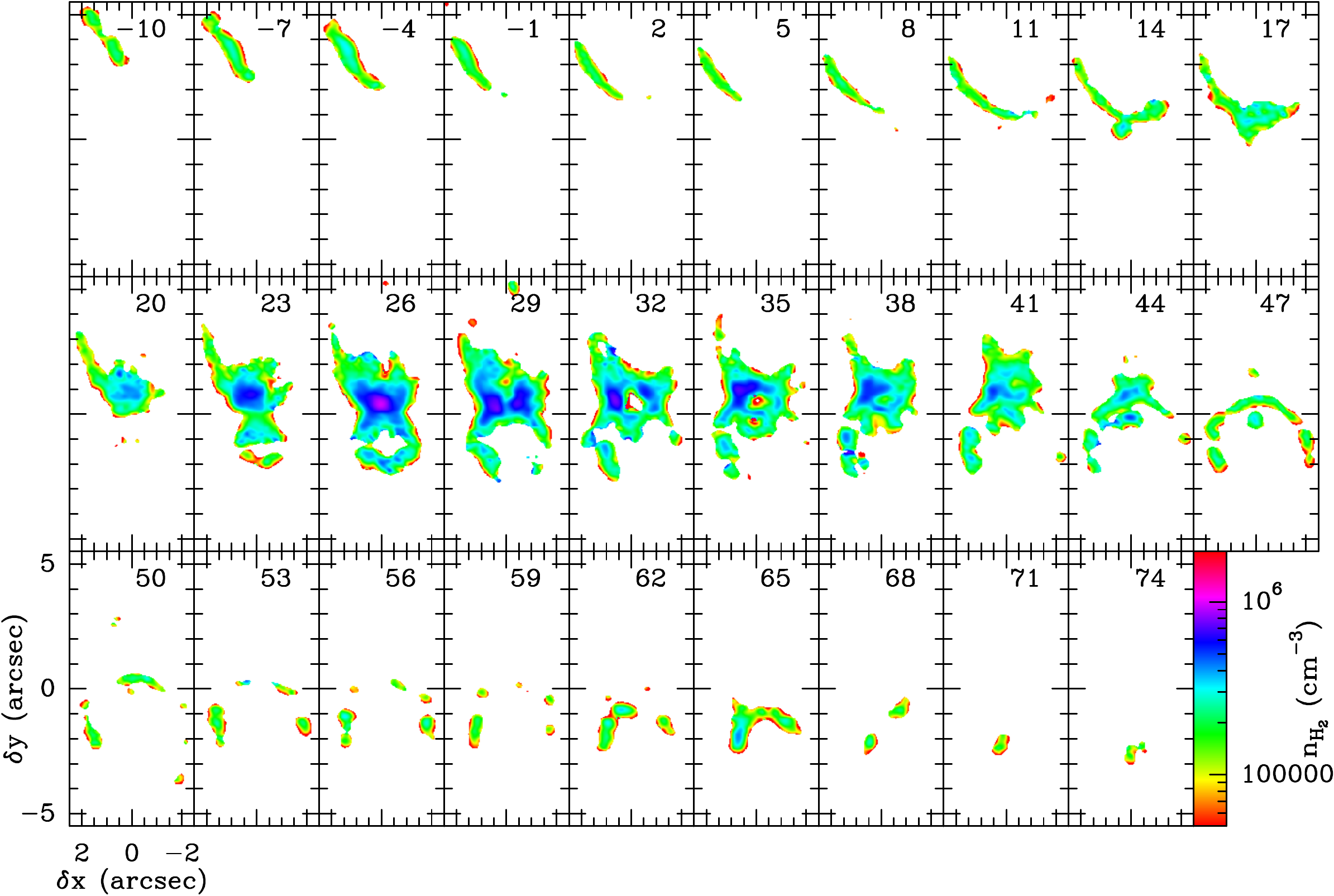}
   \caption{As in Fig.\,\ref{f-rds} but for the total (H$_2$) particle number
     density, \dens\ (\S\,\ref{s-rd}).}
   \label{f-cs-dens}
   \end{figure*}

   To compute the total (H$_2$) column density across the
   \metanol-emitting region, which is needed to estimate the
   fractional abundance of methanol, we make the assumption that this
   is roughly represented by two spherical homogeneous clumps of
   diameter $\sim$1\arcsec\ ($\sim$2\ex{16}\,cm). These regions are
   characterized by densities of \dens$\sim$[3-4]\ex{5}\cm3
   (\S\,\ref{f-rds} -- see Fig.\,\ref{f-cs-dens}, channels
   \vlsr$\sim$29 and $\sim$44\,\kms) and, therefore, the column
   density is approximately \nh$\approx$[6-8]\ex{21}\,\cm2. This
   implies a fractional abundance of methanol of the order of
   X(\metanol)$\approx$10$^{-8}$-10$^{-7}$. This value, while it is
   expected to have large errorbars, is intermediate to the methanol
   fractional abundance found in the post-AGB object HD\,101584
   \citep[$\sim$3\ex{-6},][]{olo17} and in the yellow hypergiant
   IRC+10420 \citep[$\sim$7\ex{-8},][]{quin16}.
   
   As in the post-AGB object HD101584, the detection of gas-phase
   methanol in \ohs\ in regions that have been (or are currently
   being) shocked, suggest that this molecule has been extracted from
   the icy mantles of dust gains after the passage of shocks.
   As shown by \cite{gui09}, dust destruction in slow shocks
   ($\leq$50\,\kms) is expected to be significant (at the level of a
   few percent)\footnote{Grain destruction can be produced in various
     forms, vaporizing grain-grain collisions being a dominant form for slow
     shocks.} and can return to the gas phase in a similar proportion
   molecular species that reside in their mantles (like \metanol) but
   also molecules situated in deeper layers (like SiO).

\section{Discussion}
\label{dis}

\subsection{The position of \qx\ relative to the large waist}
\label{dis-qxoffset}


One of the most puzzling
discoveries from our ALMA data is the spatial offset between some of
the nebular components identified in this work. In particular, the
mass-losing star (\qx), which is enshrouded by dust and gas inside
\cs, is clearly offset from the waist of the large hourglass 
by $\delta$y$\sim$0\farc6 (900\,AU) along the axis toward the south lobe.
The systemic
velocity of these two components is also different: \vsys$\sim$35 and
\vsys$\sim$32-33\,\kms\ for \cs\ and the large waist, respectively.
Here, we examine two scenarios that could possibly
cause the offset and \vsys\ difference, although regrettably
neither is fully satisfactory.

$a$) In principle, this offset could reflect the motion of the star
along its orbit $\sim$800-870\,yr ago, when it ejected the bulk of the
CO outflow (\S\,\ref{res-largehg} and \ref{res-hv}). This effect has
indeed been observed in the young PN \mbox{M\,2-9}, where two coaxial,
slightly off-centered CO-rings, with systemic velocities differing by
$\sim$0.6\,\kms, were formed during two brief mass-loss episodes
that occurred at different points along the orbit of the primary
post-AGB star \citep{cc12,cc17}.

In the case of \ohs, the offset observed would imply a transversal (or
sky-plane projected) velocity of \vt$\sim$5\,\kms\ and, thus, an
orbital velocity of \vorb=\vt/$\sin{(i_{\rm orb})}$=8.7\,\kms\ at the
time of the massive ejection (adopting $i_{\rm orb}$$\sim$35\degr).
Under the orbital-motion scenario, one would then expect to measure a
difference in the systemic velocities of the large waist and \cs\ of
$\Delta$\vsys=\vr=\vorb$\times$$\cos{(i_{\rm orb})}$$\sim$7\,\kms,
which is significantly larger than the difference found, of only
$\sim$2-3\,\kms. This small \vsys-difference would be consistent with
the orbital-motion hypothesis if the \los-inclination of the binary
orbit is $i_{\rm orb}$$\sim$63\degr. While not impossible, such an inclination is
improbable given that bipolar outflows in pPNe are 
nearly orthogonal to the binary orbital plane according to most
current jet-launching theories.

A secondary criticism to the orbital-motion hypothesis is that it
would require the bulk of the mass to have been ejected when the stars
were near quadrature (within about 5\degr) to explain the very small
offset of \qx\ along the equatorial direction relative to the center
of the large waist ($\delta$x$\sim$$-$0\farc05, \S\,\ref{res-largehg}).  We stress that
this is assuming (for simplicity) a circular orbit.




$b$) Alternatively, \qx\ could be a runaway
star that has been flung from the core after the strong asymmetrical
mass ejection that led to the north and south lobes of the large-scale
CO outflow.  As shown by \cite{alc01}, the linear
momentum carried by the north lobe is larger than that carried by the
south lobe
by about 2.3\,\msun\,\kms\ (with an uncertainty factor of
$\sim$\,2-3).
Assuming that the ejection was inherently asymmetric, the
jet-launching star could have recoiled in the opposite direction
(i.e., toward the south) by conservation of momentum between the star
and the ejecta.
This scenario would explain why the offset between \qx\ and the large
waist is observed predominantly along the lobes' axis. In this case,
the expected \vsys-difference between \qx\ and the large waist,
$\Delta$\,\vsys=\vt$\times$$\tan{(35\degr)}$$\sim$5$\times$$\tan{(35\degr)}$$\sim$3.5\,\kms, is closer to the observed value.

We discuss now, some important challenges
of the recoil (or jet-propulsed star) scenario. First, the fast
bipolar winds of \ohs\ (and most pPNe) are believed to be launched by
the compact companion in the binary system after trapping part of the
slow wind of the mass-losing star (\S\,\ref{intro}). This would imply
that, under this scenario, the companion would have received most of
the momentum excess, being violently pushed away from its original
position. To make this hypothesis consistent with the observed
position of \qx, the compact companion should have dragged 
\qx\ along in its journey.
Although this is not impossible, it would require a close-binary
system with a primary companion significantly more massive than \qx\ for the
latter to have remained gravitationally bound to the primary.

Second, if both stars have recoiled, the total momentum of
jet-propulsed system would be huge, at least 13\,\msun\kms, given the
value of \vt$\sim$5\,\kms\ observed and adopting a total mass of the
system $m_1$+$m_2$$\sim$2.6\,\msun. This is in the most favourable case
that \qx\ was at the tip of the AGB at the time of the ejection, i.e.,
it had a mass close to its final value in this stage
\cite[$m_2$$\sim$0.6\,\msun, given that its mass in the
  main-sequence was about 3\,\msun,][]{jur85,blo95}. The mass of the A0
main-sequence companion is $m_1$$\sim$2\,\msun. This simple
calculation, indicates that the momentum carried by the jet-propulsed
system is much larger than the momentum excess of the north lobe,
which represents a major objection to this scenario.


Perhaps the combination of orbital-motion ($a$) and recoil ($b$) could explain, at
least partially, the mysterious location of \qx\ off-centered from the
large waist. Working together, the main problems facing these two
scenarios (when they are considered individually) would be alleviated
to some extent. Both phenomena are indeed plausible even if it is not
easy to establish what is their precise connection with the observed
offset and \vsys-difference.

We note that even if a small or moderate recoil has happened, the
main argument against orbital motion, based on the small
\vsys-difference between the large waist and \qx, is not well founded 
because the systemic velocity of the binary (and probably other
orbital parameters, e.g., \iorb\ and/or \vorb) would have changed
after the massive bipolar ejection.
In addition to the effect of recoil, which may or may not have
happened, the systemic velocity and the binary orbital parameters are
likely to have changed as the consequence of the primary-to-secondary
mass ratio reversal (from $q$=$m_1$/$m_2$$\sim$0.67 to 3.3) during the
evolution of the system from the early to the late stages of the
mass-loss period that \qx\ is undergoing. This makes unreliable the
comparison of the system radial velocities at different epochs (before
and after major mass-ejections) and its interpretation.

\subsection{The binary system at the core}
\label{dis-binary}

One of the important new results from these observations is the
disclosure of the locus of the mass-losing star \qx\ inside \cs. In
light of this discovery, a key question immediately arises: Is the
companion to \qx\ also inside \cs?

Our ALMA data have uncovered a compact bipolar outflow (exclusively
traced by SiO) that is emerging from the stellar vicinity. This is an
indication that the binary companion is inside or near \cs\ assuming,
as most current PN-shaping theories do, that the bipolar SiO outflow
is launched by the companion (powered via disk-mediated accretion) and
not from the mass-losing star.
Under this hypothesis, the relative positions between the
SiO outflow's expansion centroid and \cs\ would then indicate the
separation between the two stars. Unfortunately, the angular
resolution of our current ALMA observations is not sufficient to
obtain an accurate estimate of this separation, which we measure to be
$a$=50$\pm$60\,AU, but certainly $<$150\,AU.

Below, we extend our reasoning to try to limit the range of
possible values of the orbital period (\porb) and orbital
separation ($a$) of the central binary system.

{\sl Upper limit to the orbital size.}  The inner-wind layers
of \qx\ (that is \cs) and the inner regions of the compact bipolar
SiO outflow represent the most
recent ($\la$100\,yr old) mass-ejections known in
\ohs\ (\S\,\ref{res-clumps} and \ref{res-sio}).
Curiously, although the SiO outflow is presumably launched by the companion while
\cs\ represents the inner-wind layers of \qx, we do not observe a
significant difference between the systemic velocity of these two
components.
Thus, it is likely that the
centroids of the emission lines that sample these $\la$100\,yr old
mass-ejections denote
the systemic velocity of the mass center of the binary system, which
should be determined by the orbital motion of the mass-losing or
wind-launching stars averaged over at least one complete orbit.
This would imply that \porb$<$100\,yr and, thus, $a$$<$[30-35]\,AU, assuming
a circular orbit and adopting $m_1$=2\,\msun\ and $m_2$=[0.6-2]\,\msun.

This small orbital separation is consistent with the recent VLBI
observations of the SiO and \water\ masers in \ohs\ by
\cite{dod18}. These authors find that the SiO masers, formed in the
pulsating layers of \qx\ and coincident with the position of \cs\ (to
within 0\farc01, \S\,\ref{res-cont}), lie at
$\la$10-15\,milli-arcseconds from the center of expansion of the
\water\ masers. The latter trace a young ($\sim$38\,yr) bipolar
outflow, expanding at low velocity ($\sim$19\,\kms), which represents
the innermost parts of the SiO outflow reported by us (see also
\S\,\ref{dis-recent}).  Assuming that the \water\ masers mark the
driving jets that produce the macro-scale structure of the nebula and
assuming that the jets are launched by the compact companion, the
relative separation between the SiO masing site and the
\water\ masers' expansion center suggests an orbital separation
$<$25-30\,AU (for an average angle of $\sim$45\degr\ from
conjunction).


 {\sl Lower limit to the orbital size.} The presence of SiO masers in
 \ohs, observed at distances of $\sim$6\,AU from
 \qx\ (\S\,\ref{intro}), imposes a lower limit to the orbital
 separation, since the masers will not survive the ionization by
 the hot companion if the latter is too close.
This problem was examined statistically by \cite{sch95} in a sample of
dusty symbiotics (with Mira-type stars and WD companions) with
underluminous or absent SiO masers.
In a similar way, here, we consider the influence of the UV radiation
from the A0\,V companion in the vicinity of \qx.
For an orbital separation of $a$=8\,AU, the UV photons from the 10,000\,K main-sequence companion will
reach to a Strömgren radius of about 2\,AU, that is, just to the outer
bounds of the SiO masing site in \ohs. This is for a mean nebular
density around the companion of
\dens$\sim$6\ex{8}\,\cm3 \citep[using a density law
  \dens$\sim$10$^{9}$($\frac{6\,AU}{r}$)$^2$\,\cm3\ in the inner
  wind layers of \qx; see ][]{san02}. The SiO masers
are known to be placed within the dust condensation radius in AGB
stars, therefore, 
this region could be much more difficult to penetrate by the UV
radiation from the hot companion than estimated from the simple
dust-free analysis performed. The conclusion from this crude
computation is, thus, that the presence of SiO masers is not
incompatible with small orbital separations, even as small as $\sim$8\,AU.

For an inverse-square density law as adopted, the Strömgren radius
around the companion would increase with the distance to \qx\ as
$\sim$$r^{1.33}$. Therefore, for a $\sim$25-35\,AU-wide orbit, the
Strömgren radius will be $\sim$9-15\,AU.

\subsection{Nebular formation: major events}
\label{dis-mloss}


Our ALMA observations have unveiled a series of small- and large-scale
structures previously unknown. These structures, together with the
dominant nebular macro-structures known from previous works (the
HV lobes and the large-scale equatorial waist), relate
the history of mass-loss events that have occurred in the last
$\sim$1000 years and that have resulted in the extraordinary
nebular architecture displayed by \ohs.
Although uncertain, our
estimates of the kinematical ages of the different components are
useful to reconstruct the sequence of events that resulted in the
complex array of nebular structures observed.


\subsubsection{The HV lobes and the large waist}
\label{dis-macronebula}

A comprehensive study of the HV lobes and the large waist of \ohs\ as
well as a more complete discussion of their origin is deferred to a
future publication (paper II). Here, we briefly review some basic or
new aspects that are needed to contextualise our discussion on the new
nebular structures discovered with ALMA, which are the main focus of
this paper.

In agreement (within uncertainties) with the CO\,($J$=1-0) and
($J$=2-1) interferometric maps with
$\sim$1\arcsec-3\arcsec\ resolution analyzed by \cite{alc01}, our ALMA
data suggest that the HV lobes were shaped and strongly accelerated
about 800 years ago, maybe by the action of CFWs
impinging on the dense, pre-existing AGB ambient
material.

Recent hydrodynamical simulations by \cite{bal17} show that
a combination of fast clumps and wide-angle winds launched into a
dense, slow AGB wind produces pairs of inflated lobes and thin
columnar flows along the axis (reminiscent of the H$\alpha$
bubble-like lobes and the central spine of the CO outflow,
respectively) as well as a prevailing expansive kinematics with a
linear velocity gradient.

The temperature of the shocked gas predicted by these simulations is
remarkably high, even in the columnar jet-like component that
presumably corresponds to the cold CO lobes. In this region, the
simulations leads to post-shock temperatures of
\tkin\,$>$10$^2$-10$^3$\,K but the observations indicate much lower
values at present, \tkin$\la$40\,K (\S\,\ref{s-rd}). If the CFW+`AGB
wind' interaction scenario is correct, the low temperatures of the
CO outflow add support to the hypothesis that the duration of the
CFWs or clumps ejection was short.
Cooling of the post-shocked gas to a few\,10\,K (along with molecule
reformation) is expected to happen in just a few
years after the fast winds are ``switched off'', given the high
densities in the lobes (\dens$\approx$10$^5$\,\cm3,
\S\,\ref{s-rd}; for a similar discussion, see \citealt{san15}, \citealt{san02b}, and references therein).

%

 An important difference with respect to the data presented in
 \cite{alc01} is that, in our ALMA maps, the spatio-kinematic
 structure of the large-scale equatorial waist is fully resolved and
 indicates that this component has a kinematical age of $\sim$870\,yr,
 i.e., similar to that of the HV lobes and, significantly smaller
 than the upper limit ($\sim$4000\,yr) deduced for the central core in
 \cite{alc01}. The sharp boundary at the outer edge of the equatorial
 waist points to a sudden increase of the mass-loss rate
 with respect to the surrounding material that occurred $\sim$870\,yr
 ago or less.  This implies that the bulk of the mass in the central
 low-velocity core \citep[about 0.6\,\msun,][]{alc01} was ejected at a
 rate a factor $\sim$3-4 larger than estimated before, possibly at
 \mloss$>$6\ex{-4}\,\my.
 
Our ALMA maps show that the large waist does not expand at constant
velocity, which would be expected if this component is simply the
remnant of the AGB CSE pierced along the poles by fast bipolar
ejections.
we note that the expansion velocity in the large waist varies from
\vexp$\sim$3\,\kms\ at the inner edge, at $\sim$225\,AU$\ga$100\rs, to
about \vexp$\sim$25\,\kms\ at the outer boundary, at $\sim$2700\,AU
(\S\,\ref{res-largehg}). However, in a dust-driven AGB wind, at these
large distances from the central star (well beyond the dust formation
and wind acceleration layers) the gas should flow away as a stable
wind with a constant terminal expansion velocity, in contrast to what
we observe. The velocity gradient in the large waist then suggests a
more complex formation history, perhaps involving underlying
low-latitude or flattened ejections \citep[e.g.,][]{matt04}, continuous
binary mass transfer leading to an equatorially dense circumbinary flow
\citep[e.g.\,via wind Roche-lobe overflow;][]{moh12}, etc.


\subsubsection{The fish bowls}
\label{dis-calamardos}
Perhaps the most enigmatic structures reported in this paper are the two large
($\sim$8\arcsec$\times$6\arcsec) bubble-like features that surround
the central parts of the nebula, which we call the fish bowls 
(Fig.\,\ref{f-calamardos}).  As explained in \S\,\ref{res-calamardos},
the fish bowls are probably thin-walled ellipsoids that are radially
expanding at moderate \los-velocities (|\vlsr-\vsys|$\la$30-35\,\kms)
from their respective centroids.

The fish bowls, also simply referred to as ellipsoids, resemble
inflating bubbles. Their thin-walled structure, manifested as a
notable limb-brightening in the CO maps, suggests that the ellipsoids
mainly consist of swept-up ambient circumstellar material.
As we will show next, the fish bowls are relatively old structures
although probably slightly younger than the large-scale waist and the
HV lobes.



As discussed in \S\,\ref{res-calamardos}, the kinematic age of the
fish bowls cannot be precisely determined because the size and
\los-inclination of the $c$-axis is unknown.  Following the
  principle of Ockham's Razor, a model in which
$c$=$b$$\sim$4\arcsec\ would be preferred (over models with
$c$$\neq$$b$) since fewer free parameters are needed to model the
ellipsoids, which would approach two oblate spheroids.  This is also
consistent with the overall cylindrical symmetry of most nebular
components.  Considering the radial velocity gradient between the
approaching and receding extremes,
a value of $c$$\sim$4\arcsec\ results in a growing time of
\tgro$\sim$730 and 850\,yr for the \nbw\ and the \sbw, respectively
(together with $i_{\rm c}$$\sim$35\degr, see
Fig.\,\ref{f-elipse}-right). This value represents how long ago the
ellipsoids started inflating and it could be an upper limit if the
expansion has slowed down with time (as is perhaps expected as increasing amounts of material are swept-up).




The age of the fish bowls can be estimated in an independent way by
comparing the relative distance between their centers ($\sim$2\farc3)
and their different systemic velocities
($\Delta$\vsys$\sim$22.5\,\kms).
If the offset and the \vsys-difference are both due to the
relative motion of the ellipsoids moving away from each other,
we derive \ttra$\sim$(2\farc3/22.5\kms)$\times$$\tan{(i_{\rm
    NS})}$$\sim$730$\times$$\tan{(i_{\rm NS})}$ years, where $i_{\rm
  NS}$ is the inclination of the relative-velocity vector between the
\nbw\ and the \sbw\ (also referred to as the centroids' line) to the
plane of the sky.  Adopting an average value of $i_{\rm
  NS}$$\sim$35\degr\ (under the premise that the ellipsoids are
moving away from each other along the same direction as the
HV-bipolar lobes), we derive a travel time of \ttra$\sim$510\,yr.

This value is consistent with the upper limit to the growth time of
the ellipsoids deduced above. We find that for the \nbw\ the two
time-scales above, \tgro\ and \ttra, indeed coincide for an
inclination angle of $i_{\rm c}$$\sim$$i_{\rm NS}$$\sim$40\degr\ and
$c$-axis length of $c$$\sim$3\farc8, leading to
\tgro$\sim$\ttra$\sim$650\,yr. For the \sbw, a slightly larger value of
the inclination, $i_{\rm c}$$\sim$45-50\degr, and a smaller value of
$c$$\sim$3\farc5 would be needed to match \tgro\ and \ttra\ to a
common value of $\sim$750\,yr. However, in the case of the \sbw, the
growth time is particularly uncertain since the FWZI of CO and the
|$\overline{RB}$| distance are not well determined due to
contamination with CO emission from other nebular components
(\S\,\ref{res-calamardos}). For this reason, and since the age
difference is marginal, we cannot confidently affirm that the \sbw\ is
older than the \nbw\ and, thus, we will regard the two fish bowls as
approximately coetaneous.



So, within the large uncertainties of these rough estimates of the age,
the spatio-kinematics of the fish bowls and their relative motion
suggests that these started expanding and moving away from each other
about $\sim$650$\pm$100\,yr ago, i.e., presumably some time after the
acceleration of the HV lobes and the large-scale hourglass began.

The origin of the ellipsoids is unclear. They could have
a similar origin to that of the $\sim$500\,yr-old extended lobes
sampled by \hal\ emission, including the small bubble-like structure
inner to the south lobe \citep{san00a}. The velocity field of the
shock-excited \hal-nebula is the composition of a general radial
expansion from the center and a spherical expansion originated inside
of the lobes.
The spherical expansion of the \hal-bubbles is proposed to result from
the adiabatic cooling of the shocked circumstellar gas after a
shock-regime transition from radiative to adiabatic \citep{san00a}.

The centroids of the fish bowls are aligned roughly along the major
axis of the nebula in the sky-plane, which is unlikely to be
coincidental. This could suggest that their formation was triggered by
(or somewhat related to) a ''critical'' time-period or instant of the
evolution or development of the HV lobes. For example, the moment when
the CFWs driving the HV bipolar outflow were switched off or when they
passed across a certain region of the ambient circumstellar medium
with special physical conditions (e.g., through the boundary between a
double-shell environment with an abrupt density fall; see the analytic
study of the formation of inflating bubbles at the head of the jet as
a result of its propagation from a high-density to a low-density
ambient medium by
\citealt{sok02}).

The relative offsets between the ellipsoids are also puzzling. The
centroids of the \nbw\ and \sbw\ are not equidistant to the equator of
the large waist, indeed the centroid of the ellipsoids' expansion
centers is located at about
($-$0\farc35, $+$1\farc15) -- see Figure \ref{f-calamardos}. This
offset could indicate
that the \nbw\ and the \sbw\ move along a slightly different direction
with respect to a common origin. In particular, the \nbw\ would be
travelling in a direction that is closer to the plane of the sky than
the \sbw. Additionally, the two ellipsoids may have not emerged
exactly at the same distance above and below the equator and also they
may not move away from the center at exactly the same velocity. These
differences between the \nbw\ and \sbw\ are not surprising given the
notable asymmetry of the large-scale nebula with respect to the
equatorial plane. As noted in \S\,\ref{dis-qxoffset}, the orbital
and/or recoil motion of the mass-losing star could have had an
additional effect in the observed positions and \vsys\ of the
ellipsoids, maybe partially explaining their global offset to the west
and slightly above the equator of the large waist. The composition of
different velocity vectors when the ellipsoids were formed and when
the different envelope components were ejected complicates enormously
a detailed reconstruction of the nebular formation and shaping history of
\ohs.


\subsubsection{Recent mass-loss history} 
\label{dis-recent}

-- {\sl The SiO outflow.} The compact ($\sim$1\arcsec$\times$4\arcsec)
bipolar SiO outflow that emerges from \cs\ (\S\,\ref{res-sio}) and the
hourglass-shaped shell adjacent to the outflow (i.e.,the mini-hg,
\S\,\ref{res-minihg}) denote recent bipolar mass ejections that
started about 500\,yr ago (or less if strong deceleration has occurred).
The spatio-kinematics of these components suggests that the
SiO outflow (perhaps jointly with an underlying powering CFW not
directly seen in these data) is sculpting the ambient material as it
propagates outwards. The SiO outflow and the neighboring mini-hg
probably sample different adjacent layers of the shell of
shock-compressed material that results from the interaction between a
fast light wind and a slow high-density environment
\citep[e.g.,][]{lee03,san04,bal17}. In this scenario, the mini-hg and
the SiO outflow are both probably made of a mixture (in different
proportions) of ambient material displaced by forward-shocks and fast
wind material compressed by backward shocks.
Given that SiO is a shock-tracer, it is unlikely that
the SiO outflow represents the {\emph unshocked} CFW itself preserved in
a pristine stage.

The two SiO bright knots observed along the axis within $\pm$150\,AU
from \cs\ (\S\,\ref{res-sio}) have kinematical ages of
$\sim$50-80\,yr. The abrupt velocity rise from the center to the
compact SiO knots, where the largest velocity spread is observed,
suggests that they may represent bow-shocks. Since the shock velocity
at the knots is moderate (\vs$\sim$23\,\kms) we do not expect
significant molecular destruction but we do anticipate SiO (and maybe
other refractory materials) to be extracted from dust grains,
consistent with the observations.

The SiO knots, if bow-shocks, could be associated
with internal working surfaces produced by a recent variation of the
speed and/or directionality of the underlying CFW.
A (small) change in the direction of the CFW's axis is consistent with
the fact that the SiO knots do not lie at the tips of the
mini-lobes. It is also accordant with the slightly different (by a few
degrees) sky-plane orientation and \los-inclination of the SiO  and
the CO outflow, and, finally, with the {\tt S}-shaped
emission distribution of the SiO lobes (\S\,\ref{res-sio}). 
There are other large-scale nebular structures in \ohs\ that are
unequally oriented (e.g.\, the two fish bowls), however, there is not
a clear periodic or regular pattern in these large-scale components
that we can relate to a gradual change in the orientation.
This probably denotes episodic (erratic?)
wobbling or swing of the CFWs rather than smooth uninterrupted precession. 

The spatio-kinematics, the location, and the orientation of the
SiO outflow are consistent with this being the large-scale counterpart
(or outermost parts) of the bi-lobe $\pm$90\,AU-long outflowing
structure traced by \water\ masers (\S\,\ref{intro}). In their recent
VLBI study of the \water\ 22\,GHz-maser spots in \ohs, \cite{dod18}
find bulk motions with radial velocities of about 19\,\kms,
leading to an expansion age of 38$\pm$2\,yr. Since their discovery in
1980 by \cite{bow84}, the \water\ 22\,GHz-masers in \ohs\ have
preserved a relatively stable profile, composed on two major peaks
centered near \vlsr$\sim$25 and $\sim$\,40\,\kms, suggesting that the
structure of the underlying bipolar wind on which the maser spots
lie\footnote{we note that \water\ maser spots form only in certain
  specific areas or clumps with suitable pumping conditions.}
has not undergone substantial structural or dynamical changes. This is
consistent with a steady bipolar outflow that has been ongoing for a
period longer than a few decades (and maybe during the last
$\sim$500\,yr) but with at least one time-varying speed and/or orientation
event about $\la$50-80\,yr ago leading to the SiO knots.

-- {\sl The mini-waist.} The equatorial waist of the small-scale
hourglass that surrounds the SiO outflow is another component unveiled
by ALMA with an unclear origin. The are two major aspects that are
difficult to explain in the frame of our current understanding of PN
shaping. One is the presence of a linear velocity gradient along the
equatorial direction, indicative of equatorial flows that originated
at the same time as the polar ejections (i.e., the mini-lobes).  A
single sudden mass-ejection event was originally proposed by
\cite{alc07} to explain a similar structure in the pPN M\,1-92; these
authors speculate on the possibility that the whole nebula was formed
as a result of a magneto-rotational explosion in a common-envelope
system.


Another one is the extremely low expansion velocity observed in
the mini-waist. As is best seen in the axial PV of the CS\,(6-5)
transition (Fig.\,\ref{f-miniw}), the maximum expansion velocity,
which is reached in the outer edge of the waist (\rout$\sim$500\,AU),
is only \vexp$\sim$\,4-5\,\kms.  The velocity is progressively smaller
in regions closer to the center. Although the inner edge of the waist
is not well resolved, we identify a bright inner region at about
$\pm$150\,AU that expands at a remarkably low velocity,
\vexp$\sim$\,1-1.2\,\kms. This value is exceedingly small compared
with the expansion velocity of the inner-wind layers of \qx, within
$\sim$60-100\,AU, where we deduce bulk motions of \vexp=5-7\,\kms, and
up to $\sim$10\,\kms\ (given the full width of the high-excitation
lines that sample these regions, \S\,\ref{res-clumps}).  

We believe that a plausible explanation for the insignificant
expansion velocities in the inner boundary of the mini-waist (even
smaller than, or comparable to, the typical turbulent motions in AGB
and post-AGB envelopes, see, e.e., \citealt{dec18,SanG15}) is that
this extremely slow equatorial outflow lies in, an possibly emerges
from, a stable structure orbiting around a central mass, perhaps
analog to the rotating and expanding disks observed around a number
of post-AGB stars \cite[e.g., the Red Rectangle,][]{buj16} and, to
date, in one AGB star \citep[L$_2$Pup,][]{ker16}. The expansion in
these disks is indeed quite slow, with velocities down to
$\sim$1\,\kms\ in some regions.
The spatio-kinematic structure of the mini-waist is 
not inconsistent with slow rotation, \vrot$\la$1.5\,\kms, in its
innermost regions (in addition to the expansion), however, the spatial
resolution in our maps is insufficient to confirm it. We note that,
for a central mass of 0.6-2.5\,\msun\ and assuming keplerian rotation
we expect rather low values of
\vrot$\sim$(2-4)$\times$$\cos{30\degr}$\,\kms\
at distances of $\sim$150\,AU.
The fact that there is expansion in the mini-waist suggest that
rotation could be in a sub-Keplerian regime, that is, with a steeper
fall of the rotation speed with the distance to the center, as
observed, e.g.\, in the outer disk of L$_2$ Pup, 
where \vrot$\propto$$r^{-0.85}$ \citep{ker16} or 
the Red Rectangle, where \vrot$\propto$$r^{-1}$ \citep{buj16}.

\subsection{Final remarks}

The complex array of small- and large-scale nebular structures in
\ohs\ traced by these ALMA observations indicate that a series of
bipolar mass-ejection events have occurred during the last
$\sim$800-900\,yr.
The origin of the oldest and most prominent bipolar structures remains
unknown, but the presence of much younger bipolar ejections
indicate that the CFW-engine is still active at the core of \ohs.
This fact renders a common envelope (CE) origin improbable, at least,
for the range of input parameters explored by \cite{sta16}.
As show by these authors, once the close companion has been swallowed
by the AGB star (in less than $\sim$30\,yr after the first Roche lobe
overflow mass-transfer event), it is difficult to produce subsequent
jet-launching episodes that are in turn needed to explain the
present-day bipolar ejections.

There are two notable differences between the large-scale
$\sim$800\,yr old CO outflow and the most current ($\la$80\,yr old)
bipolar ejections traced by SiO and \water\ masers
(\S\,\ref{dis-recent}). The moderate expansion velocities of the
present-day bipolar outflows ($\la$35\,\kms) are in marked contrast
with the much larger expansion velocities reached in the massive
HV lobes, of up to $\sim$400\,\kms.
Since the velocity of the outflows must be similar to the escape
velocity of the launching site, the A0\,V companion of \qx\ is
more likely to have blown the $\sim$400\,\kms\ winds (or the
underlying driving CFWs, as already discussed in more detail by
\citealt{san04}).  However, which star is blowing the current
SiO outflow? Is \qx\ launching a bipolar outflow with or without the
assistance of a companion?

Another glaring difference between the SiO  and CO outflow is that the
former is rather symmetric about the equatorial plane, whereas the
south lobe of the large-scale nebula almost doubles the north in size.

These notable differences could indicate that the jet-launching
mechanism itself has changed or that the initial
conditions under which the jets formed (presumably, involving
disk-mediated mass transfer from the AGB to the companion star that
blows the jets) have been modified in such a way that they now
produce slower and more symmetric bipolar ejections than in the
past. These changes may be intimately linked to major adjustments in
the binary system configuration, which may have occurred after the
powerful
ejections that resulted in the formation of the massive and
fast CO outflow.

\section{Summary}
\label{summ}
\begin{figure}[t]
     \centering
     \includegraphics*[bb=70 10 710 500,width=1.015\hsize]{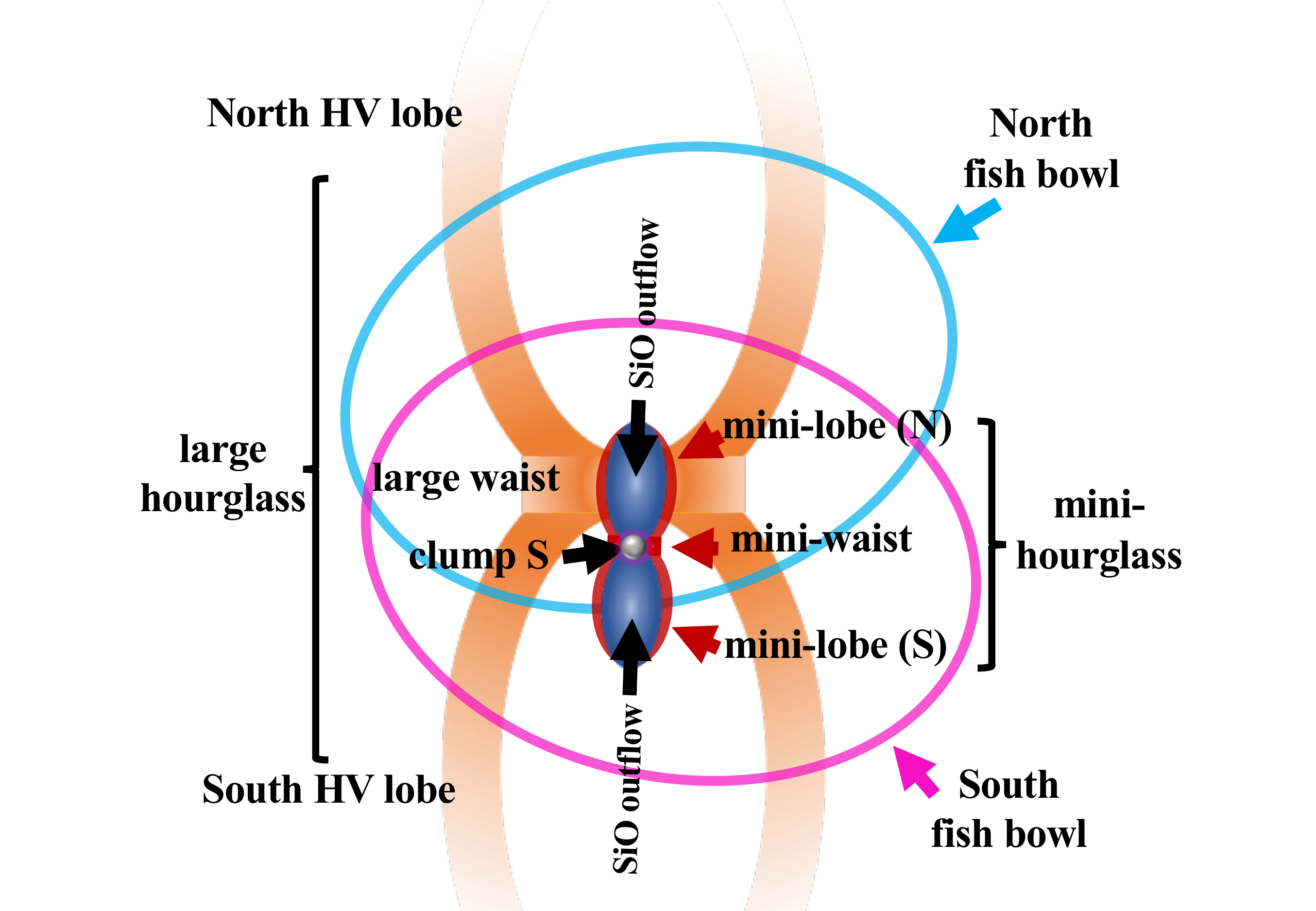}
           \caption{Schematic view of the main nebular components
             traced by ALMA; the outer parts of the high-velocity (HV)
             lobes and the central spine are not represented. A
             summary of the main properties of these structures is
             given in Table\,\ref{t-sketch}. The vertical axis is
             aligned along PA=21\degr.}
         \label{f-sketch}
   \end{figure}

We present new ALMA continuum and molecular line observations in band
7 of the bipolar molecular outflow of the pre-PNe candidate \oh. The
high angular resolution ($\sim$0\farc2-0\farc3) and sensitivity of our
ALMA maps provide the most detailed and accurate description of the
global nebular structure and kinematics of this object to date. We
find an extravagant array of nested (but not always co-axial)
small-to-large scale structures previously unknown (schematically
depicted in Fig.\,\ref{f-sketch} and listed in
Table\,\ref{t-sketch}). Here we offer a summary of our main results:

\begin{itemize}

\item[-] Our continuum maps, at 294, 304, 330 and 344\,GHz, show a
  $\sim$8\arcsec$\times$4\arcsec\ hourglass-like structure with its long axis 
  oriented along PA$\sim$21\degr.\,The peak of the continuum maps has
  coordinates R.A.=07\h42\m16\fs915 and
  Dec.=$-$14\degr42\arcmin50\farc06 (J2000) and marks the position of
  the central star \qx\ (for epoch 2016.6). The latter is 
  enshrouded within a compact ($\la$60\,AU) clump of dust (and gas)
  referred to as \mbox{\sl clump S}.

  \item[-] Unexpectedly, \cs\ does not lie on the equatorial plane of
    the hourglass but it is clearly displaced by $\sim$0\farc6 to the
    south along the nebular axis. 

  \item[-] Except for \cs, the continuum follows a \snu{3.3} power-law
    throughout the hourglass. This implies a dominant contribution by
    optically thin $\sim$75\,K dust with an emissivity index of
    $\alpha\sim$1.3, consistent with previous studies.  At \cs, the
    continuum obeys a \snu{2.1} dependence consistent with a
    population of warm (\td$\sim$300-400\,K), large
    ($>$100\,$\mu$m-sized) grains in the vicinity of \qx.


\item[-] The close surroundings of \qx\ (\cs) are sampled by
  high-excitation transitions of \nacl, SiS, and \tsio, which show
  very compact emission (implying a small emitting volume of radius
  $\sim$60\,AU) and narrow profiles consistent with low expansion
  velocities of \vexp$\sim$5-7\,\kms. 

\item[-] We discover a compact ($\sim$1\arcsec$\times$4\arcsec)
  bipolar outflow that emerges from \cs\ and that is exclusively traced by
  \siot\ and \vsiot. This outflow is oriented similarly to the
  large-scale nebula but is significantly more symmetric about the
  equatorial plane and slower. The observed radial velocity increases
  abruptly from the center to two bright and compact regions located
  at offsets $\delta$y$\sim$$\pm$0\farcs11 about \cs, where the
  maximum radial velocities ($\sim$35\,\kms) and full line widths
  ($\sim$23\,\kms) are observed. These regions may represent
  bow-shocks (viewed at a sky-plane inclination angle of
  $i$$\sim$30\degr) denoting recent bipolar mass ejections.

\item[-] We deduce short kinematical ages for the SiO outflow, ranging
  from \tdyn$\sim$50-80\,yr, in the inner regions ($\sim$150\,AU), to
  \tdyn$\sim$400-500\,yr at the tips ($\sim$3500\,AU). This outflow
  seems to be the large scale counterpart of the \water\ maser,
  $\sim$38\,yr-old bipolar flow seen at $\la$100 milliarcsecond scales.

\item[-] Adjacent to the SiO outflow, we identify a small-scale
  hourglass-shaped structure ({\sl mini-hourglass}) that is probably
  made of compressed material formed as the SiO outflow
  penetrates the dense, central regions of the nebula. The lobes and
  the equatorial waist of the mini-hourglass \emph{both} are radially
  expanding with a constant velocity gradient (\vexp$\propto$\,$r$).
  The dimensions of the mini-waist are consistent with a torus-like
  structure with an outer radius of about 500\,AU orthogonal to, and
  coeval with, the SiO outflow. The expansion velocity at the inner
  edge of the mini-waist ($\sim$150\,AU) is extremely low,
  \vexp$\sim$1\,\kms, which tentatively suggest the presence of a
  stable (orbiting?)  structure.

  \item[-] The high-velocity, $\approx$0\farc1\,pc-long bipolar lobes ({\sl
    HV lobes}) and the dense equatorial waist ({\sl large waist})
  known from previous works are highly structured, specially the
  former. The sharp outer boundary of the large waist is now
  accurately determined (at a radius of $\sim$2700\,AU) as well as the
  internal velocity field, indicating that the waist was shaped nearly
  simultaneously with the HV lobes, about 800-900\,yr ago.

  \item[-] We report the discovery of two large
    ($\sim$8\arcsec$\times$6\arcsec) faint bubble-like structures
    surrounding the central parts of the nebula ({\sl
      fish bowls}). The fish bowls, which are probably made of
    swept-up ambient circumstellar material, are relatively old
    structures although probably slightly ($\sim$100-200\,yr) younger
    than the large-scale waist and the HV lobes. The origin of these
    enigmatic structures is unknown, but it may be linked to a
    critical time period or instant of the evolution or development of the
    HV lobes. 

\item[-] As part of the data analysis, and using the two CS
  transitions observed with ALMA in this work ($J$=6-5 and $J$=7-6),
  we have created velocity-channel maps of the rotational temperature,
  CS column density and total (H$_2$) number density in the hourglass
  nebula of \ohs\ (\S\,\ref{s-rd}). We find large densities in the
  walls of the lobes (\dens$\approx$10$^5$\,\cm3) and higher, as
  expected, in the equatorial waist of the nebula where the lobes
  join (\dens$\ga$10$^6$\,\cm3). The lobes are predominantly cold
  (with \trot$\sim$20-40\,K). The warmest molecular gas (with
  \trot$\approx$100\,K) is found in the dense, central nebular parts.
  
\item[-] We report first detections of \nacl\ and \metanol\ in \ohs,
  with \metanol\ being also a first detection in an AGB star.
  High-excitation lines of \nacl\ and other refractory species,
  namely, SiS and \tsio, arise in the inner-wind regions of
  \qx\ (\cs), where we estimate fractional abundances of
  \mbox{X(\nacl)$\approx$[1-3]\ex{-9}}, X(SiS)$\approx$[2-3]\ex{-7},
  and X(\tsio)$\approx$[0.5-1.5]\ex{-7} (\S\,\ref{abun-refr}). The
  \metanol\ emission arises solely from the bipolar outflow,
  predominantly from the periphery of the compact bipolar outflow
  traced by SiO, where the fractional abundance is
  X(\metanol)$\approx$10$^{-8}$-10$^{-7}$ (\S\,\ref{abun-ch3oh}).

\item[-] The position of \qx\ 
  off-centered from the waist of the large-hourglass is one of the
  most puzzling discoveries from our ALMA data. Perhaps the
  combination of orbital-motion and recoil of the binary system after
  strong asymmetrical mass ejections could explain, at least
  partially, its mysterious location. 

\item[-] The orbital parameters of the central binary system are very
  poorly known. A loose upper limit to the orbital separation of
  $a$$<$150\,AU is deduced from the relative separation between
  \cs\ and the centroid of the SiO outflow (presumably collimated by
  the companion). We present some arguments in favour of a smaller
  orbital separations, $a$$<$35\,AU. 
  
\item[-] Our ALMA observations unveil a series of new substructures
  that point to a nebular formation and shaping history significantly more
  complex than previously thought, in particular, indicative of
  multiple non-spherical mass ejections. The speed and directionality
  of the driving CFWs seem to have changed over time, although the
  lack of periodic or regular pattern in the large-scale components
  probably denotes episodic (erratic?) wobbling or swing of the CFWs
  rather than smooth uninterrupted precession.

\item[-] The origin of the bipolar ejections that led to the formation
  of the outstanding nebular architecture of \oh\ remains unknown, but
  the presence of present-day bipolar ejections indicate that the CFW
  engine is still active at its core. This is hardly reconcilable with
  a common envelope ejection scenario, at least in a simplified
  version of it, where the AGB star has already swollen a compact
  companion after the premier ejection of the fast collimated
  winds that produced the large-scale CO outflow $\sim$800-900\,yr
  ago.

\end{itemize}

\begin{acknowledgements}
We are grateful to the referee of this paper, Joel Kastner, for his
detailed report and very useful remarks and suggestions to improve the
manuscript. We thank Bruce Balick for fruitful discussions on winds
interaction hydrodynamics. This paper makes use of the following ALMA
data: ADS/JAO.ALMA\#2015.1.00256.S.  ALMA is a partnership of ESO
(representing its member states), NSF (USA) and NINS (Japan), together
with NRC (Canada), NSC and ASIAA (Taiwan), and KASI (Republic of
Korea), in cooperation with the Republic of Chile. The Joint ALMA
Observatory is operated by ESO, AUI/NRAO and NAOJ.  The data here
presented have been reduced by CASA (ALMA default calibration
software; {\tt https://casa.nrao.edu}); data analysis was made using
the GILDAS software ({\tt http://www.iram.fr/IRAMFR/GILDAS)}. This
work has been partially supported by the Spanish MINECO through grants
AYA2012-32032, AYA2016-75066-C2-1-P, and AYA2016-78994-P and by the
European Research Council through ERC grant 610256: NANOCOSMOS.  This
research has made use of the The JPL Molecular Spectroscopy catalog,
The Cologne Database for Molecular Spectroscopy, the SIMBAD database
operated at CDS (Strasbourg, France), the NASA’s Astrophysics Data
System  and Aladin.

\end{acknowledgements}

%

%

%

   %

%
%

%
%
   %
   %
   %
   %

   %
   %

%
%


\begin{appendix}
\section{ALMA vs. IRAM\,30m line profiles}
\label{res-30m}

   \begin{figure*}[htbp]
   \centering 
   \includegraphics[width=0.33\hsize]{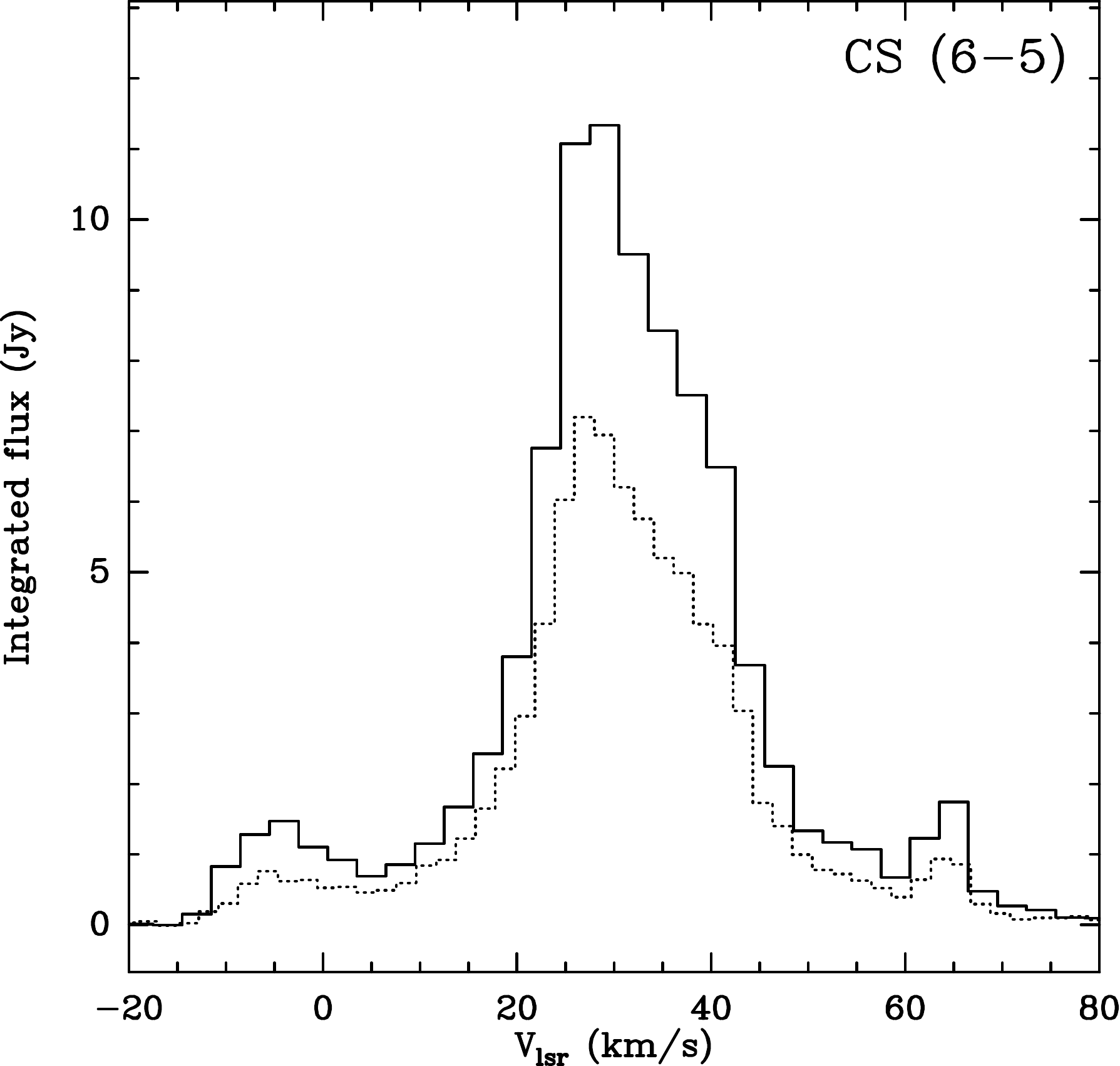}
   \includegraphics[width=0.33\hsize]{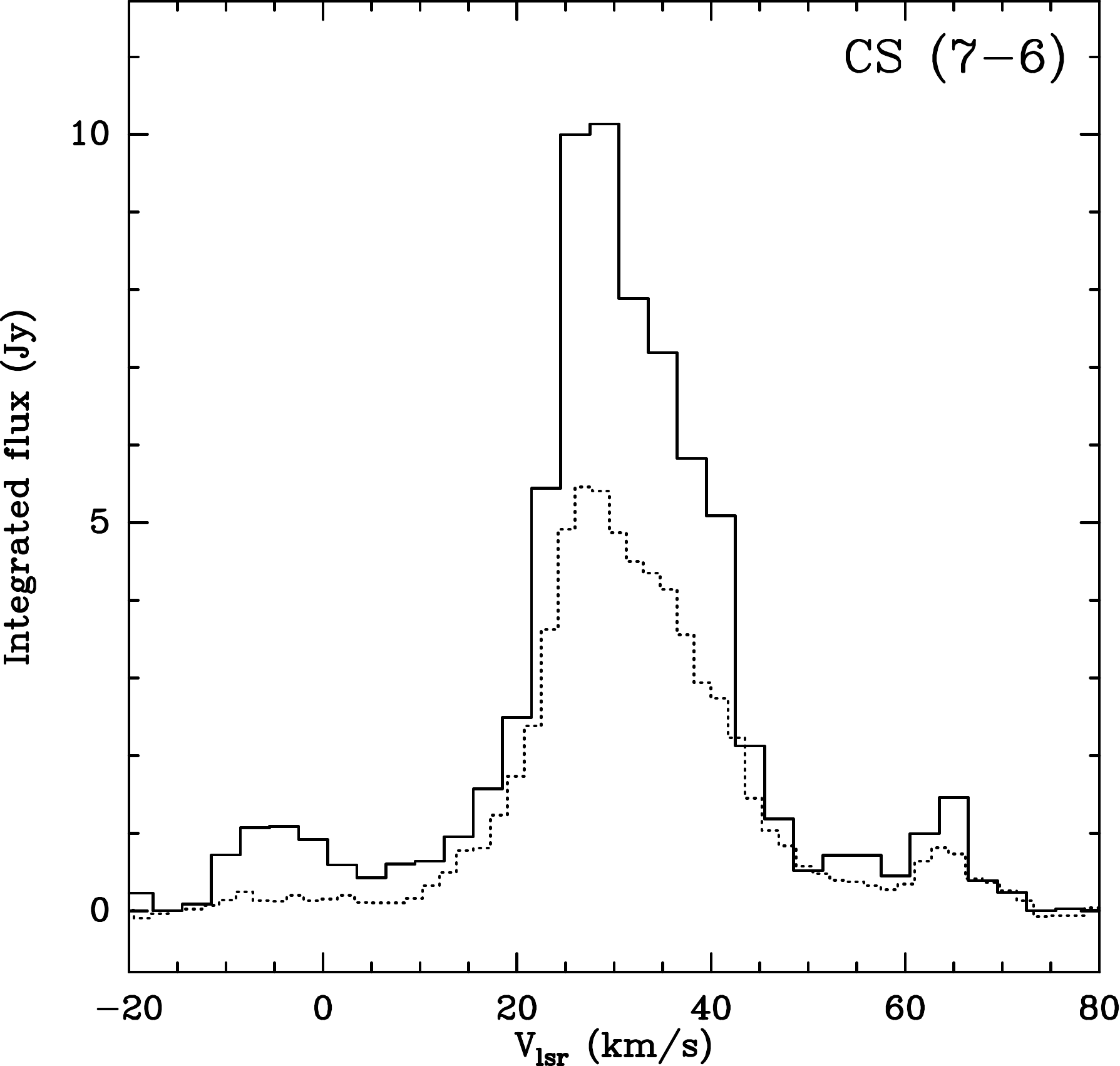}
   \includegraphics[width=0.33\hsize]{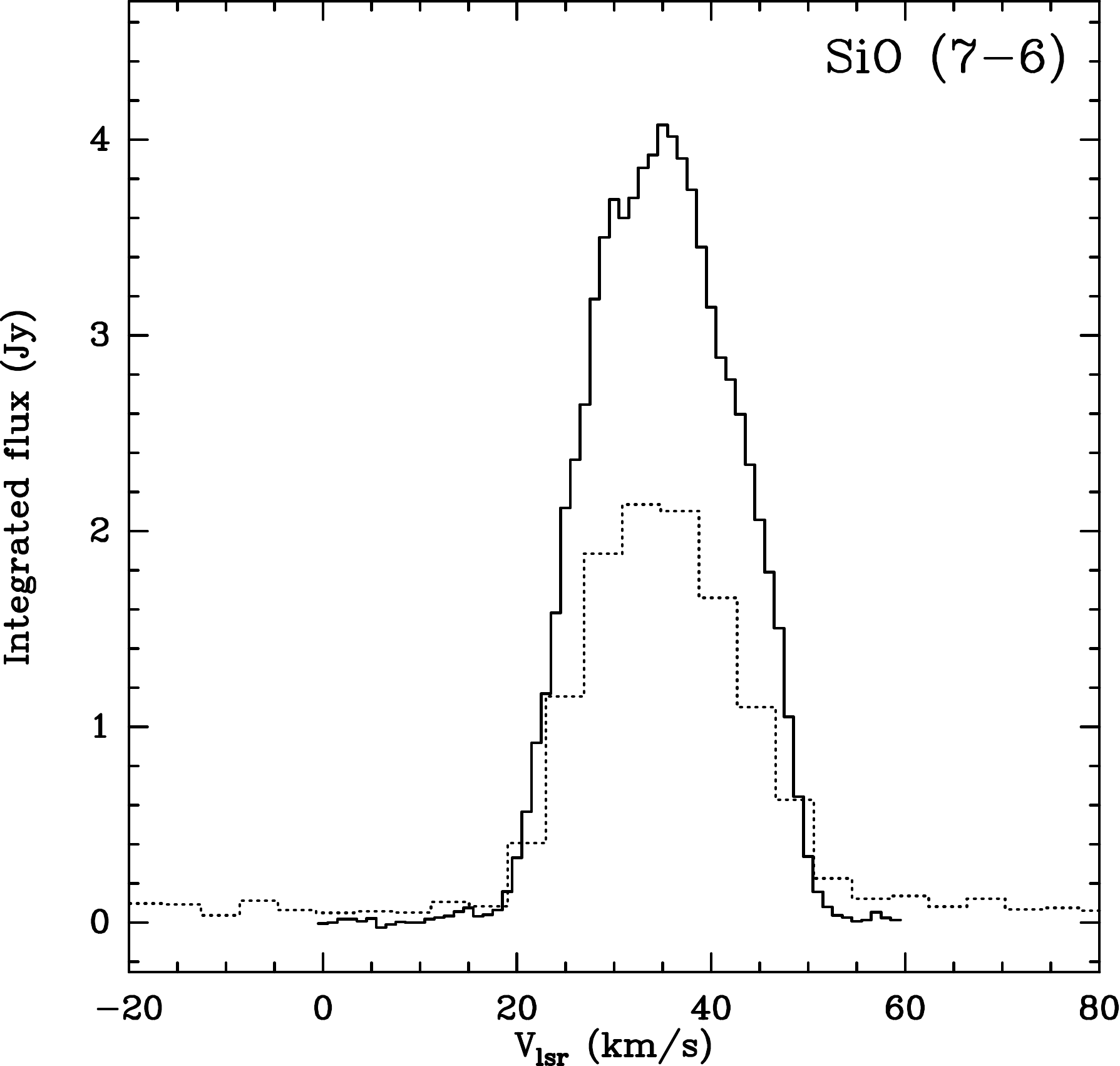}
   \includegraphics[width=0.33\hsize]{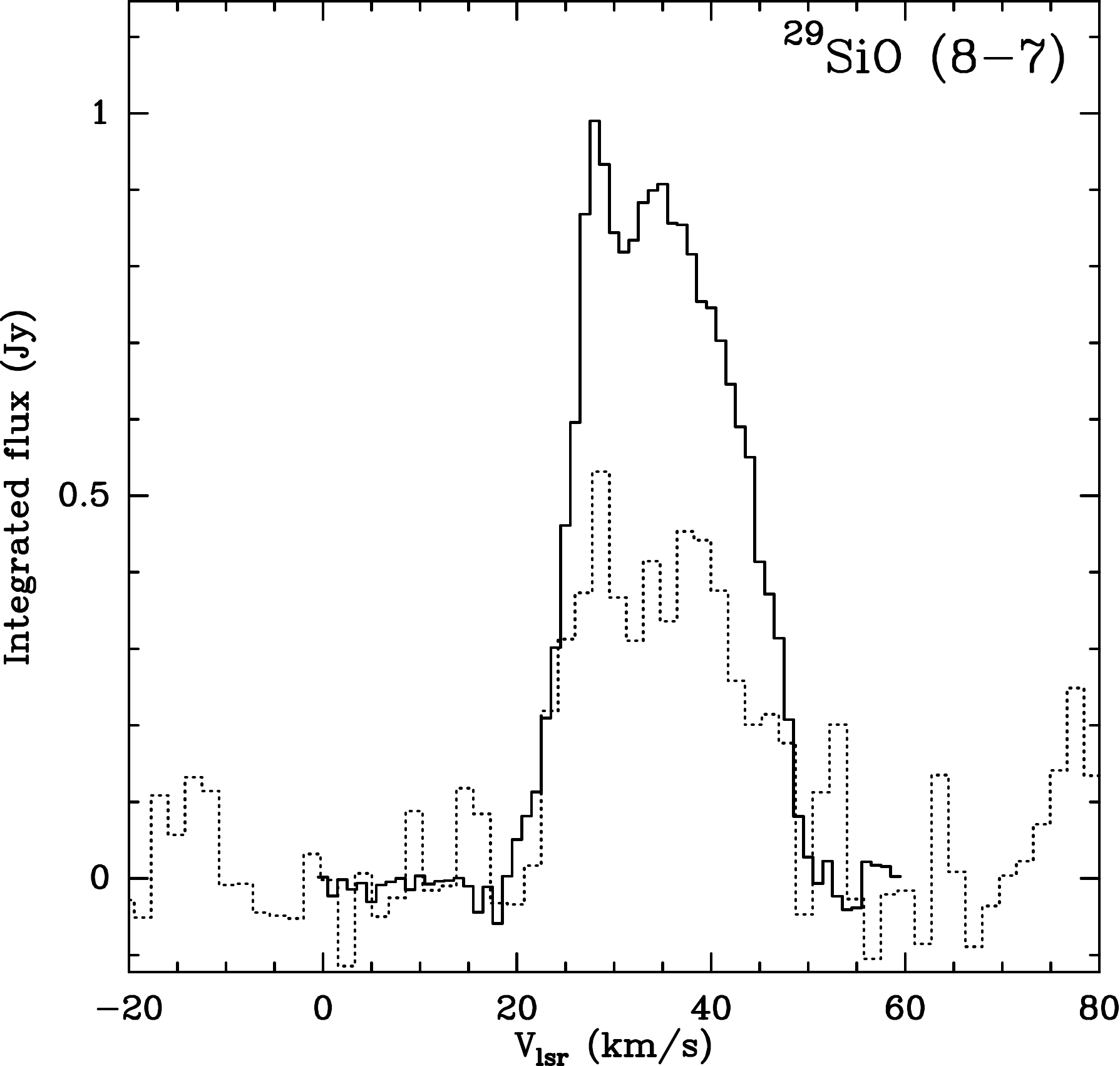}
   \includegraphics[width=0.33\hsize]{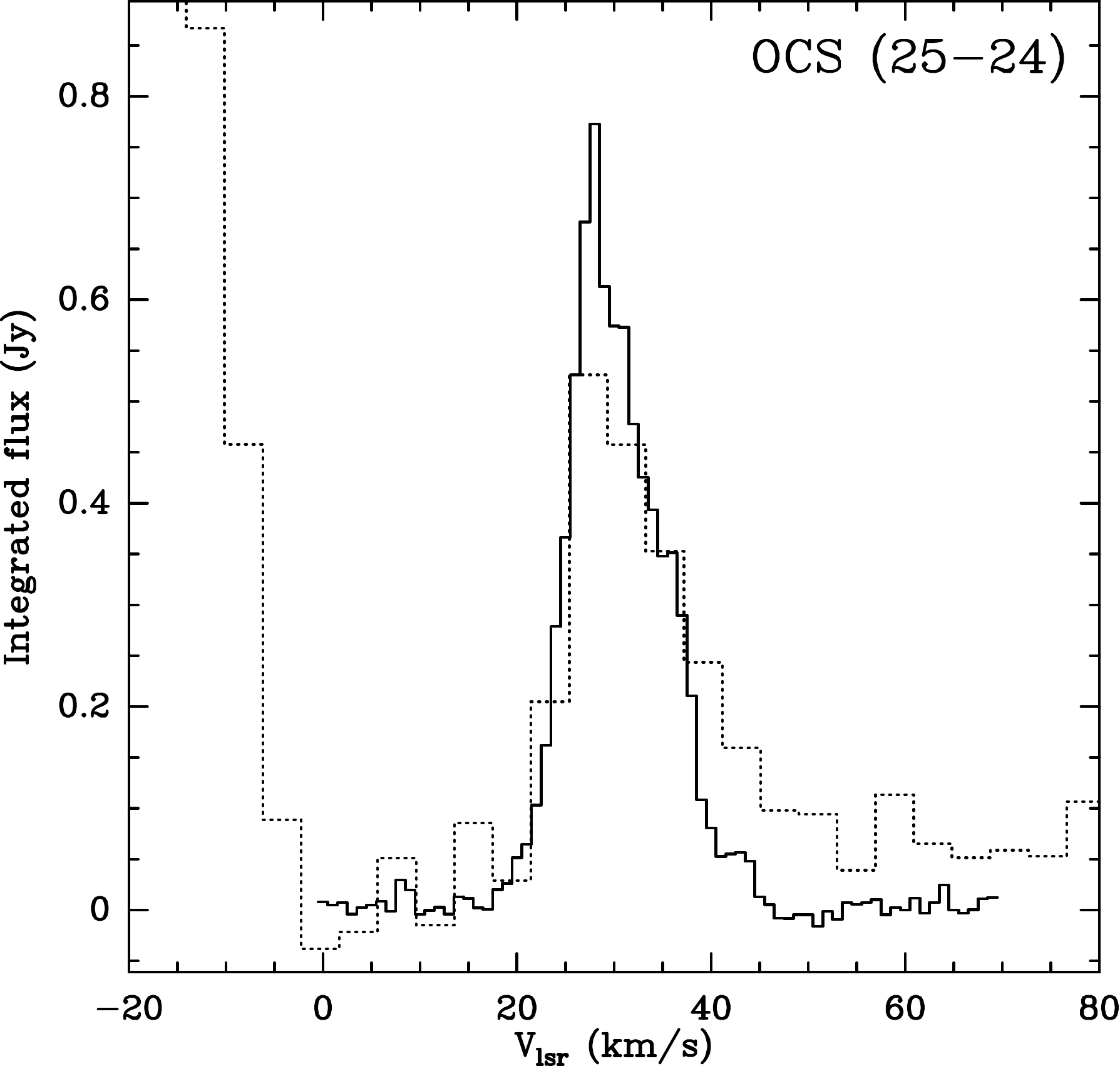}
   \includegraphics[width=0.33\hsize]{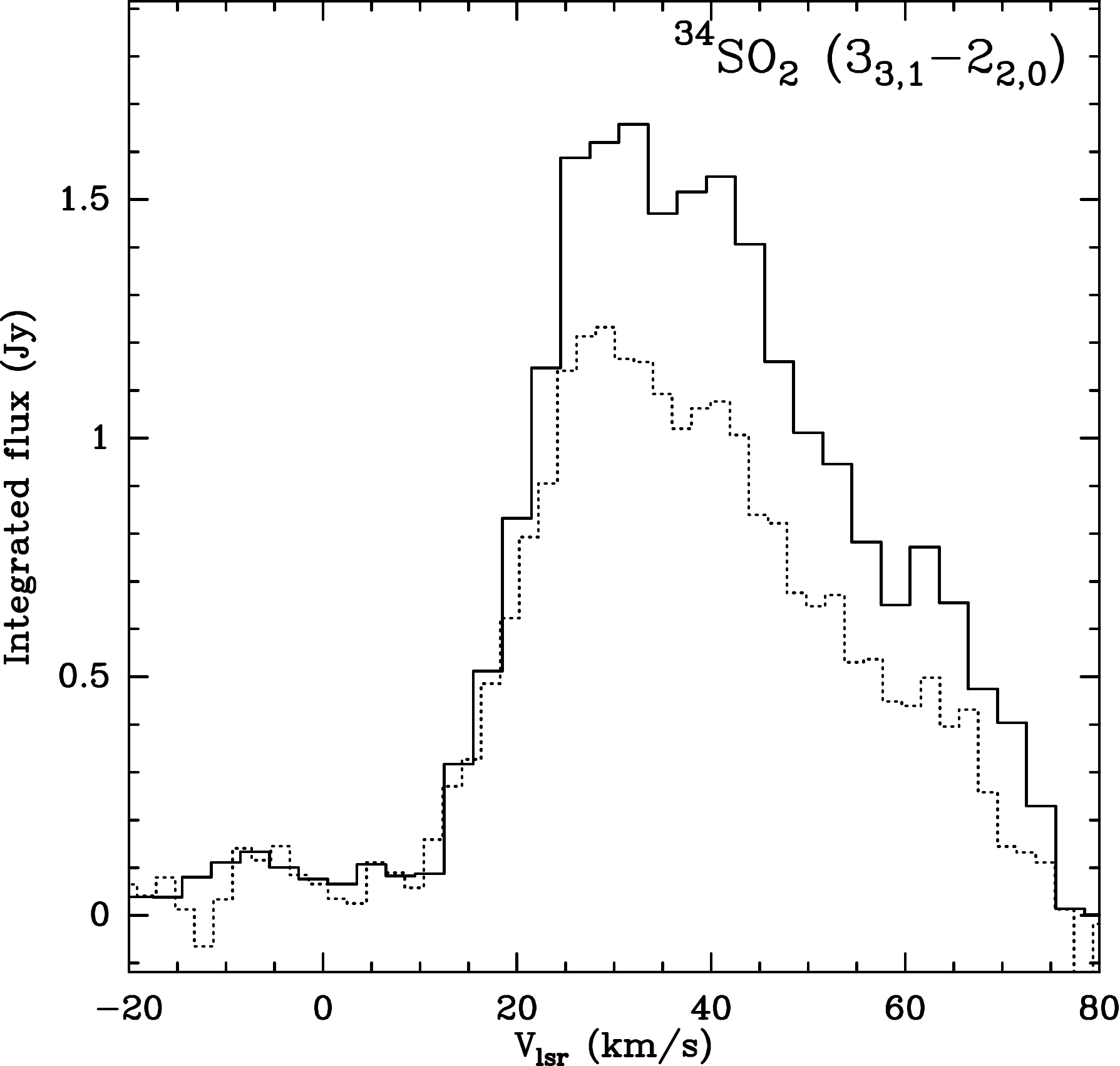}

   \caption{Comparison of ALMA and \iram\ single-dish spectra (solid
     and dotted lines, respectively). See \S\,\ref{floss}.}
         \label{f-30m}
   \end{figure*}
%

In Fig.\,\ref{f-30m}, we show the profiles of representative molecular
transitions mapped with ALMA (integrated over the emitting region)
together with single-dish spectra obtained with the IRAM\,30m
telescope as part of our mm-wavelength spectroscopic survey of
\ohs\ (\citealt{vel15}, \citealt{san15}, and Velilla Prieto et al., in
prep.).

The half-power beam width (HPBW) and point source sensitivity
($S$/\ta, i.e., the K-to-Jy conversion factor) of the \iram\ data are
described to a good accuracy as a function of the frequency by
HPBW(\arcsec)=2460/$\nu$[GHz] and $S/$\ta=5.44+($\frac{\nu[{\rm
      GHz}]}{147.148})^2$, respectively, according to measurement
updates performed in August 2013\footnote{\tt
  http://www.iram.es/IRAMES/mainWiki/EmirforAstronomers}. Additional
observational details on the IRAM\,30m data are provided in the
references above.

\section{Velocity-channel maps}
\label{ap-extramaps}

\begin{figure*}[htpb]   
   \centering 
   \includegraphics[width=0.95\hsize]{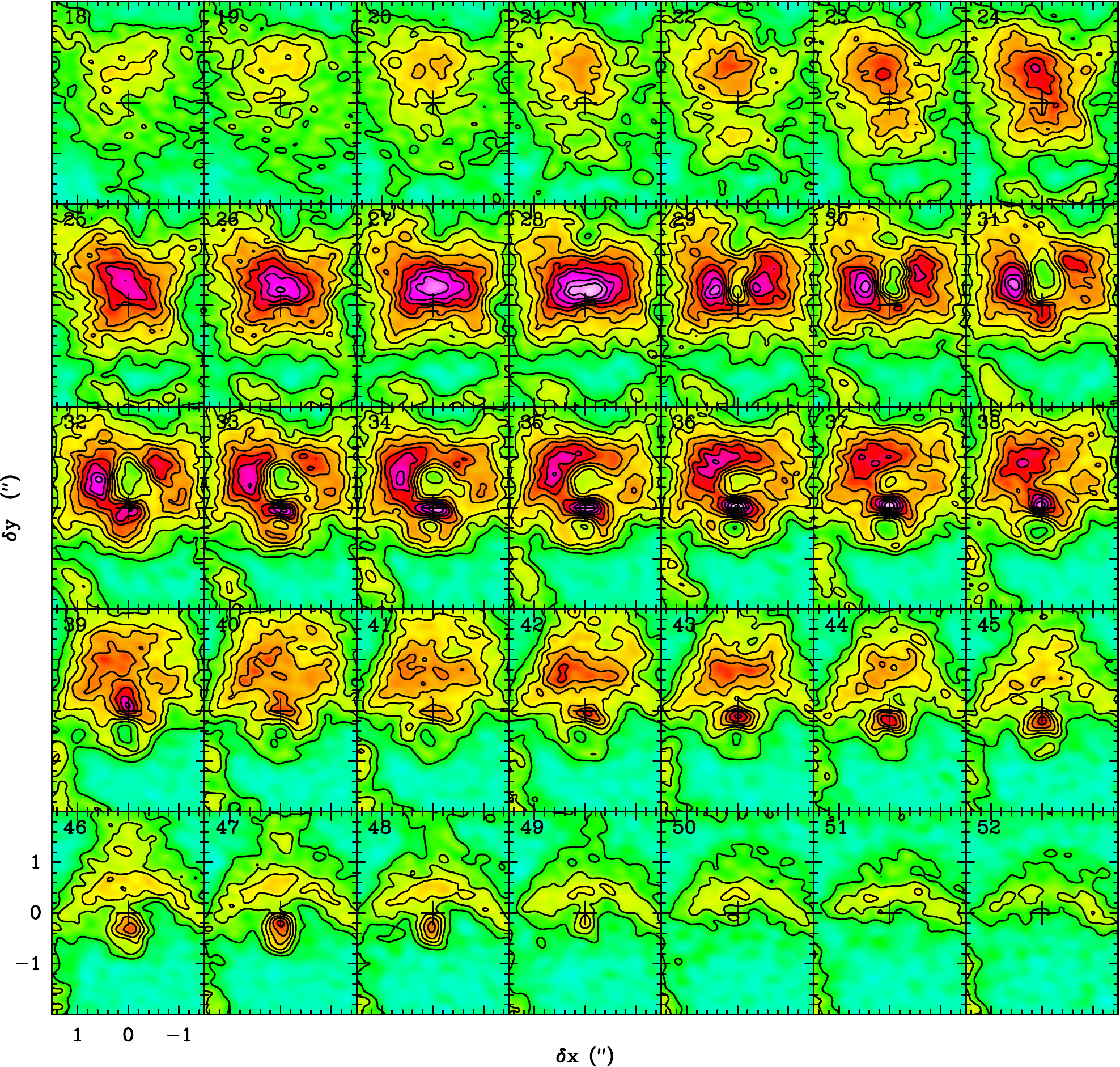}
   \caption{Velocity-channel maps of the \trecet\ transition for the
     same \vlsr\ range as in Fig.\,\ref{f-pv-SiO-13co}. Beam size is
     HPBW=0\farc29$\times$0\farc23 (PA=$-$83\degr) and the level
     spacing is 40\,mJy/beam.
   }
   \label{f-13co}
   \end{figure*}

\begin{figure*}[htpb]
   \centering 
   \includegraphics[width=0.95\hsize]{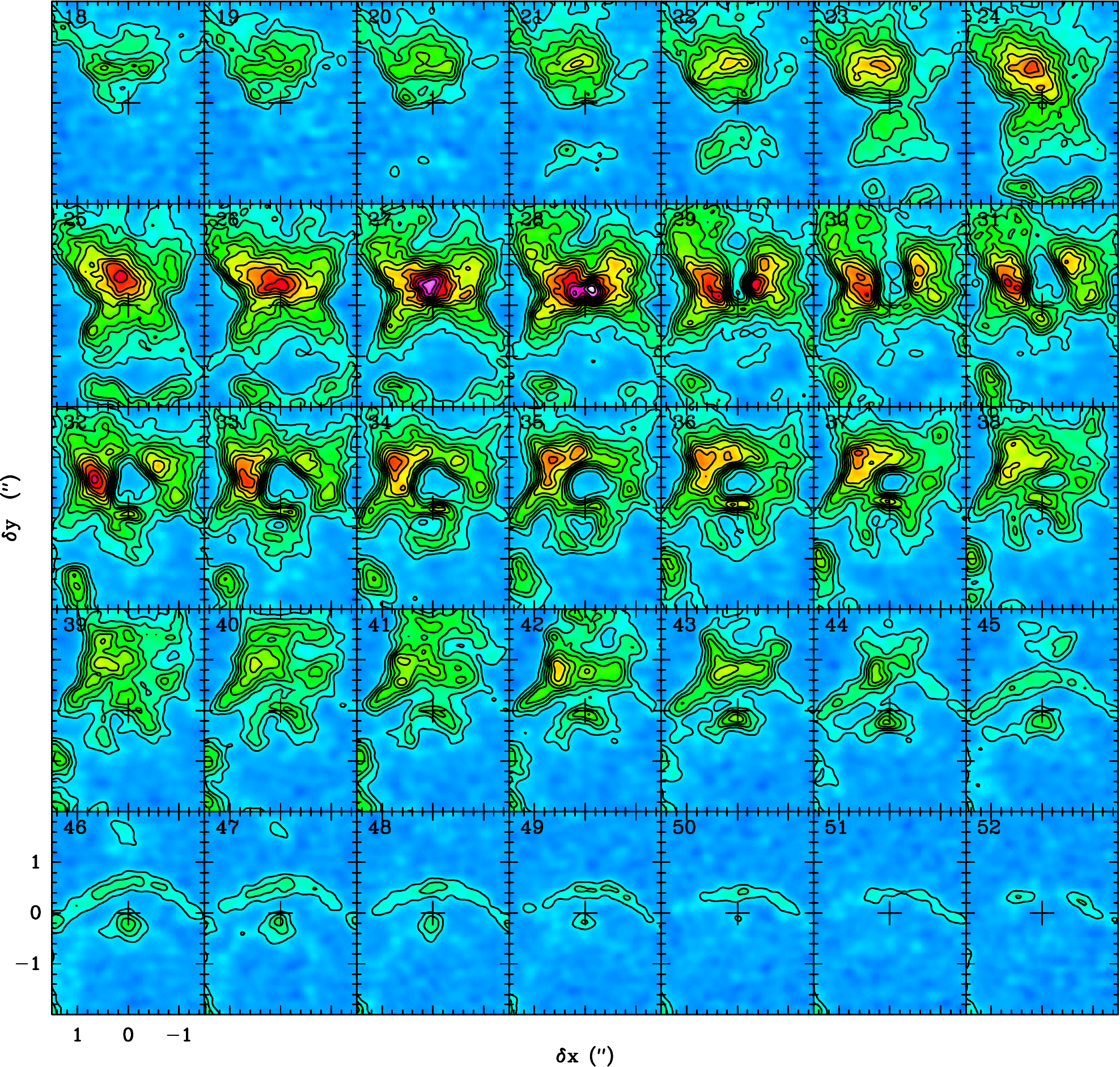} 
   \caption{Velocity-channel maps of the CS\,(6-5) transition for the
     same \vlsr\ range as in Fig.\,\ref{f-pv-SiO-13co}. Beam size is
     HPBW=0\farc24$\times$0\farc20 (PA=$-$78\degr) and the level
     spacing is 15\,mJy/beam.
   }
   \label{f-cs65}
   \end{figure*}

\begin{figure*}[htbp] 
   \centering 
   \includegraphics[width=0.95\hsize]{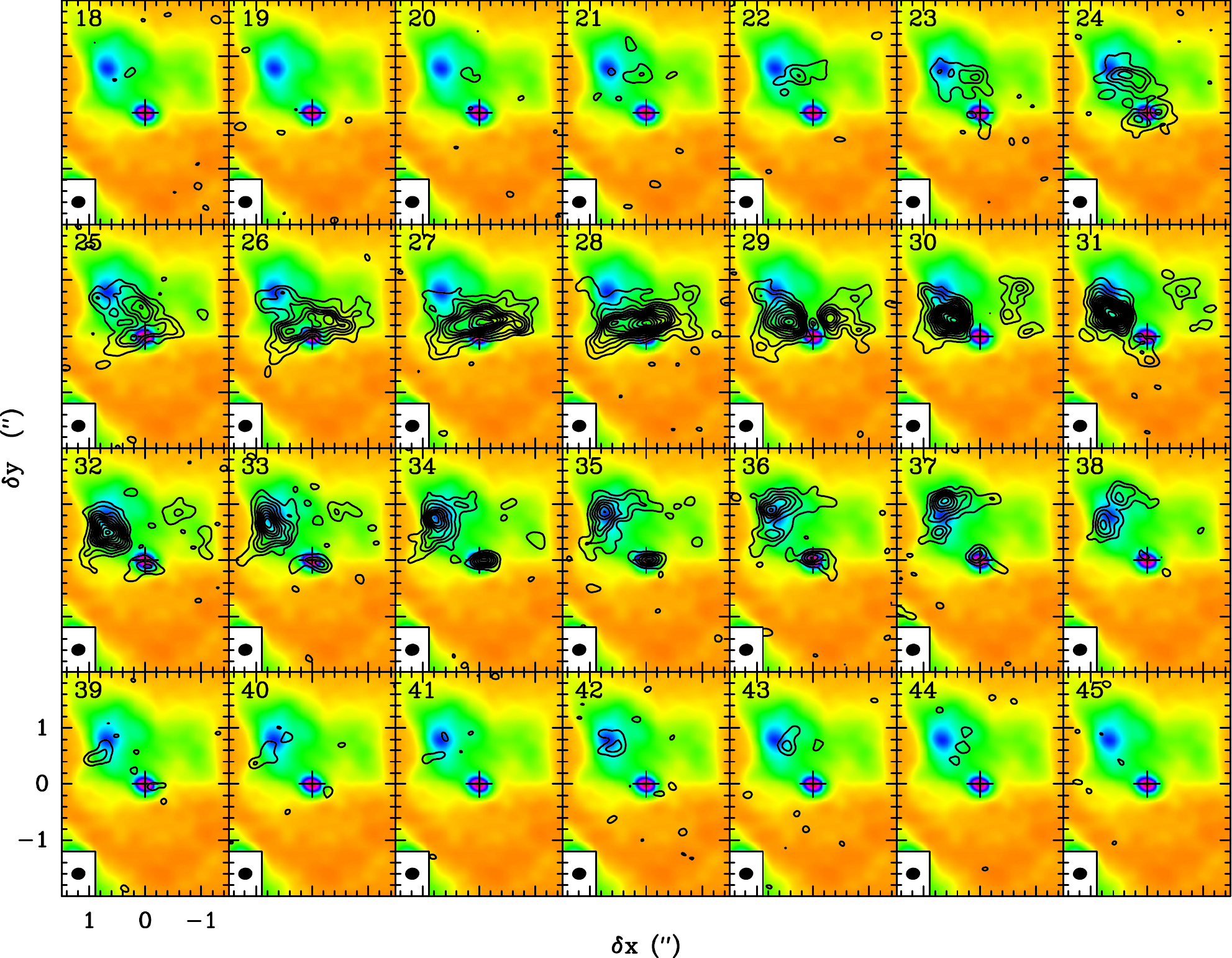} 
   \caption{Velocity-channel maps of OCS\,($J$=25-24) atop the
     330\,GHz-continuum map (contours and color, respectively). The
     beam has HPBW=0\farc22$\times$0\farc18 (PA=$-$79\degr) and level
     spacing is 5\,mJy/beam.}
   \label{f-ocs}
   \end{figure*}

Here we show velocity-channel maps of some the molecular transitions
discussed in the main body of the paper that further help visualizing
some of the nebular components of \ohs\ unveiled by our ALMA
observations. In particular, \trecet\ and CS\,(6-5) emission maps of
the central (3\arcsec$\times$4\arcsec) regions in the LSR velocity
range where the small-scale hourglass that surrounds the compact SiO
outflow is showing up (\vlsr=[18-52]\,\kms) -- Fig.\ref{f-13co} and
Fig.\ref{f-cs65}, respectively. Velocity-channel maps of the
OCS\,(25-24) transition in a similar (somewhat narrower) \vlsr\ range
are plotted in Fig.\,\ref{f-ocs}. This transition selectively traces
the dense equatorial waist of both the large-scale and the small-scale
hourglass nebula (dubbed as the large hg and mini-hg,
respectively). See Sections \ref{res-minihg} and \ref{res-largehg} for additional details on
these nebular components.

%

%

%

\end{appendix}

\end{document}